\documentclass[aps, prl, twocolumn, superscriptaddress, longbibliography]{revtex4-2}
\usepackage[colorlinks, linkcolor=blue, anchorcolor=blue, citecolor=blue]{hyperref}
\usepackage{amsmath}
\usepackage{amssymb}
\usepackage{graphicx}
\usepackage{braket}
\usepackage{multirow}
\usepackage{color}
\usepackage{soul}
\usepackage{diagbox}
\usepackage{rotating}
\usepackage{gensymb}
\usepackage{makecell}

\usepackage{ulem}%only for the command \sout = scrap
\begin{document}

\title{Many-Body Amplification of Ligand Instabilities Driving Verwey-Type Transitions in Altermagnet CsCr$_2$S$_2$O}
\author{Xiuhua Chen}
\affiliation{School of Emerging Technology, University of Science and Technology of China, Hefei 230026, China}
\affiliation{Hefei National Laboratory, Hefei 230088, China}
\affiliation{New Cornerstone Science Laboratory, University of Science and Technology of China, Hefei, 230026, China}

\author{Yilin Wang}\email{yilinwang@ustc.edu.cn}
\affiliation{School of Emerging Technology, University of Science and Technology of China, Hefei 230026, China}
\affiliation{Hefei National Laboratory, Hefei 230088, China}
\affiliation{New Cornerstone Science Laboratory, University of Science and Technology of China, Hefei, 230026, China}

\date{\today}

\begin{abstract}
Altermagnet CsCr$_2$S$_2$O undergoes a Verwey-type metal-to-insulator transition (MIT), accompanied by a lattice distortion and stripe charge order on the Cr sublattice. Intriguingly, the atomic distortions occur exclusively at the ligand sites rather than the Cr sites, leaving the origin of the pronounced Cr charge disproportionation unresolved. Using density functional theory plus dynamical mean-field theory (DFT+DMFT) calculations, we identify an orbital-selective Mott state in which the correlated metallic $d_{yz}$ orbital governs the low-energy physics. We show that S-site distortions produces only a small bare charge asymmetry between the Cr sublattices through Cr-$d_{yz}$--S-$p$ hybridization. Dynamical electronic correlations, however, dramatically amplify this asymmetry, producing pronounced differentiation in both the charge occupancy and correlation strength of the $d_{yz}$ orbitals. The resulting site-dependent spin polarization ultimately drives the MIT. In contrast, replacing S with Te substantially weakens this correlation-amplification effect and preserves a metallic state. Our work reveals a hidden, correlation-amplified feedback loop between ligand instabilities and electronic symmetry breaking in this altermagnetic family, highlighting ligand engineering as an effective route toward robust metallic altermagnetism.
\end{abstract} 

\maketitle

\textit{Introduction.} Altermagnets (AMs) have rapidly emerged as a frontier of condensed matter physics, combing vanishing net magnetization with pronounced non-relativistic spin splitting in momentum space~\cite{MnO2_DFT,2019_THEORY,V2Se2O_theory,theory_RbVTeO,theory_prx_1,theory_prx_2,Anomalous_Hall,review_am,largescale_DFT,review_yao,Landa_Theory_of_Altermagnetism,theory_AM,Magnetic_geometry}. This unconventional form of magnetic symmetry breaking has stimulated broad interest in spintronics~\cite{Electricalspinsplitter,GMR_theory,GateFieldControl,180switchingneelvector,altermagneticorder,controlSST}, multifunctional materials~\cite{ZhouTong:2025,LiuQihang:2025,MaYanming:2025} and unconventional superconductivity~\cite{NSM,TSC_1,FFLO,FFLO_3,TSC_2,FFLO_2,pwavesc,FFLO_4,TSC_3}. Guided by spin space groups~\cite{spingroup1,spingroup2,spingroup3,spingroup4,spingroup5,spingroup6,spingroup_lqh,luo2026spingroup_1,luo2026spingroup_2}, numerous candidate materials have been proposed and experimentally explored~\cite{ruo2_theory,ruo2_anomalous_Hall,Transport_ruo2,ruo_arpes_2,ruo_arpes_1,ruo2_Magnetoresistance,ruo2_Transport,theory_RbVTeO,KV2Se2O_ARPES,RbVTeO_arpes,CsVTeO,KVSeO_stm,KVSeO_stm2,CsV2Se2O_stm,SanDongGuo:2026,CrSb_arpes_3,CrSb_arpes_1,CrSb_arpes_2,CrSb_arpes_4,CrSb_arpes5,MnTe_arpes_3,MnTe_arpes_1,MnTe_arpes_2,MnTe_arpes_4,MnTe_arpes5,PPEM_MnTe,komoto:1963,Mandal:2025}. Among them, the quasi-two-dimensional \textit{AT}$_2$\textit{X}$_2$O family (\textit{A}=K, Rb, Cs; \textit{T}=transition-metal ions, \textit{X}=S, Se, Te)~\cite{V2Se2O_theory,theory_RbVTeO,SanDongGuo:2026} is particularly attractive because metallic AM is desirable for transport-based spintronic functionalities and for engineering Majorana states~\cite{NSM,TSC_1,FFLO,FFLO_3,TSC_2,FFLO_2,pwavesc,FFLO_4,TSC_3}. Indeed, room-temperature metallic AM with pronounced spin splitting has recently been established in KV$_{2}$Se$_{2}$O~\cite{KV2Se2O_ARPES}, Rb$_{1-\delta}$V$_2$Te$_2$O~\cite{RbVTeO_arpes} and Cs$_{1-\delta}$V$_2$Te$_2$O~\cite{Liuyang:2025}. 

\begin{figure*}
    \centering
    \includegraphics[width=1.0\textwidth]{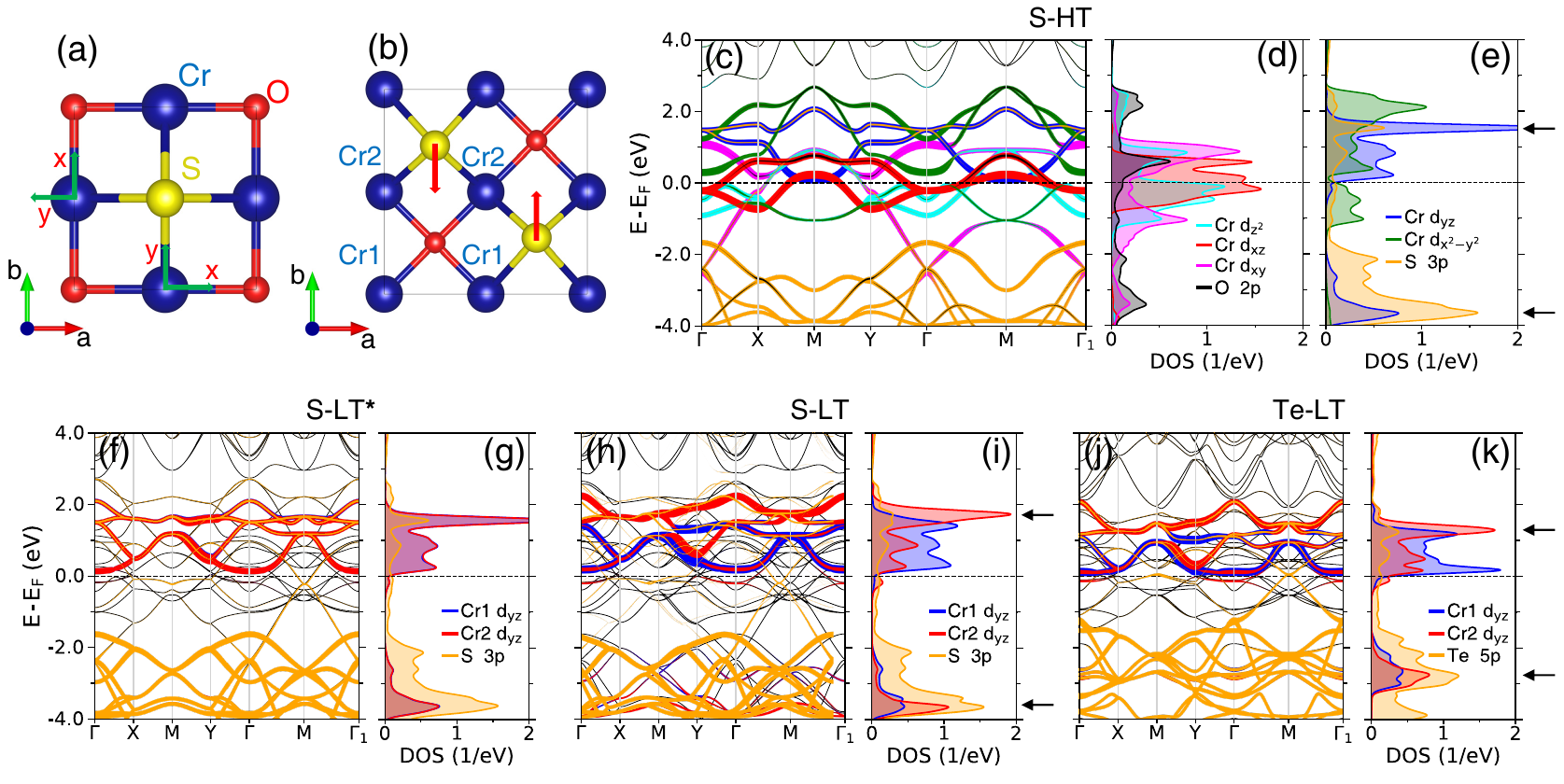}
    \caption{(a),(b) Top view of the high-temperature tetragonal and low-temperature orthorhombic structures of CsCr$_2$S$_2$O, respectively. The green arrows indicate the local $xy$-axis, and the red arrows illustrate the S-site distortions within $ab$-plane. (c)-(k) DFT calculated band structure and density of states in the nonmagnetic state for the S-HT, S-LT$^*$, S-LT and Te-LT structures, as defined in the main text.
     }
    \label{fig:dftband}
\end{figure*}

Recently, CsCr$_2$S$_2$O has been successfully synthesized by Cao's group~\cite{Caoguanghan:2026}, providing a unique platform in which altermagnetism coexists with multiple intertwined electronic and structural orders. Upon cooling, the system undergoes an altermagnetic (C-type AFM) transition at $T_{\text{N}}$ = 326 K, followed by a coupled tetragonal-to-orthorhombic structural transition and metal-to-insulator transition (MIT) at $T_{\text{MIT}}$ = 305 K. The latter is accompanied by a stripe charge ordering that differentiates the nominal Cr$^{2.5+}$ sites into two inequivalent sublattices, Cr1 and Cr2, with estimated valences of Cr$^{2.22+}$ and Cr$^{2.81+}$, respectively--a behavior reminiscent of the Verwey transition in Fe$_3$O$_4$~\cite{VERWEY:1939,Senn:2012}. Remarkably, despite this substantial charge disproportionation, X-ray diffraction reveals essentially no structural displacement of the magnetic Cr ions; instead, the lattice distortions reside entirely on the Cs, O, and S sites. This raises fundamental questions: how can ligand-site distortions generate such a large electronic disproportionation between structurally intact Cr sites? how do many-body effects bridge the ligand distortion and the electronic transition? From an application perspective, understanding this mechanism is equally important because the insulating ground state suppresses the transport functionality expected from metallic AMs.

In this Letter, we address these questions using first-principles density functional theory plus dynamical mean-field theory (DFT+DMFT)~\cite{Georges:1996,lichtenstein:2001,kotliar:2006,Haule:2010} calculations for CsCr$_2$S$_2$O and its Te-substituted analogue. We identify an orbital-selective Mott transition (OSMT)~\cite{Anisimov2002,Liebsch:2003,Koga:2004,Medici:2005,Ferrero:2005,Arita:2005,Knecht:2005,Inaba:2007,Werner:2007,Jakobi:2009,Medici:2009,Werner:2009,Kita:2011,HuangLi:2012,HuangLi_pd_2012,Capone:2013,georges:2013,Rincon:2014,Wang_2016,Yurong:2017,Kugler:2019,Kugler:2022,Minjae:2024} that leaves the correlated metallic Cr-$d_{yz}$ orbital as the dominant low-energy electronic channel. Crucially, we find that the S-site distortion itself produces only a weak charge asymmetry between the Cr1 and Cr2 sites through Cr-$d_{yz}$--S-$p$ hybridization at DFT level. Dynamical electronic correlations dramatically amplify this asymmetry, generating pronounced differentiation in the charge occupancy and correlation strength of the Cr-$d_{yz}$ orbitals between the two Cr sublattices. This differentiation further induces strongly site-dependent spin polarizations and ultimately opens the insulating gap. By contrast, replacing S with Te suppresses this correlation amplification and stabilizes a metallic AM state. Our work reveals a hidden, correlation-amplified feedback loop between the ligand instabilities and electronic symmetry breaking in this AM family, highlighting the paramount importance of ligand selection for achieving robust metallic AMs.

 \begin{table}
     \centering
     \caption{The DFT and DFT+DMFT calculated Cr-$3d$ occupancies of S-LT structure in the NM and PM state, respectively. $U=12$ eV, $J_H=1$ eV for DFT+DMFT.}
     \begin{ruledtabular}
     \begin{tabular}{cccccccc}
    & & $d_{z^2}$  & $d_{xz}$  & $d_{xy}$ & $d_{x^2-y^2}$ & $d_{yz}$ & total \\
          \hline
          \multirow{3}{*}{\makecell{DFT}} 
          & $n_{\text{Cr1}}$ & 1.34 & 0.91 & 0.89 & 0.46 & 0.36 & 3.96 \\
          & $n_{\text{Cr2}}$ & 1.18 & 0.96 & 0.85 & 0.45 & 0.49 & 3.93 \\
          & $n_{\text{Cr1}}$-$n_{\text{Cr2}}$ & 0.16 & -0.05 & 0.04 & 0.01 & -0.13 & 0.03\\
          \hline
          \multirow{3}{*}{\makecell{DMFT}}
          & $n_{\text{Cr1}}$ & 0.92 & 0.98 & 0.95 & 0.21 & 0.94 & 4.00 \\
          & $n_{\text{Cr2}}$ & 0.93 & 0.99 & 0.97 & 0.31 & 0.48 & 3.68 \\
          & $n_{\text{Cr1}}$-$n_{\text{Cr2}}$ & -0.01 & -0.01 & -0.02 & 0.10 & 0.46 & 0.32\\         
     \end{tabular}
 \end{ruledtabular}
     \label{tab:occu-nm}
 \end{table}

\textit{Results.} To isolate the microscopic origin of the charge order, we consider four crystal structures: (i) the experimental high-temperature tetragonal phase of CsCr$_2$S$_2$O (S-HT) [Fig.~\ref{fig:dftband}(a)]; (ii) its experimental low-temperature orthorhombic phase (S-LT) [Fig.~\ref{fig:dftband}(b)]; (iii) a hypothetical S-LT$^*$ structure obtained from S-LT by restoring the S sites to their undistorted positions while retaining the Cs and O displacements; and (iv) a hypothetical orthorhombic CsCr$_2$Te$_2$O structure (Te-LT), constructed by imposing the same fractional in-plane ligand displacements as in S-LT (see the Supplementary Materials~\cite{suppl}). The Cr-$3d$ orbitals are defined in a local coordinate system where the $x$-axis aligns along the O-Cr-O direction [Fig.~\ref{fig:dftband}(a)].

Figs.~\ref{fig:dftband}(c)-(e) show the nonmagnetic (NM) DFT band structure and orbital-resolved density of states (DOS) of S-HT. Near the Fermi level $E_F$, the Cr-$d_{z^{2}}$, $d_{xz}$, and $d_{xy}$ orbitals are nearly half-filled, whereas the $d_{x^{2}-y^{2}}$ and $d_{yz}$ orbitals are predominantly unoccupied (Table~\ref{tab:occu-nm}). Importantly, the S-$p$ states exhibits strong and highly selective hybridization with the Cr-$d_{yz}$ states, as evidenced by the bonding and anti-bonding features around -3.5 eV and 1.5 eV, respectively [Fig.~\ref{fig:dftband}(e)].

\begin{figure*}
    \centering
    \includegraphics[width=1.0\textwidth]{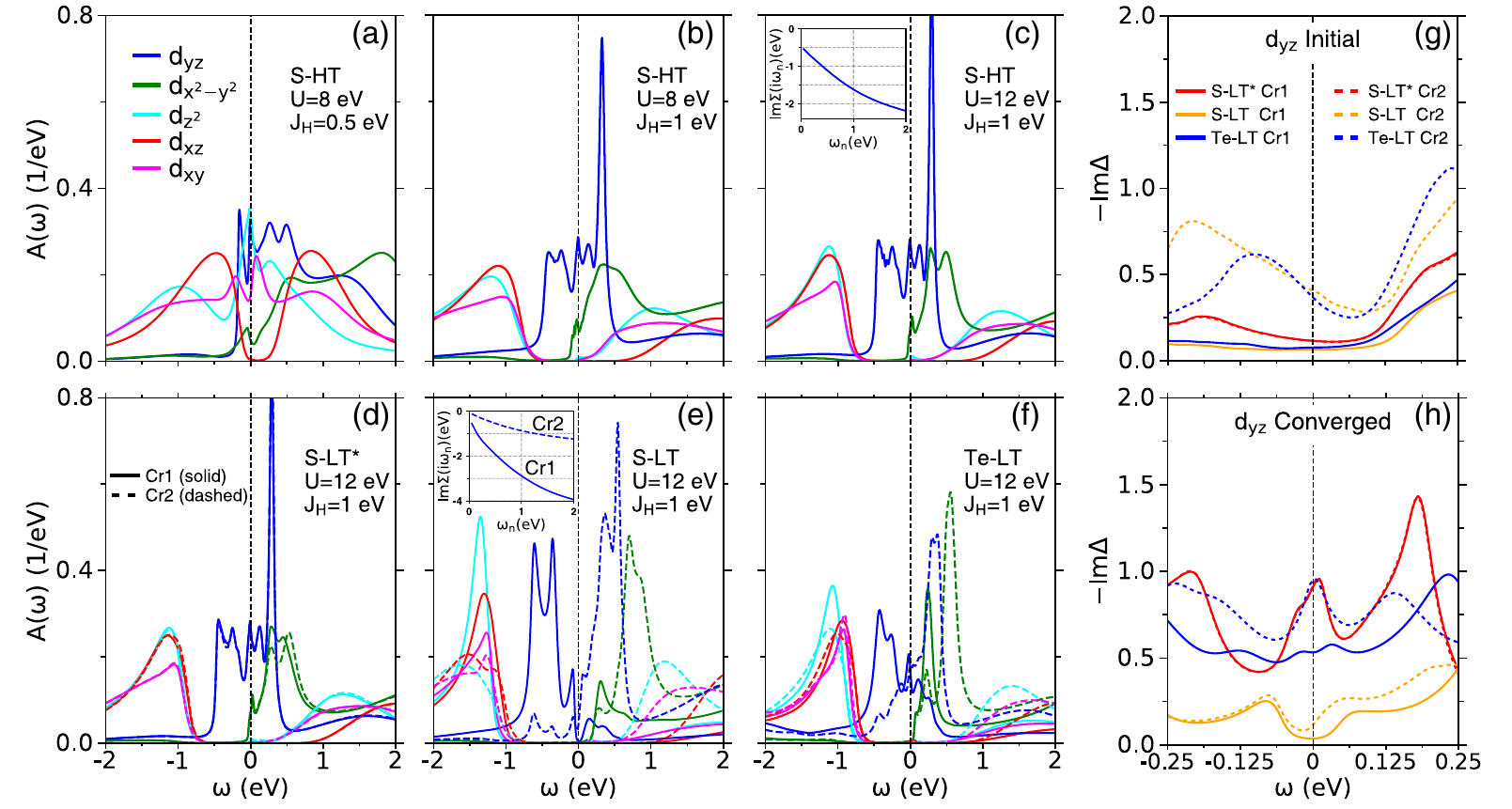}
    \caption{(a)-(f) The DFT+DMFT calculated spectral functions in the PM phase. (a)-(c) S-HT, (d) S-LT$^*$, (e) S-LT, (f) Te-LT. The insets in (c) and (e) are the imaginary part of self-energy Im$\Sigma(i\omega_n)$ of $d_{yz}$ orbital. (g),(h) The initial and DMFT converged hybridization functions of Cr-$d_{yz}$ orbital for S-LT$^*$, S-LT and Te-LT structures.   
     }
    \label{fig:dmftdos}
\end{figure*}

In the S-LT phase, the S ions undergo a pronounced in-plane displacement away from Cr1 and toward Cr2 [Fig.~\ref{fig:dftband}(b)]. This distortion lifts the equivalence between the Cr1-$d_{yz}$ and Cr2-$d_{yz}$ states, producing a clear band splitting [Figs.~\ref{fig:dftband}(h),(i)]. By contrast, in the S-LT$^{*}$, where the S distortions are removed, the two $d_{yz}$ bands nearly overlap [Figs.~\ref{fig:dftband}(f),(g)]. These results identify the S ligand distortion, mediated by strong Cr-$d_{yz}$--S-$p$ hybridization, as the primary structural channel that breaks the electronic equivalence of the two Cr sublattices. An analogous Te distortion produces a comparable bare splitting of the Cr1- and Cr2-$d_{yz}$ bands in Te-LT, while the Te-$5p$ states shift substantially closer to $E_F$ [Figs.~\ref{fig:dftband}(j),(k)]. As shown in Table.~\ref{tab:occu-nm}, the total Cr-$3d$ occupancy in S-LT is close to 4 electrons because of substantial Cr-ligand covalency, despite the nominal $d^{3.5}$ configuration. Importantly, at the DFT level the S distortion produces a total charge difference of only 0.03 electrons between Cr1 and Cr2. Such a tiny difference is far too small to account for the experimentally inferred Verwey-type charge disproportionation~\cite{Caoguanghan:2026}. The lattice distortion therefore acts only as a symmetry-breaking seed rather than the dominant driving force of the charge order. This points to many-body electronic correlations as the key ingredient that amplifies the ligand-induced electronic response.

To elucidate the role of electronic correlations, we first perform DFT+DMFT calculations in the paramagnetic (PM) phase. The corresponding spectral functions $A(\omega)$ are shown in Figs.~\ref{fig:dmftdos}(a)-(f). We find a pronounced orbital-selective Mott transition (OSMT) that is strongly controlled by Hund's coupling $J_H$~\cite{Medici:2009,georges:2013}. For the S-HT at $U=8$ eV and $J_H=0.5$ eV [Fig.~\ref{fig:dmftdos}(a)], only the Cr-$d_{xz}$ orbital is Mott localized, whereas the remaining orbitals retain itinerant spectral weight. Increasing $J_H$ to 1 eV drives the $d_{z^{2}}$ and $d_{xy}$ orbitals into the Mott-insulating state and pushes $d_{x^{2}-y^{2}}$ close to a band-insulating state. In striking contrast, the $d_{yz}$ orbital remains  metallic and retains substantial spectral weight at $E_F$ [Fig.~\ref{fig:dmftdos}(b)], even up to $U=12$ eV [Fig.~\ref{fig:dmftdos}(c)]. The low-energy electronic physics is therefore effectively reduced to a single, strongly correlated $d_{yz}$ channel. The large finite value of Im$\Sigma(i\omega_n\rightarrow 0)$, approximately 0.5 eV, further indicates strongly incoherent non-Fermi-liquid behavior [inset of Fig.~\ref{fig:dmftdos}(c)], consistent with the weakly negative temperature dependence of the resistivity observed experimentally in the high-temperature PM phase~\cite{Caoguanghan:2026}. 

In the S-LT$^*$ structure [Fig.~\ref{fig:dmftdos}(d)], the spectral functions remain nearly identical to those of S-HT, demonstrating that the Cs and O displacements alone have little effect on the correlated low-energy electronic structure. The situation changes qualitatively once the full S distortion is introduced. In S-LT [Fig.~\ref{fig:dmftdos}(e)], the $d_{x^{2}-y^{2}}$ orbital also becomes fully insulating, leaving $d_{yz}$ as the only active low-energy channel. Most importantly, dynamical correlations dramatically amplify the weak charge asymmetry initially generated by the S distortion. The Cr1- and Cr2-$d_{yz}$ states become strongly differentiated, with occupancies of 0.94 and 0.48 electrons, respectively [Fig.~\ref{fig:dmftdos}(e) and Table.~\ref{tab:occu-nm}]. As a result, the total Cr-sublattice charge disproportionation increases from only 0.03 electrons at the DFT level to approximately 0.32 electrons in DFT+DMFT. At the same time, the two Cr sites acquire markedly different correlation strengths. Because the Cr1-$d_{yz}$ orbital is driven close to half filling, it becomes substantially more correlated than its Cr2 counterpart, as directly reflected in their self-energies [inset of Fig.~\ref{fig:dmftdos}(e)]. Thus, dynamical correlations convert the ligand-induced symmetry-breaking field into a profound differentiation of the correlated $d_{yz}$ states. The resulting charge disproportionation produces a pseudogap at $E_F$, placing the system close to a global insulating state. Notably, the pronounced Cr-sublattice differentiation already emerges in the PM calculation, demonstrating that long-range magnetic order is not a prerequisite for the correlation-amplified effect. In contrast, the hypothetical Te-LT structure remains substantially more metallic than S-LT. Although the imposed Te distortion generates a bare Cr-sublattice splitting comparable to that produced by S [Fig.~\ref{fig:dftband}], the many-body amplification is strongly suppressed. The $d_{yz}$ orbital retains robust spectral weight at $E_F$, and the occupancies of Cr1- and Cr2-$d_{yz}$ remain 0.88 and 0.65, respectively~\cite{suppl}. Correspondingly, the gaps in the other Mott-localized orbitals are substantially reduced [Fig.~\ref{fig:dmftdos}(f)], indicating weaker effective correlation effects in the Te compound.

\begin{figure}
    \centering
    \includegraphics[width=0.48\textwidth]{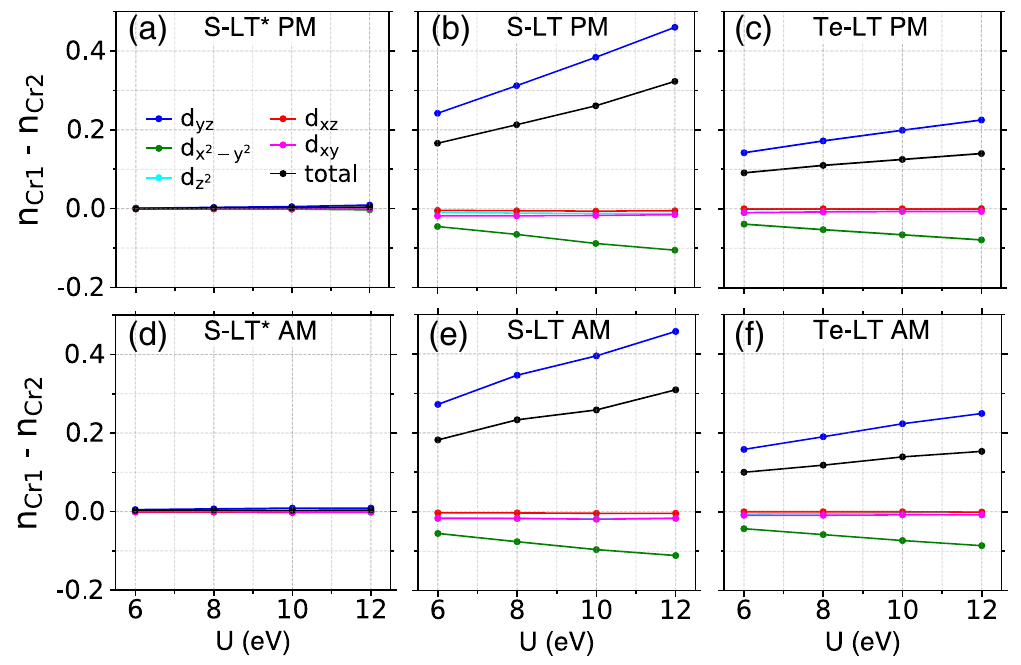}
    \caption{The DFT+DMFT calculated orbital occupancy difference between Cr1 and Cr2 sites in the S-LT structure vs. Hubbard $U$. (a)-(c) PM phase. (d)-(f) AM phase with $n_{\uparrow}+n_{\downarrow}$. 
     }
    \label{fig:occu}
\end{figure}
\begin{table}
     \centering
     \caption{The DFT+DMFT calculated spin-resolved Cr-$3d$ occupancies of S-LT in the AM state. $U=12$, $J_H=1$ eV.}
     \begin{ruledtabular}
     \begin{tabular}{cccccccc}
    & & $d_{z^2}$  & $d_{xz}$  & $d_{xy}$ & $d_{x^2-y^2}$ & $d_{yz}$ & total \\
          \hline
       \multirow{3}{*}{\makecell{Majority \\spin}}  
          & $n_{\text{Cr1}}$ & 0.88 & 0.96 & 0.90 & 0.17 & 0.90 & 3.81 \\
          & $n_{\text{Cr2}}$ & 0.83 & 0.90 & 0.86 & 0.25 & 0.39 & 3.23 \\
          & $n_{\text{Cr1}}$-$n_{\text{Cr2}}$ & 0.05 & 0.06 & 0.04 & -0.08 & 0.51 & 0.58 \\
          \hline
        \multirow{3}{*}{\makecell{Minority \\spin}}  
          & $n_{\text{Cr1}}$ & 0.03 & 0.03 & 0.04 & 0.04 & 0.04 & 0.18 \\
          & $n_{\text{Cr2}}$ & 0.09 & 0.09 & 0.10 & 0.07 & 0.09 & 0.44 \\
          & $n_{\text{Cr1}}$-$n_{\text{Cr2}}$ & -0.06 & -0.06 & -0.06 & -0.03 & -0.05 & -0.26 \\
     \end{tabular}
 \end{ruledtabular}
     \label{tab:occu-am}
 \end{table}

Within DFT+DMFT, the microscopic origin of this amplification can be traced to the self-consistent hybridization functions $\Delta(\omega)$. Figs.~\ref{fig:dmftdos}(g) and (h) compare the initial and converged -Im$\Delta(\omega)$ of Cr-$d_{yz}$ orbital near $E_F$ (in the PM phase). In S-LT$^{*}$, the hybridization functions of the Cr1 and Cr2 sublattices are essentially identical (red curves), confirming that the Cs and O displacements alone leave the local electronic environments nearly equivalent. In both the S-LT and Te-LT, ligand displacements break this equivalence. Because the displaced S/Te ligands move toward Cr2, the Cr2 site initially experiences stronger hybridization near $E_F$ than Cr1. After including electron correlations, the DFT+DMFT self-consistency strongly enhances this initial difference and generates a stronger effective correlation in the Cr1-$d_{yz}$ orbital than in Cr2-$d_{yz}$. Crucially, it also drives distinct localization behaviors in the two systems, despite their similar initial hybridization functions [Fig.~\ref{fig:dmftdos}(g)]. For S-LT, the converged -Im$\Delta(\omega)$ for both Cr sublattices are strongly suppressed toward $E_F$, signaling proximity to Mott localization and consistent with the pseudo-gapped spectrum shown in Fig.~\ref{fig:dmftdos}(e). In contrast, the hybridization functions in Te-LT remain large and finite at $E_F$, characteristic of a correlated metallic state well away from the Mott boundary. The contrast between S-LT and Te-LT can be understood from their different ligand electronic structures. The more spatially extended Te-$5p$ orbitals provide an enhanced electronic screening that effectively reduces the local Coulomb repulsion. Consequently, the effective electronic correlation strength is weakened in Te-LT, thereby weakening the correlation-driven amplification of the local charge disproportionation and the resulting MIT.

\begin{figure*}
    \centering
    \includegraphics[width=1.0\textwidth]{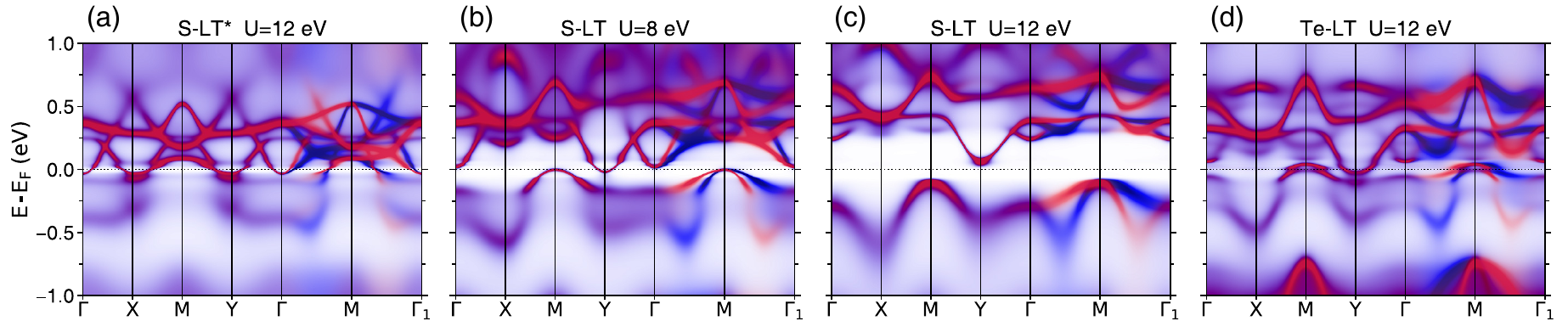}
    \caption{The DFT+DMFT calculated band structure in the AM phase. The red/blue indicate spin up/down, respectively.
     }
    \label{fig:afmband}
\end{figure*}

To establish the robustness of this correlation-amplification mechanism in the magnetic ground state, Figs.~\ref{fig:occu}(d)-(f) show the DFT+DMFT charge-occupancy difference $n_{\text{Cr}1}-n_{\text{Cr}2}$ as a function of $U$ ($J_H=1$ eV) in the AM state, together with the PM results in Fig.~\ref{fig:occu}(a)-(c). The strong charge disproportionation obtained in the PM calculations persists in the AM phase and exhibits nearly identical scaling with $U$. For $\text{S-LT}^{*}$, the charge asymmetry remains negligible over the entire range of $U$, despite the formal structural inequivalence introduced by the Cs and O displacements [Figs.~\ref{fig:occu}(a) and (d)]. In sharp contrast, the physical S-LT structure develops a pronounced correlation-driven charge imbalance dominated by the $d_{yz}$ orbital [Figs.~\ref{fig:occu}(b) and (e)]. At $U=12$ eV, the $d_{yz}$ occupancy difference reaches approximately 0.46 electrons (blue curves) and produces a total charge difference of about 0.3 electrons between Cr1 and Cr2 (black curves). This value captures the experimentally observed tendency toward strong Verwey-type charge disproportionation, although the experimental bond-valence-sum analysis yields a somewhat larger value of approximately 0.6 electrons~\cite{Caoguanghan:2026}. Such a quantitative difference may partly reflect the approximate nature of extracting ionic charges from bond-valence analysis~\cite{Brese:st0462}.
By comparison, static mean-field DFT+$U$ approach, although capable of reproducing an insulating state at $U=2.5$ eV~\cite{Caoguanghan:2026}, yields a total charge difference of only about 0.08 electrons. The much stronger disproportionation obtained within DFT+DMFT therefore underscores the essential role of dynamical correlations in the Verwey-type transition.
For Te-LT, the total charge difference is remarkably suppressed to $\sim0.14$ electrons at $U=12$ eV [Figs.~\ref{fig:occu}(c) and (f)], further demonstrating that the many-body amplification is strongly suppressed in the Te compound.

The spin-resolved Cr-$3d$ occupancy of S-LT at $U=12$ eV and $J_H=1$ eV are summarized in Table~\ref{tab:occu-am}. In the majority-spin channel, the total charge difference reaches $n_{\text{Cr1}}$-$n_{\text{Cr2}}$=0.58 electrons and is again dominated by the $d_{yz}$ orbital. The minority-spin channel exhibits the opposite trend, with $n_{\text{Cr1}}$-$n_{\text{Cr2}}$=-0.26 electrons. Thus, the spin polarization strength at Cr1 is significantly stronger than that at Cr2, consistent with that Cr1 experiences much stronger electronic correlation than Cr2. While the compensation between the two spin channels yields a net total charge difference of $0.58-0.26=0.32$ electrons, it produces a large differentiation in local spin polarization of $0.58+0.26=0.84$ $\mu_B$ between Cr1 (3.63 $\mu_B$) and Cr2 (2.79 $\mu_B$). This pronounced site selectivity substantially reconstructs the low-energy electronic structure and promotes the opening of charge gap. In Te-LT, the corresponding spin-polarization difference is reduced to $0.73$ $\mu_B$ at $U=12$ eV [Cr1 (3.34 $\mu_B$) and Cr2 (2.61 $\mu_B$)]~\cite{suppl}, consistent with its weaker electron correlations and charge disproportionation.

The DFT+DMFT spin-resolved band structures in the AM state at $k_z = 0$ are shown in Fig.~\ref{fig:afmband}. The low-energy states are dominated by the $d_{yz}$ orbitals. In S-LT$^{*}$ [Fig.~\ref{fig:afmband}(a)], the system remains a robust metallic AM state even at $U = 12$ eV, confirming that the Cs and O distortions alone are insufficient to drive an MIT. For S-LT at $U = 8$ eV [Fig.~\ref{fig:afmband}(b)], the system lies close to the MIT: the total charge disproportionation is only 0.2 electrons [Fig.~\ref{fig:occu}(b)], and the local spin-polarization difference is 0.69 $\mu_B$, insufficient to open a full gap. Increasing $U$ to 12 eV substantially enhances both charge and spin differentiation and produces a global insulating gap [Fig.~\ref{fig:afmband}(c)], reproducing the experimentally observed insulating state~\cite{Caoguanghan:2026}. In contrast, Te-LT remains metallic even at $U = 12\text{ eV}$ [Fig.~\ref{fig:afmband}(d)]. These comparative results establish that sufficiently strong correlation-amplified electronic differentiation between the two Cr sublattices is indispensable for the Verwey-type MIT in this AM family.

\textit{Conclusion.} To conclude, we have elucidated the microscopic mechanism underlying the Verwey-type MIT in the recently synthesized room-temperature altermagnet CsCr$_2$S$_2$O. An orbital-selective Mott state isolates the correlated $d_{yz}$ orbital as the dominant low-energy electronic channel. Although the pronounced S-ligand displacement breaks the equivalence of the two Cr sublattices through Cr-$d_{yz}$--S-$p$ hybridization, its bare effect on the Cr charge disproportionation is surprisingly small. Dynamical correlations strongly amplify this ligand-induced electronic asymmetry, producing pronounced charge disproportionation, site-selective correlation strengths, and strongly differentiated local spin polarizations that ultimately open the insulating gap, which naturally explains the experimental observations~\cite{Caoguanghan:2026}. We also predict that this correlation-amplified effect is effectively suppressed upon substituting S with Te, offering a viable route to quench the MIT. Our work highlights how ligand dynamics are
deeply intertwined with strong electron correlations--a crucial effect that is largely inaccessible to conventional DFT-based high-throughput screenings of AM materials. These findings introduce a practical avenue to manipulate AM phases, thereby preserving the metallic transport functionalities required for future spintronic devices.

\textit{Acknowledgement.} This project was supported by National Key R\&D Program of China (Grant No. 2023YFA1406304), the Quantum Science and Technology-National Science and Technology Major Project (No. 2021ZD0302803), HFNL Self-Deployed Project (No. ZB2602000301) and the New Cornerstone Science Foundation. The calculations were performed in Hefei Advanced Computing Center, China.

%\pagebreak
\bibliography{main}

\begin{thebibliography}{109}%
\makeatletter
\providecommand \@ifxundefined [1]{%
 \@ifx{#1\undefined}
}%
\providecommand \@ifnum [1]{%
 \ifnum #1\expandafter \@firstoftwo
 \else \expandafter \@secondoftwo
 \fi
}%
\providecommand \@ifx [1]{%
 \ifx #1\expandafter \@firstoftwo
 \else \expandafter \@secondoftwo
 \fi
}%
\providecommand \natexlab [1]{#1}%
\providecommand \enquote  [1]{``#1''}%
\providecommand \bibnamefont  [1]{#1}%
\providecommand \bibfnamefont [1]{#1}%
\providecommand \citenamefont [1]{#1}%
\providecommand \href@noop [0]{\@secondoftwo}%
\providecommand \href [0]{\begingroup \@sanitize@url \@href}%
\providecommand \@href[1]{\@@startlink{#1}\@@href}%
\providecommand \@@href[1]{\endgroup#1\@@endlink}%
\providecommand \@sanitize@url [0]{\catcode `\\12\catcode `\$12\catcode `\&12\catcode `\#12\catcode `\^12\catcode `\_12\catcode `\%12\relax}%
\providecommand \@@startlink[1]{}%
\providecommand \@@endlink[0]{}%
\providecommand \url  [0]{\begingroup\@sanitize@url \@url }%
\providecommand \@url [1]{\endgroup\@href {#1}{\urlprefix }}%
\providecommand \urlprefix  [0]{URL }%
\providecommand \Eprint [0]{\href }%
\providecommand \doibase [0]{https://doi.org/}%
\providecommand \selectlanguage [0]{\@gobble}%
\providecommand \bibinfo  [0]{\@secondoftwo}%
\providecommand \bibfield  [0]{\@secondoftwo}%
\providecommand \translation [1]{[#1]}%
\providecommand \BibitemOpen [0]{}%
\providecommand \bibitemStop [0]{}%
\providecommand \bibitemNoStop [0]{.\EOS\space}%
\providecommand \EOS [0]{\spacefactor3000\relax}%
\providecommand \BibitemShut  [1]{\csname bibitem#1\endcsname}%
\let\auto@bib@innerbib\@empty
%</preamble>
\bibitem [{\citenamefont {Noda}\ \emph {et~al.}(2016)\citenamefont {Noda}, \citenamefont {Ohno},\ and\ \citenamefont {Nakamura}}]{MnO2_DFT}%
  \BibitemOpen
  \bibfield  {author} {\bibinfo {author} {\bibfnamefont {Y.}~\bibnamefont {Noda}}, \bibinfo {author} {\bibfnamefont {K.}~\bibnamefont {Ohno}},\ and\ \bibinfo {author} {\bibfnamefont {S.}~\bibnamefont {Nakamura}},\ }\bibfield  {title} {\bibinfo {title} {Momentum-dependent band spin splitting in semiconducting {MnO$_2$}: a density functional calculation},\ }\href {https://doi.org/10.1039/C5CP07806G} {\bibfield  {journal} {\bibinfo  {journal} {Phys. Chem. Chem. Phys.}\ }\textbf {\bibinfo {volume} {18}},\ \bibinfo {pages} {13294} (\bibinfo {year} {2016})}\BibitemShut {NoStop}%
\bibitem [{\citenamefont {Hayami}\ \emph {et~al.}(2019)\citenamefont {Hayami}, \citenamefont {Yanagi},\ and\ \citenamefont {Kusunose}}]{2019_THEORY}%
  \BibitemOpen
  \bibfield  {author} {\bibinfo {author} {\bibfnamefont {S.}~\bibnamefont {Hayami}}, \bibinfo {author} {\bibfnamefont {Y.}~\bibnamefont {Yanagi}},\ and\ \bibinfo {author} {\bibfnamefont {H.}~\bibnamefont {Kusunose}},\ }\bibfield  {title} {\bibinfo {title} {Momentum-dependent spin splitting by collinear antiferromagnetic ordering},\ }\href {https://doi.org/10.7566/JPSJ.88.123702} {\bibfield  {journal} {\bibinfo  {journal} {Journal of the Physical Society of Japan}\ }\textbf {\bibinfo {volume} {88}},\ \bibinfo {pages} {123702} (\bibinfo {year} {2019})}\BibitemShut {NoStop}%
\bibitem [{\citenamefont {Ma}\ \emph {et~al.}(2021)\citenamefont {Ma}, \citenamefont {Hu}, \citenamefont {Li}, \citenamefont {Liu}, \citenamefont {Yao}, \citenamefont {Jia},\ and\ \citenamefont {Liu}}]{V2Se2O_theory}%
  \BibitemOpen
  \bibfield  {author} {\bibinfo {author} {\bibfnamefont {H.-Y.}\ \bibnamefont {Ma}}, \bibinfo {author} {\bibfnamefont {M.}~\bibnamefont {Hu}}, \bibinfo {author} {\bibfnamefont {N.}~\bibnamefont {Li}}, \bibinfo {author} {\bibfnamefont {J.}~\bibnamefont {Liu}}, \bibinfo {author} {\bibfnamefont {W.}~\bibnamefont {Yao}}, \bibinfo {author} {\bibfnamefont {J.-F.}\ \bibnamefont {Jia}},\ and\ \bibinfo {author} {\bibfnamefont {J.}~\bibnamefont {Liu}},\ }\bibfield  {title} {\bibinfo {title} {Multifunctional antiferromagnetic materials with giant piezomagnetism and noncollinear spin current},\ }\href {https://doi.org/10.1038/s41467-021-23127-7} {\bibfield  {journal} {\bibinfo  {journal} {Nature Communications}\ }\textbf {\bibinfo {volume} {12}},\ \bibinfo {pages} {2846} (\bibinfo {year} {2021})}\BibitemShut {NoStop}%
\bibitem [{\citenamefont {Hu}\ \emph {et~al.}(2025{\natexlab{a}})\citenamefont {Hu}, \citenamefont {Cheng}, \citenamefont {Huang},\ and\ \citenamefont {Liu}}]{theory_RbVTeO}%
  \BibitemOpen
  \bibfield  {author} {\bibinfo {author} {\bibfnamefont {M.}~\bibnamefont {Hu}}, \bibinfo {author} {\bibfnamefont {X.}~\bibnamefont {Cheng}}, \bibinfo {author} {\bibfnamefont {Z.}~\bibnamefont {Huang}},\ and\ \bibinfo {author} {\bibfnamefont {J.}~\bibnamefont {Liu}},\ }\bibfield  {title} {\bibinfo {title} {Catalog of $c$-paired spin-momentum locking in antiferromagnetic systems},\ }\href {https://doi.org/10.1103/PhysRevX.15.021083} {\bibfield  {journal} {\bibinfo  {journal} {Phys. Rev. X}\ }\textbf {\bibinfo {volume} {15}},\ \bibinfo {pages} {021083} (\bibinfo {year} {2025}{\natexlab{a}})}\BibitemShut {NoStop}%
\bibitem [{\citenamefont {\ifmmode~\check{S}\else \v{S}\fi{}mejkal}\ \emph {et~al.}(2022{\natexlab{a}})\citenamefont {\ifmmode~\check{S}\else \v{S}\fi{}mejkal}, \citenamefont {Sinova},\ and\ \citenamefont {Jungwirth}}]{theory_prx_1}%
  \BibitemOpen
  \bibfield  {author} {\bibinfo {author} {\bibfnamefont {L.}~\bibnamefont {\ifmmode~\check{S}\else \v{S}\fi{}mejkal}}, \bibinfo {author} {\bibfnamefont {J.}~\bibnamefont {Sinova}},\ and\ \bibinfo {author} {\bibfnamefont {T.}~\bibnamefont {Jungwirth}},\ }\bibfield  {title} {\bibinfo {title} {Emerging research landscape of altermagnetism},\ }\href {https://doi.org/10.1103/PhysRevX.12.040501} {\bibfield  {journal} {\bibinfo  {journal} {Phys. Rev. X}\ }\textbf {\bibinfo {volume} {12}},\ \bibinfo {pages} {040501} (\bibinfo {year} {2022}{\natexlab{a}})}\BibitemShut {NoStop}%
\bibitem [{\citenamefont {\ifmmode~\check{S}\else \v{S}\fi{}mejkal}\ \emph {et~al.}(2022{\natexlab{b}})\citenamefont {\ifmmode~\check{S}\else \v{S}\fi{}mejkal}, \citenamefont {Sinova},\ and\ \citenamefont {Jungwirth}}]{theory_prx_2}%
  \BibitemOpen
  \bibfield  {author} {\bibinfo {author} {\bibfnamefont {L.}~\bibnamefont {\ifmmode~\check{S}\else \v{S}\fi{}mejkal}}, \bibinfo {author} {\bibfnamefont {J.}~\bibnamefont {Sinova}},\ and\ \bibinfo {author} {\bibfnamefont {T.}~\bibnamefont {Jungwirth}},\ }\bibfield  {title} {\bibinfo {title} {Beyond conventional ferromagnetism and antiferromagnetism: A phase with nonrelativistic spin and crystal rotation symmetry},\ }\href {https://doi.org/10.1103/PhysRevX.12.031042} {\bibfield  {journal} {\bibinfo  {journal} {Phys. Rev. X}\ }\textbf {\bibinfo {volume} {12}},\ \bibinfo {pages} {031042} (\bibinfo {year} {2022}{\natexlab{b}})}\BibitemShut {NoStop}%
\bibitem [{\citenamefont {{\v{S}}mejkal}\ \emph {et~al.}(2022)\citenamefont {{\v{S}}mejkal}, \citenamefont {MacDonald}, \citenamefont {Sinova}, \citenamefont {Nakatsuji},\ and\ \citenamefont {Jungwirth}}]{Anomalous_Hall}%
  \BibitemOpen
  \bibfield  {author} {\bibinfo {author} {\bibfnamefont {L.}~\bibnamefont {{\v{S}}mejkal}}, \bibinfo {author} {\bibfnamefont {A.~H.}\ \bibnamefont {MacDonald}}, \bibinfo {author} {\bibfnamefont {J.}~\bibnamefont {Sinova}}, \bibinfo {author} {\bibfnamefont {S.}~\bibnamefont {Nakatsuji}},\ and\ \bibinfo {author} {\bibfnamefont {T.}~\bibnamefont {Jungwirth}},\ }\bibfield  {title} {\bibinfo {title} {Anomalous hall antiferromagnets},\ }\href {https://doi.org/10.1038/s41578-022-00430-3} {\bibfield  {journal} {\bibinfo  {journal} {Nature Reviews Materials}\ }\textbf {\bibinfo {volume} {7}},\ \bibinfo {pages} {482} (\bibinfo {year} {2022})}\BibitemShut {NoStop}%
\bibitem [{\citenamefont {Jungwirth}\ \emph {et~al.}(2024)\citenamefont {Jungwirth}, \citenamefont {Fernandes}, \citenamefont {Sinova},\ and\ \citenamefont {Smejkal}}]{review_am}%
  \BibitemOpen
  \bibfield  {author} {\bibinfo {author} {\bibfnamefont {T.}~\bibnamefont {Jungwirth}}, \bibinfo {author} {\bibfnamefont {R.~M.}\ \bibnamefont {Fernandes}}, \bibinfo {author} {\bibfnamefont {J.}~\bibnamefont {Sinova}},\ and\ \bibinfo {author} {\bibfnamefont {L.}~\bibnamefont {Smejkal}},\ }\href {https://arxiv.org/abs/2409.10034} {\bibinfo {title} {Altermagnets and beyond: Nodal magnetically-ordered phases}} (\bibinfo {year} {2024}),\ \Eprint {https://arxiv.org/abs/2409.10034} {arXiv:2409.10034 [cond-mat.mtrl-sci]} \BibitemShut {NoStop}%
\bibitem [{\citenamefont {Guo}\ \emph {et~al.}(2023)\citenamefont {Guo}, \citenamefont {Liu}, \citenamefont {Janson}, \citenamefont {Fulga}, \citenamefont {{van den Brink}},\ and\ \citenamefont {Facio}}]{largescale_DFT}%
  \BibitemOpen
  \bibfield  {author} {\bibinfo {author} {\bibfnamefont {Y.}~\bibnamefont {Guo}}, \bibinfo {author} {\bibfnamefont {H.}~\bibnamefont {Liu}}, \bibinfo {author} {\bibfnamefont {O.}~\bibnamefont {Janson}}, \bibinfo {author} {\bibfnamefont {I.~C.}\ \bibnamefont {Fulga}}, \bibinfo {author} {\bibfnamefont {J.}~\bibnamefont {{van den Brink}}},\ and\ \bibinfo {author} {\bibfnamefont {J.~I.}\ \bibnamefont {Facio}},\ }\bibfield  {title} {\bibinfo {title} {Spin-split collinear antiferromagnets: A large-scale ab-initio study},\ }\href {https://doi.org/https://doi.org/10.1016/j.mtphys.2023.100991} {\bibfield  {journal} {\bibinfo  {journal} {Materials Today Physics}\ }\textbf {\bibinfo {volume} {32}},\ \bibinfo {pages} {100991} (\bibinfo {year} {2023})}\BibitemShut {NoStop}%
\bibitem [{\citenamefont {Bai}\ \emph {et~al.}(2024)\citenamefont {Bai}, \citenamefont {Feng}, \citenamefont {Liu}, \citenamefont {Šmejkal}, \citenamefont {Mokrousov},\ and\ \citenamefont {Yao}}]{review_yao}%
  \BibitemOpen
  \bibfield  {author} {\bibinfo {author} {\bibfnamefont {L.}~\bibnamefont {Bai}}, \bibinfo {author} {\bibfnamefont {W.}~\bibnamefont {Feng}}, \bibinfo {author} {\bibfnamefont {S.}~\bibnamefont {Liu}}, \bibinfo {author} {\bibfnamefont {L.}~\bibnamefont {Šmejkal}}, \bibinfo {author} {\bibfnamefont {Y.}~\bibnamefont {Mokrousov}},\ and\ \bibinfo {author} {\bibfnamefont {Y.}~\bibnamefont {Yao}},\ }\bibfield  {title} {\bibinfo {title} {Altermagnetism: Exploring new frontiers in magnetism and spintronics},\ }\href {https://doi.org/https://doi.org/10.1002/adfm.202409327} {\bibfield  {journal} {\bibinfo  {journal} {Advanced Functional Materials}\ }\textbf {\bibinfo {volume} {34}},\ \bibinfo {pages} {2409327} (\bibinfo {year} {2024})}\BibitemShut {NoStop}%
\bibitem [{\citenamefont {McClarty}\ and\ \citenamefont {Rau}(2024)}]{Landa_Theory_of_Altermagnetism}%
  \BibitemOpen
  \bibfield  {author} {\bibinfo {author} {\bibfnamefont {P.~A.}\ \bibnamefont {McClarty}}\ and\ \bibinfo {author} {\bibfnamefont {J.~G.}\ \bibnamefont {Rau}},\ }\bibfield  {title} {\bibinfo {title} {Landau theory of altermagnetism},\ }\href {https://doi.org/10.1103/PhysRevLett.132.176702} {\bibfield  {journal} {\bibinfo  {journal} {Phys. Rev. Lett.}\ }\textbf {\bibinfo {volume} {132}},\ \bibinfo {pages} {176702} (\bibinfo {year} {2024})}\BibitemShut {NoStop}%
\bibitem [{\citenamefont {Liu}\ \emph {et~al.}(2025{\natexlab{a}})\citenamefont {Liu}, \citenamefont {Dai},\ and\ \citenamefont {Bl{\"u}gel}}]{theory_AM}%
  \BibitemOpen
  \bibfield  {author} {\bibinfo {author} {\bibfnamefont {Q.}~\bibnamefont {Liu}}, \bibinfo {author} {\bibfnamefont {X.}~\bibnamefont {Dai}},\ and\ \bibinfo {author} {\bibfnamefont {S.}~\bibnamefont {Bl{\"u}gel}},\ }\bibfield  {title} {\bibinfo {title} {Different facets of unconventional magnetism},\ }\href {https://doi.org/10.1038/s41567-024-02750-3} {\bibfield  {journal} {\bibinfo  {journal} {Nature Physics}\ }\textbf {\bibinfo {volume} {21}},\ \bibinfo {pages} {329} (\bibinfo {year} {2025}{\natexlab{a}})}\BibitemShut {NoStop}%
\bibitem [{\citenamefont {Zhu}\ \emph {et~al.}(2025)\citenamefont {Zhu}, \citenamefont {Li}, \citenamefont {Chen}, \citenamefont {Yu},\ and\ \citenamefont {Liu}}]{Magnetic_geometry}%
  \BibitemOpen
  \bibfield  {author} {\bibinfo {author} {\bibfnamefont {H.}~\bibnamefont {Zhu}}, \bibinfo {author} {\bibfnamefont {J.}~\bibnamefont {Li}}, \bibinfo {author} {\bibfnamefont {X.}~\bibnamefont {Chen}}, \bibinfo {author} {\bibfnamefont {Y.}~\bibnamefont {Yu}},\ and\ \bibinfo {author} {\bibfnamefont {Q.}~\bibnamefont {Liu}},\ }\bibfield  {title} {\bibinfo {title} {Magnetic geometry induced quantum geometry and nonlinear transports},\ }\href {https://doi.org/10.1038/s41467-025-60128-2} {\bibfield  {journal} {\bibinfo  {journal} {Nature Communications}\ }\textbf {\bibinfo {volume} {16}},\ \bibinfo {pages} {4882} (\bibinfo {year} {2025})}\BibitemShut {NoStop}%
\bibitem [{\citenamefont {Gonz\'alez-Hern\'andez}\ \emph {et~al.}(2021)\citenamefont {Gonz\'alez-Hern\'andez}, \citenamefont {\ifmmode~\check{S}\else \v{S}\fi{}mejkal}, \citenamefont {V\'yborn\'y}, \citenamefont {Yahagi}, \citenamefont {Sinova}, \citenamefont {Jungwirth},\ and\ \citenamefont {\ifmmode~\check{Z}\else \v{Z}\fi{}elezn\'y}}]{Electricalspinsplitter}%
  \BibitemOpen
  \bibfield  {author} {\bibinfo {author} {\bibfnamefont {R.}~\bibnamefont {Gonz\'alez-Hern\'andez}}, \bibinfo {author} {\bibfnamefont {L.}~\bibnamefont {\ifmmode~\check{S}\else \v{S}\fi{}mejkal}}, \bibinfo {author} {\bibfnamefont {K.}~\bibnamefont {V\'yborn\'y}}, \bibinfo {author} {\bibfnamefont {Y.}~\bibnamefont {Yahagi}}, \bibinfo {author} {\bibfnamefont {J.}~\bibnamefont {Sinova}}, \bibinfo {author} {\bibfnamefont {T.~c.~v.}\ \bibnamefont {Jungwirth}},\ and\ \bibinfo {author} {\bibfnamefont {J.}~\bibnamefont {\ifmmode~\check{Z}\else \v{Z}\fi{}elezn\'y}},\ }\bibfield  {title} {\bibinfo {title} {Efficient electrical spin splitter based on nonrelativistic collinear antiferromagnetism},\ }\href {https://doi.org/10.1103/PhysRevLett.126.127701} {\bibfield  {journal} {\bibinfo  {journal} {Phys. Rev. Lett.}\ }\textbf {\bibinfo {volume} {126}},\ \bibinfo {pages} {127701} (\bibinfo {year} {2021})}\BibitemShut {NoStop}%
\bibitem [{\citenamefont {\ifmmode~\check{S}\else \v{S}\fi{}mejkal}\ \emph {et~al.}(2022{\natexlab{c}})\citenamefont {\ifmmode~\check{S}\else \v{S}\fi{}mejkal}, \citenamefont {Hellenes}, \citenamefont {Gonz\'alez-Hern\'andez}, \citenamefont {Sinova},\ and\ \citenamefont {Jungwirth}}]{GMR_theory}%
  \BibitemOpen
  \bibfield  {author} {\bibinfo {author} {\bibfnamefont {L.}~\bibnamefont {\ifmmode~\check{S}\else \v{S}\fi{}mejkal}}, \bibinfo {author} {\bibfnamefont {A.~B.}\ \bibnamefont {Hellenes}}, \bibinfo {author} {\bibfnamefont {R.}~\bibnamefont {Gonz\'alez-Hern\'andez}}, \bibinfo {author} {\bibfnamefont {J.}~\bibnamefont {Sinova}},\ and\ \bibinfo {author} {\bibfnamefont {T.}~\bibnamefont {Jungwirth}},\ }\bibfield  {title} {\bibinfo {title} {Giant and tunneling magnetoresistance in unconventional collinear antiferromagnets with nonrelativistic spin-momentum coupling},\ }\href {https://doi.org/10.1103/PhysRevX.12.011028} {\bibfield  {journal} {\bibinfo  {journal} {Phys. Rev. X}\ }\textbf {\bibinfo {volume} {12}},\ \bibinfo {pages} {011028} (\bibinfo {year} {2022}{\natexlab{c}})}\BibitemShut {NoStop}%
\bibitem [{\citenamefont {Zhang}\ \emph {et~al.}(2024{\natexlab{a}})\citenamefont {Zhang}, \citenamefont {Cui}, \citenamefont {Li}, \citenamefont {Duan}, \citenamefont {Li}, \citenamefont {Yu},\ and\ \citenamefont {Yao}}]{GateFieldControl}%
  \BibitemOpen
  \bibfield  {author} {\bibinfo {author} {\bibfnamefont {R.-W.}\ \bibnamefont {Zhang}}, \bibinfo {author} {\bibfnamefont {C.}~\bibnamefont {Cui}}, \bibinfo {author} {\bibfnamefont {R.}~\bibnamefont {Li}}, \bibinfo {author} {\bibfnamefont {J.}~\bibnamefont {Duan}}, \bibinfo {author} {\bibfnamefont {L.}~\bibnamefont {Li}}, \bibinfo {author} {\bibfnamefont {Z.-M.}\ \bibnamefont {Yu}},\ and\ \bibinfo {author} {\bibfnamefont {Y.}~\bibnamefont {Yao}},\ }\bibfield  {title} {\bibinfo {title} {Predictable gate-field control of spin in altermagnets with spin-layer coupling},\ }\href {https://doi.org/10.1103/PhysRevLett.133.056401} {\bibfield  {journal} {\bibinfo  {journal} {Phys. Rev. Lett.}\ }\textbf {\bibinfo {volume} {133}},\ \bibinfo {pages} {056401} (\bibinfo {year} {2024}{\natexlab{a}})}\BibitemShut {NoStop}%
\bibitem [{\citenamefont {Han}\ \emph {et~al.}(2024)\citenamefont {Han}, \citenamefont {Fu}, \citenamefont {Peng}, \citenamefont {Cheng}, \citenamefont {Dai}, \citenamefont {Liu}, \citenamefont {Li}, \citenamefont {Zhang}, \citenamefont {Zhu}, \citenamefont {Bai}, \citenamefont {Zhou}, \citenamefont {Liang}, \citenamefont {Chen}, \citenamefont {Wang}, \citenamefont {Chen}, \citenamefont {Yang}, \citenamefont {Zhang}, \citenamefont {Song}, \citenamefont {Liu},\ and\ \citenamefont {Pan}}]{180switchingneelvector}%
  \BibitemOpen
  \bibfield  {author} {\bibinfo {author} {\bibfnamefont {L.}~\bibnamefont {Han}}, \bibinfo {author} {\bibfnamefont {X.}~\bibnamefont {Fu}}, \bibinfo {author} {\bibfnamefont {R.}~\bibnamefont {Peng}}, \bibinfo {author} {\bibfnamefont {X.}~\bibnamefont {Cheng}}, \bibinfo {author} {\bibfnamefont {J.}~\bibnamefont {Dai}}, \bibinfo {author} {\bibfnamefont {L.}~\bibnamefont {Liu}}, \bibinfo {author} {\bibfnamefont {Y.}~\bibnamefont {Li}}, \bibinfo {author} {\bibfnamefont {Y.}~\bibnamefont {Zhang}}, \bibinfo {author} {\bibfnamefont {W.}~\bibnamefont {Zhu}}, \bibinfo {author} {\bibfnamefont {H.}~\bibnamefont {Bai}}, \bibinfo {author} {\bibfnamefont {Y.}~\bibnamefont {Zhou}}, \bibinfo {author} {\bibfnamefont {S.}~\bibnamefont {Liang}}, \bibinfo {author} {\bibfnamefont {C.}~\bibnamefont {Chen}}, \bibinfo {author} {\bibfnamefont {Q.}~\bibnamefont {Wang}}, \bibinfo {author} {\bibfnamefont {X.}~\bibnamefont {Chen}}, \bibinfo {author} {\bibfnamefont {L.}~\bibnamefont {Yang}}, \bibinfo {author} {\bibfnamefont {Y.}~\bibnamefont {Zhang}}, \bibinfo {author} {\bibfnamefont {C.}~\bibnamefont {Song}}, \bibinfo {author} {\bibfnamefont {J.}~\bibnamefont {Liu}},\ and\ \bibinfo {author} {\bibfnamefont {F.}~\bibnamefont {Pan}},\ }\bibfield  {title} {\bibinfo {title} {Electrical 180° switching of néel vector in spin-splitting antiferromagnet},\ }\href {https://doi.org/10.1126/sciadv.adn0479} {\bibfield  {journal} {\bibinfo  {journal} {Science Advances}\ }\textbf {\bibinfo {volume} {10}},\ \bibinfo {pages} {eadn0479} (\bibinfo {year} {2024})}\BibitemShut {NoStop}%
\bibitem [{\citenamefont {Zhou}\ \emph {et~al.}(2025)\citenamefont {Zhou}, \citenamefont {Cheng}, \citenamefont {Hu}, \citenamefont {Chu}, \citenamefont {Bai}, \citenamefont {Han}, \citenamefont {Liu}, \citenamefont {Pan},\ and\ \citenamefont {Song}}]{altermagneticorder}%
  \BibitemOpen
  \bibfield  {author} {\bibinfo {author} {\bibfnamefont {Z.}~\bibnamefont {Zhou}}, \bibinfo {author} {\bibfnamefont {X.}~\bibnamefont {Cheng}}, \bibinfo {author} {\bibfnamefont {M.}~\bibnamefont {Hu}}, \bibinfo {author} {\bibfnamefont {R.}~\bibnamefont {Chu}}, \bibinfo {author} {\bibfnamefont {H.}~\bibnamefont {Bai}}, \bibinfo {author} {\bibfnamefont {L.}~\bibnamefont {Han}}, \bibinfo {author} {\bibfnamefont {J.}~\bibnamefont {Liu}}, \bibinfo {author} {\bibfnamefont {F.}~\bibnamefont {Pan}},\ and\ \bibinfo {author} {\bibfnamefont {C.}~\bibnamefont {Song}},\ }\bibfield  {title} {\bibinfo {title} {Manipulation of the altermagnetic order in {CrSb} via crystal symmetry},\ }\href {https://doi.org/10.1038/s41586-024-08436-3} {\bibfield  {journal} {\bibinfo  {journal} {Nature}\ }\textbf {\bibinfo {volume} {638}},\ \bibinfo {pages} {645} (\bibinfo {year} {2025})}\BibitemShut {NoStop}%
\bibitem [{\citenamefont {Zhang}\ \emph {et~al.}(2025{\natexlab{a}})\citenamefont {Zhang}, \citenamefont {Bai}, \citenamefont {Dai}, \citenamefont {Han}, \citenamefont {Chen}, \citenamefont {Liang}, \citenamefont {Cao}, \citenamefont {Zhang}, \citenamefont {Wang}, \citenamefont {Zhu}, \citenamefont {Pan},\ and\ \citenamefont {Song}}]{controlSST}%
  \BibitemOpen
  \bibfield  {author} {\bibinfo {author} {\bibfnamefont {Y.}~\bibnamefont {Zhang}}, \bibinfo {author} {\bibfnamefont {H.}~\bibnamefont {Bai}}, \bibinfo {author} {\bibfnamefont {J.}~\bibnamefont {Dai}}, \bibinfo {author} {\bibfnamefont {L.}~\bibnamefont {Han}}, \bibinfo {author} {\bibfnamefont {C.}~\bibnamefont {Chen}}, \bibinfo {author} {\bibfnamefont {S.}~\bibnamefont {Liang}}, \bibinfo {author} {\bibfnamefont {Y.}~\bibnamefont {Cao}}, \bibinfo {author} {\bibfnamefont {Y.}~\bibnamefont {Zhang}}, \bibinfo {author} {\bibfnamefont {Q.}~\bibnamefont {Wang}}, \bibinfo {author} {\bibfnamefont {W.}~\bibnamefont {Zhu}}, \bibinfo {author} {\bibfnamefont {F.}~\bibnamefont {Pan}},\ and\ \bibinfo {author} {\bibfnamefont {C.}~\bibnamefont {Song}},\ }\bibfield  {title} {\bibinfo {title} {Electrical manipulation of spin splitting torque in altermagnetic {RuO$_2$}},\ }\href {https://doi.org/10.1038/s41467-025-60891-2} {\bibfield  {journal} {\bibinfo  {journal} {Nature Communications}\ }\textbf {\bibinfo {volume} {16}},\ \bibinfo {pages} {5646} (\bibinfo {year} {2025}{\natexlab{a}})}\BibitemShut {NoStop}%
\bibitem [{\citenamefont {Duan}\ \emph {et~al.}(2025)\citenamefont {Duan}, \citenamefont {Zhang}, \citenamefont {Zhu}, \citenamefont {Liu}, \citenamefont {Zhang}, \citenamefont {\ifmmode \check{Z}\else \v{Z}\fi{}uti\ifmmode~\acute{c}\else \'{c}\fi{}},\ and\ \citenamefont {Zhou}}]{ZhouTong:2025}%
  \BibitemOpen
  \bibfield  {author} {\bibinfo {author} {\bibfnamefont {X.}~\bibnamefont {Duan}}, \bibinfo {author} {\bibfnamefont {J.}~\bibnamefont {Zhang}}, \bibinfo {author} {\bibfnamefont {Z.}~\bibnamefont {Zhu}}, \bibinfo {author} {\bibfnamefont {Y.}~\bibnamefont {Liu}}, \bibinfo {author} {\bibfnamefont {Z.}~\bibnamefont {Zhang}}, \bibinfo {author} {\bibfnamefont {I.}~\bibnamefont {\ifmmode \check{Z}\else \v{Z}\fi{}uti\ifmmode~\acute{c}\else \'{c}\fi{}}},\ and\ \bibinfo {author} {\bibfnamefont {T.}~\bibnamefont {Zhou}},\ }\bibfield  {title} {\bibinfo {title} {Antiferroelectric altermagnets: Antiferroelectricity alters magnets},\ }\href {https://doi.org/10.1103/PhysRevLett.134.106801} {\bibfield  {journal} {\bibinfo  {journal} {Phys. Rev. Lett.}\ }\textbf {\bibinfo {volume} {134}},\ \bibinfo {pages} {106801} (\bibinfo {year} {2025})}\BibitemShut {NoStop}%
\bibitem [{\citenamefont {Gu}\ \emph {et~al.}(2025)\citenamefont {Gu}, \citenamefont {Liu}, \citenamefont {Zhu}, \citenamefont {Yananose}, \citenamefont {Chen}, \citenamefont {Hu}, \citenamefont {Stroppa},\ and\ \citenamefont {Liu}}]{LiuQihang:2025}%
  \BibitemOpen
  \bibfield  {author} {\bibinfo {author} {\bibfnamefont {M.}~\bibnamefont {Gu}}, \bibinfo {author} {\bibfnamefont {Y.}~\bibnamefont {Liu}}, \bibinfo {author} {\bibfnamefont {H.}~\bibnamefont {Zhu}}, \bibinfo {author} {\bibfnamefont {K.}~\bibnamefont {Yananose}}, \bibinfo {author} {\bibfnamefont {X.}~\bibnamefont {Chen}}, \bibinfo {author} {\bibfnamefont {Y.}~\bibnamefont {Hu}}, \bibinfo {author} {\bibfnamefont {A.}~\bibnamefont {Stroppa}},\ and\ \bibinfo {author} {\bibfnamefont {Q.}~\bibnamefont {Liu}},\ }\bibfield  {title} {\bibinfo {title} {Ferroelectric switchable altermagnetism},\ }\href {https://doi.org/10.1103/PhysRevLett.134.106802} {\bibfield  {journal} {\bibinfo  {journal} {Phys. Rev. Lett.}\ }\textbf {\bibinfo {volume} {134}},\ \bibinfo {pages} {106802} (\bibinfo {year} {2025})}\BibitemShut {NoStop}%
\bibitem [{\citenamefont {Yu}\ \emph {et~al.}(2025)\citenamefont {Yu}, \citenamefont {Zhao}, \citenamefont {Bellaiche},\ and\ \citenamefont {Ma}}]{MaYanming:2025}%
  \BibitemOpen
  \bibfield  {author} {\bibinfo {author} {\bibfnamefont {L.}~\bibnamefont {Yu}}, \bibinfo {author} {\bibfnamefont {H.~J.}\ \bibnamefont {Zhao}}, \bibinfo {author} {\bibfnamefont {L.}~\bibnamefont {Bellaiche}},\ and\ \bibinfo {author} {\bibfnamefont {Y.}~\bibnamefont {Ma}},\ }\bibfield  {title} {\bibinfo {title} {Electrically switchable nonrelativistic zeeman spin splittings in collinear antiferromagnets},\ }\href {https://doi.org/10.1103/96gy-sn83} {\bibfield  {journal} {\bibinfo  {journal} {Phys. Rev. Lett.}\ }\textbf {\bibinfo {volume} {135}},\ \bibinfo {pages} {256704} (\bibinfo {year} {2025})}\BibitemShut {NoStop}%
\bibitem [{\citenamefont {Hu}\ \emph {et~al.}(2025{\natexlab{b}})\citenamefont {Hu}, \citenamefont {Matsyshyn},\ and\ \citenamefont {Song}}]{NSM}%
  \BibitemOpen
  \bibfield  {author} {\bibinfo {author} {\bibfnamefont {J.-X.}\ \bibnamefont {Hu}}, \bibinfo {author} {\bibfnamefont {O.}~\bibnamefont {Matsyshyn}},\ and\ \bibinfo {author} {\bibfnamefont {J.~C.~W.}\ \bibnamefont {Song}},\ }\bibfield  {title} {\bibinfo {title} {Nonlinear superconducting magnetoelectric effect},\ }\href {https://doi.org/10.1103/PhysRevLett.134.026001} {\bibfield  {journal} {\bibinfo  {journal} {Phys. Rev. Lett.}\ }\textbf {\bibinfo {volume} {134}},\ \bibinfo {pages} {026001} (\bibinfo {year} {2025}{\natexlab{b}})}\BibitemShut {NoStop}%
\bibitem [{\citenamefont {Zhu}\ \emph {et~al.}(2023)\citenamefont {Zhu}, \citenamefont {Zhuang}, \citenamefont {Wu},\ and\ \citenamefont {Yan}}]{TSC_1}%
  \BibitemOpen
  \bibfield  {author} {\bibinfo {author} {\bibfnamefont {D.}~\bibnamefont {Zhu}}, \bibinfo {author} {\bibfnamefont {Z.-Y.}\ \bibnamefont {Zhuang}}, \bibinfo {author} {\bibfnamefont {Z.}~\bibnamefont {Wu}},\ and\ \bibinfo {author} {\bibfnamefont {Z.}~\bibnamefont {Yan}},\ }\bibfield  {title} {\bibinfo {title} {Topological superconductivity in two-dimensional altermagnetic metals},\ }\href {https://doi.org/10.1103/PhysRevB.108.184505} {\bibfield  {journal} {\bibinfo  {journal} {Phys. Rev. B}\ }\textbf {\bibinfo {volume} {108}},\ \bibinfo {pages} {184505} (\bibinfo {year} {2023})}\BibitemShut {NoStop}%
\bibitem [{\citenamefont {Sumita}\ \emph {et~al.}(2023)\citenamefont {Sumita}, \citenamefont {Naka},\ and\ \citenamefont {Seo}}]{FFLO}%
  \BibitemOpen
  \bibfield  {author} {\bibinfo {author} {\bibfnamefont {S.}~\bibnamefont {Sumita}}, \bibinfo {author} {\bibfnamefont {M.}~\bibnamefont {Naka}},\ and\ \bibinfo {author} {\bibfnamefont {H.}~\bibnamefont {Seo}},\ }\bibfield  {title} {\bibinfo {title} {Fulde-ferrell-larkin-ovchinnikov state induced by antiferromagnetic order in $\ensuremath{\kappa}$-type organic conductors},\ }\href {https://doi.org/10.1103/PhysRevResearch.5.043171} {\bibfield  {journal} {\bibinfo  {journal} {Phys. Rev. Res.}\ }\textbf {\bibinfo {volume} {5}},\ \bibinfo {pages} {043171} (\bibinfo {year} {2023})}\BibitemShut {NoStop}%
\bibitem [{\citenamefont {Zhang}\ \emph {et~al.}(2024{\natexlab{b}})\citenamefont {Zhang}, \citenamefont {Hu},\ and\ \citenamefont {Neupert}}]{FFLO_3}%
  \BibitemOpen
  \bibfield  {author} {\bibinfo {author} {\bibfnamefont {S.-B.}\ \bibnamefont {Zhang}}, \bibinfo {author} {\bibfnamefont {L.-H.}\ \bibnamefont {Hu}},\ and\ \bibinfo {author} {\bibfnamefont {T.}~\bibnamefont {Neupert}},\ }\bibfield  {title} {\bibinfo {title} {Finite-momentum cooper pairing in proximitized altermagnets},\ }\href {https://doi.org/10.1038/s41467-024-45951-3} {\bibfield  {journal} {\bibinfo  {journal} {Nature Communications}\ }\textbf {\bibinfo {volume} {15}},\ \bibinfo {pages} {1801} (\bibinfo {year} {2024}{\natexlab{b}})}\BibitemShut {NoStop}%
\bibitem [{\citenamefont {Ghorashi}\ \emph {et~al.}(2024)\citenamefont {Ghorashi}, \citenamefont {Hughes},\ and\ \citenamefont {Cano}}]{TSC_2}%
  \BibitemOpen
  \bibfield  {author} {\bibinfo {author} {\bibfnamefont {S.~A.~A.}\ \bibnamefont {Ghorashi}}, \bibinfo {author} {\bibfnamefont {T.~L.}\ \bibnamefont {Hughes}},\ and\ \bibinfo {author} {\bibfnamefont {J.}~\bibnamefont {Cano}},\ }\bibfield  {title} {\bibinfo {title} {Altermagnetic routes to majorana modes in zero net magnetization},\ }\href {https://doi.org/10.1103/PhysRevLett.133.106601} {\bibfield  {journal} {\bibinfo  {journal} {Phys. Rev. Lett.}\ }\textbf {\bibinfo {volume} {133}},\ \bibinfo {pages} {106601} (\bibinfo {year} {2024})}\BibitemShut {NoStop}%
\bibitem [{\citenamefont {Chakraborty}\ and\ \citenamefont {Black-Schaffer}(2024)}]{FFLO_2}%
  \BibitemOpen
  \bibfield  {author} {\bibinfo {author} {\bibfnamefont {D.}~\bibnamefont {Chakraborty}}\ and\ \bibinfo {author} {\bibfnamefont {A.~M.}\ \bibnamefont {Black-Schaffer}},\ }\bibfield  {title} {\bibinfo {title} {Zero-field finite-momentum and field-induced superconductivity in altermagnets},\ }\href {https://doi.org/10.1103/PhysRevB.110.L060508} {\bibfield  {journal} {\bibinfo  {journal} {Phys. Rev. B}\ }\textbf {\bibinfo {volume} {110}},\ \bibinfo {pages} {L060508} (\bibinfo {year} {2024})}\BibitemShut {NoStop}%
\bibitem [{\citenamefont {Hong}\ \emph {et~al.}(2025)\citenamefont {Hong}, \citenamefont {Park},\ and\ \citenamefont {Kim}}]{pwavesc}%
  \BibitemOpen
  \bibfield  {author} {\bibinfo {author} {\bibfnamefont {S.}~\bibnamefont {Hong}}, \bibinfo {author} {\bibfnamefont {M.~J.}\ \bibnamefont {Park}},\ and\ \bibinfo {author} {\bibfnamefont {K.-M.}\ \bibnamefont {Kim}},\ }\bibfield  {title} {\bibinfo {title} {Unconventional $p$-wave and finite-momentum superconductivity induced by altermagnetism through the formation of bogoliubov fermi surface},\ }\href {https://doi.org/10.1103/PhysRevB.111.054501} {\bibfield  {journal} {\bibinfo  {journal} {Phys. Rev. B}\ }\textbf {\bibinfo {volume} {111}},\ \bibinfo {pages} {054501} (\bibinfo {year} {2025})}\BibitemShut {NoStop}%
\bibitem [{\citenamefont {Mukasa}\ and\ \citenamefont {Masaki}(2025)}]{FFLO_4}%
  \BibitemOpen
  \bibfield  {author} {\bibinfo {author} {\bibfnamefont {K.}~\bibnamefont {Mukasa}}\ and\ \bibinfo {author} {\bibfnamefont {Y.}~\bibnamefont {Masaki}},\ }\bibfield  {title} {\bibinfo {title} {Finite-momentum superconductivity in two-dimensional altermagnets with a rashba-type spin–orbit coupling},\ }\href {https://doi.org/10.7566/JPSJ.94.064705} {\bibfield  {journal} {\bibinfo  {journal} {Journal of the Physical Society of Japan}\ }\textbf {\bibinfo {volume} {94}},\ \bibinfo {pages} {064705} (\bibinfo {year} {2025})}\BibitemShut {NoStop}%
\bibitem [{\citenamefont {Luo}\ \emph {et~al.}(2026{\natexlab{a}})\citenamefont {Luo}, \citenamefont {Sun}, \citenamefont {Feng}, \citenamefont {Tian},\ and\ \citenamefont {Law}}]{TSC_3}%
  \BibitemOpen
  \bibfield  {author} {\bibinfo {author} {\bibfnamefont {X.-J.}\ \bibnamefont {Luo}}, \bibinfo {author} {\bibfnamefont {Z.-T.}\ \bibnamefont {Sun}}, \bibinfo {author} {\bibfnamefont {X.}~\bibnamefont {Feng}}, \bibinfo {author} {\bibfnamefont {M.}~\bibnamefont {Tian}},\ and\ \bibinfo {author} {\bibfnamefont {K.~T.}\ \bibnamefont {Law}},\ }\href {https://arxiv.org/abs/2603.15147} {\bibinfo {title} {Hidden zeeman field in odd-parity magnets: An ideal platform for topological superconductivity}} (\bibinfo {year} {2026}{\natexlab{a}}),\ \Eprint {https://arxiv.org/abs/2603.15147} {arXiv:2603.15147 [cond-mat.supr-con]} \BibitemShut {NoStop}%
\bibitem [{\citenamefont {Brinkman}\ and\ \citenamefont {Elliott}(1966)}]{spingroup1}%
  \BibitemOpen
  \bibfield  {author} {\bibinfo {author} {\bibfnamefont {W.~F.}\ \bibnamefont {Brinkman}}\ and\ \bibinfo {author} {\bibfnamefont {R.~J.}\ \bibnamefont {Elliott}},\ }\bibfield  {title} {\bibinfo {title} {Theory of spin-space groups},\ }\href {https://doi.org/10.1098/rspa.1966.0211} {\bibfield  {journal} {\bibinfo  {journal} {Proceedings of the Royal Society of London. A. Mathematical and Physical Sciences}\ }\textbf {\bibinfo {volume} {294}},\ \bibinfo {pages} {343} (\bibinfo {year} {1966})}\BibitemShut {NoStop}%
\bibitem [{\citenamefont {Litvin}\ and\ \citenamefont {Opechowski}(1974)}]{spingroup2}%
  \BibitemOpen
  \bibfield  {author} {\bibinfo {author} {\bibfnamefont {D.}~\bibnamefont {Litvin}}\ and\ \bibinfo {author} {\bibfnamefont {W.}~\bibnamefont {Opechowski}},\ }\bibfield  {title} {\bibinfo {title} {Spin groups},\ }\href {https://doi.org/https://doi.org/10.1016/0031-8914(74)90157-8} {\bibfield  {journal} {\bibinfo  {journal} {Physica}\ }\textbf {\bibinfo {volume} {76}},\ \bibinfo {pages} {538} (\bibinfo {year} {1974})}\BibitemShut {NoStop}%
\bibitem [{\citenamefont {Liu}\ \emph {et~al.}(2022)\citenamefont {Liu}, \citenamefont {Li}, \citenamefont {Han}, \citenamefont {Wan},\ and\ \citenamefont {Liu}}]{spingroup3}%
  \BibitemOpen
  \bibfield  {author} {\bibinfo {author} {\bibfnamefont {P.}~\bibnamefont {Liu}}, \bibinfo {author} {\bibfnamefont {J.}~\bibnamefont {Li}}, \bibinfo {author} {\bibfnamefont {J.}~\bibnamefont {Han}}, \bibinfo {author} {\bibfnamefont {X.}~\bibnamefont {Wan}},\ and\ \bibinfo {author} {\bibfnamefont {Q.}~\bibnamefont {Liu}},\ }\bibfield  {title} {\bibinfo {title} {Spin-group symmetry in magnetic materials with negligible spin-orbit coupling},\ }\href {https://doi.org/10.1103/PhysRevX.12.021016} {\bibfield  {journal} {\bibinfo  {journal} {Phys. Rev. X}\ }\textbf {\bibinfo {volume} {12}},\ \bibinfo {pages} {021016} (\bibinfo {year} {2022})}\BibitemShut {NoStop}%
\bibitem [{\citenamefont {Chen}\ \emph {et~al.}(2024)\citenamefont {Chen}, \citenamefont {Ren}, \citenamefont {Zhu}, \citenamefont {Yu}, \citenamefont {Zhang}, \citenamefont {Liu}, \citenamefont {Li}, \citenamefont {Liu}, \citenamefont {Li},\ and\ \citenamefont {Liu}}]{spingroup4}%
  \BibitemOpen
  \bibfield  {author} {\bibinfo {author} {\bibfnamefont {X.}~\bibnamefont {Chen}}, \bibinfo {author} {\bibfnamefont {J.}~\bibnamefont {Ren}}, \bibinfo {author} {\bibfnamefont {Y.}~\bibnamefont {Zhu}}, \bibinfo {author} {\bibfnamefont {Y.}~\bibnamefont {Yu}}, \bibinfo {author} {\bibfnamefont {A.}~\bibnamefont {Zhang}}, \bibinfo {author} {\bibfnamefont {P.}~\bibnamefont {Liu}}, \bibinfo {author} {\bibfnamefont {J.}~\bibnamefont {Li}}, \bibinfo {author} {\bibfnamefont {Y.}~\bibnamefont {Liu}}, \bibinfo {author} {\bibfnamefont {C.}~\bibnamefont {Li}},\ and\ \bibinfo {author} {\bibfnamefont {Q.}~\bibnamefont {Liu}},\ }\bibfield  {title} {\bibinfo {title} {Enumeration and representation theory of spin space groups},\ }\href {https://doi.org/10.1103/PhysRevX.14.031038} {\bibfield  {journal} {\bibinfo  {journal} {Phys. Rev. X}\ }\textbf {\bibinfo {volume} {14}},\ \bibinfo {pages} {031038} (\bibinfo {year} {2024})}\BibitemShut {NoStop}%
\bibitem [{\citenamefont {Jiang}\ \emph {et~al.}(2024)\citenamefont {Jiang}, \citenamefont {Song}, \citenamefont {Zhu}, \citenamefont {Fang}, \citenamefont {Weng}, \citenamefont {Liu}, \citenamefont {Yang},\ and\ \citenamefont {Fang}}]{spingroup5}%
  \BibitemOpen
  \bibfield  {author} {\bibinfo {author} {\bibfnamefont {Y.}~\bibnamefont {Jiang}}, \bibinfo {author} {\bibfnamefont {Z.}~\bibnamefont {Song}}, \bibinfo {author} {\bibfnamefont {T.}~\bibnamefont {Zhu}}, \bibinfo {author} {\bibfnamefont {Z.}~\bibnamefont {Fang}}, \bibinfo {author} {\bibfnamefont {H.}~\bibnamefont {Weng}}, \bibinfo {author} {\bibfnamefont {Z.-X.}\ \bibnamefont {Liu}}, \bibinfo {author} {\bibfnamefont {J.}~\bibnamefont {Yang}},\ and\ \bibinfo {author} {\bibfnamefont {C.}~\bibnamefont {Fang}},\ }\bibfield  {title} {\bibinfo {title} {Enumeration of spin-space groups: Toward a complete description of symmetries of magnetic orders},\ }\href {https://doi.org/10.1103/PhysRevX.14.031039} {\bibfield  {journal} {\bibinfo  {journal} {Phys. Rev. X}\ }\textbf {\bibinfo {volume} {14}},\ \bibinfo {pages} {031039} (\bibinfo {year} {2024})}\BibitemShut {NoStop}%
\bibitem [{\citenamefont {Xiao}\ \emph {et~al.}(2024)\citenamefont {Xiao}, \citenamefont {Zhao}, \citenamefont {Li}, \citenamefont {Shindou},\ and\ \citenamefont {Song}}]{spingroup6}%
  \BibitemOpen
  \bibfield  {author} {\bibinfo {author} {\bibfnamefont {Z.}~\bibnamefont {Xiao}}, \bibinfo {author} {\bibfnamefont {J.}~\bibnamefont {Zhao}}, \bibinfo {author} {\bibfnamefont {Y.}~\bibnamefont {Li}}, \bibinfo {author} {\bibfnamefont {R.}~\bibnamefont {Shindou}},\ and\ \bibinfo {author} {\bibfnamefont {Z.-D.}\ \bibnamefont {Song}},\ }\bibfield  {title} {\bibinfo {title} {Spin space groups: Full classification and applications},\ }\href {https://doi.org/10.1103/PhysRevX.14.031037} {\bibfield  {journal} {\bibinfo  {journal} {Phys. Rev. X}\ }\textbf {\bibinfo {volume} {14}},\ \bibinfo {pages} {031037} (\bibinfo {year} {2024})}\BibitemShut {NoStop}%
\bibitem [{\citenamefont {Chen}\ \emph {et~al.}(2025{\natexlab{a}})\citenamefont {Chen}, \citenamefont {Liu}, \citenamefont {Liu}, \citenamefont {Yu}, \citenamefont {Ren}, \citenamefont {Li}, \citenamefont {Zhang},\ and\ \citenamefont {Liu}}]{spingroup_lqh}%
  \BibitemOpen
  \bibfield  {author} {\bibinfo {author} {\bibfnamefont {X.}~\bibnamefont {Chen}}, \bibinfo {author} {\bibfnamefont {Y.}~\bibnamefont {Liu}}, \bibinfo {author} {\bibfnamefont {P.}~\bibnamefont {Liu}}, \bibinfo {author} {\bibfnamefont {Y.}~\bibnamefont {Yu}}, \bibinfo {author} {\bibfnamefont {J.}~\bibnamefont {Ren}}, \bibinfo {author} {\bibfnamefont {J.}~\bibnamefont {Li}}, \bibinfo {author} {\bibfnamefont {A.}~\bibnamefont {Zhang}},\ and\ \bibinfo {author} {\bibfnamefont {Q.}~\bibnamefont {Liu}},\ }\bibfield  {title} {\bibinfo {title} {Unconventional magnons in collinear magnets dictated by spin space groups},\ }\href {https://doi.org/10.1038/s41586-025-08715-7} {\bibfield  {journal} {\bibinfo  {journal} {Nature}\ }\textbf {\bibinfo {volume} {640}},\ \bibinfo {pages} {349} (\bibinfo {year} {2025}{\natexlab{a}})}\BibitemShut {NoStop}%
\bibitem [{\citenamefont {Luo}\ \emph {et~al.}(2026{\natexlab{b}})\citenamefont {Luo}, \citenamefont {Hu}, \citenamefont {Hu},\ and\ \citenamefont {Law}}]{luo2026spingroup_1}%
  \BibitemOpen
  \bibfield  {author} {\bibinfo {author} {\bibfnamefont {X.-J.}\ \bibnamefont {Luo}}, \bibinfo {author} {\bibfnamefont {J.-X.}\ \bibnamefont {Hu}}, \bibinfo {author} {\bibfnamefont {M.-L.}\ \bibnamefont {Hu}},\ and\ \bibinfo {author} {\bibfnamefont {K.~T.}\ \bibnamefont {Law}},\ }\href {https://arxiv.org/abs/2510.05512} {\bibinfo {title} {Spin group symmetry criteria for odd-parity magnets}} (\bibinfo {year} {2026}{\natexlab{b}}),\ \Eprint {https://arxiv.org/abs/2510.05512} {arXiv:2510.05512 [cond-mat.other]} \BibitemShut {NoStop}%
\bibitem [{\citenamefont {Luo}\ \emph {et~al.}(2026{\natexlab{c}})\citenamefont {Luo}, \citenamefont {Hu}, \citenamefont {Hu},\ and\ \citenamefont {Law}}]{luo2026spingroup_2}%
  \BibitemOpen
  \bibfield  {author} {\bibinfo {author} {\bibfnamefont {X.-J.}\ \bibnamefont {Luo}}, \bibinfo {author} {\bibfnamefont {J.-X.}\ \bibnamefont {Hu}}, \bibinfo {author} {\bibfnamefont {M.}~\bibnamefont {Hu}},\ and\ \bibinfo {author} {\bibfnamefont {K.~T.}\ \bibnamefont {Law}},\ }\href {https://arxiv.org/abs/2603.07643} {\bibinfo {title} {Spin group symmetry criteria for unconventional magnetism}} (\bibinfo {year} {2026}{\natexlab{c}}),\ \Eprint {https://arxiv.org/abs/2603.07643} {arXiv:2603.07643 [cond-mat.str-el]} \BibitemShut {NoStop}%
\bibitem [{\citenamefont {Šmejkal}\ \emph {et~al.}(2020)\citenamefont {Šmejkal}, \citenamefont {González-Hernández}, \citenamefont {Jungwirth},\ and\ \citenamefont {Sinova}}]{ruo2_theory}%
  \BibitemOpen
  \bibfield  {author} {\bibinfo {author} {\bibfnamefont {L.}~\bibnamefont {Šmejkal}}, \bibinfo {author} {\bibfnamefont {R.}~\bibnamefont {González-Hernández}}, \bibinfo {author} {\bibfnamefont {T.}~\bibnamefont {Jungwirth}},\ and\ \bibinfo {author} {\bibfnamefont {J.}~\bibnamefont {Sinova}},\ }\bibfield  {title} {\bibinfo {title} {Crystal time-reversal symmetry breaking and spontaneous hall effect in collinear antiferromagnets},\ }\href {https://doi.org/10.1126/sciadv.aaz8809} {\bibfield  {journal} {\bibinfo  {journal} {Science Advances}\ }\textbf {\bibinfo {volume} {6}},\ \bibinfo {pages} {eaaz8809} (\bibinfo {year} {2020})}\BibitemShut {NoStop}%
\bibitem [{\citenamefont {Feng}\ \emph {et~al.}(2022)\citenamefont {Feng}, \citenamefont {Zhou}, \citenamefont {{\v{S}}mejkal}, \citenamefont {Wu}, \citenamefont {Zhu}, \citenamefont {Guo}, \citenamefont {Gonz{\'a}lez-Hern{\'a}ndez}, \citenamefont {Wang}, \citenamefont {Yan}, \citenamefont {Qin}, \citenamefont {Zhang}, \citenamefont {Wu}, \citenamefont {Chen}, \citenamefont {Meng}, \citenamefont {Liu}, \citenamefont {Xia}, \citenamefont {Sinova}, \citenamefont {Jungwirth},\ and\ \citenamefont {Liu}}]{ruo2_anomalous_Hall}%
  \BibitemOpen
  \bibfield  {author} {\bibinfo {author} {\bibfnamefont {Z.}~\bibnamefont {Feng}}, \bibinfo {author} {\bibfnamefont {X.}~\bibnamefont {Zhou}}, \bibinfo {author} {\bibfnamefont {L.}~\bibnamefont {{\v{S}}mejkal}}, \bibinfo {author} {\bibfnamefont {L.}~\bibnamefont {Wu}}, \bibinfo {author} {\bibfnamefont {Z.}~\bibnamefont {Zhu}}, \bibinfo {author} {\bibfnamefont {H.}~\bibnamefont {Guo}}, \bibinfo {author} {\bibfnamefont {R.}~\bibnamefont {Gonz{\'a}lez-Hern{\'a}ndez}}, \bibinfo {author} {\bibfnamefont {X.}~\bibnamefont {Wang}}, \bibinfo {author} {\bibfnamefont {H.}~\bibnamefont {Yan}}, \bibinfo {author} {\bibfnamefont {P.}~\bibnamefont {Qin}}, \bibinfo {author} {\bibfnamefont {X.}~\bibnamefont {Zhang}}, \bibinfo {author} {\bibfnamefont {H.}~\bibnamefont {Wu}}, \bibinfo {author} {\bibfnamefont {H.}~\bibnamefont {Chen}}, \bibinfo {author} {\bibfnamefont {Z.}~\bibnamefont {Meng}}, \bibinfo {author} {\bibfnamefont {L.}~\bibnamefont {Liu}}, \bibinfo {author} {\bibfnamefont {Z.}~\bibnamefont {Xia}}, \bibinfo {author} {\bibfnamefont {J.}~\bibnamefont {Sinova}}, \bibinfo {author} {\bibfnamefont {T.}~\bibnamefont {Jungwirth}},\ and\ \bibinfo {author} {\bibfnamefont {Z.}~\bibnamefont {Liu}},\ }\bibfield  {title} {\bibinfo {title} {An anomalous hall effect in altermagnetic ruthenium dioxide},\ }\href {https://doi.org/10.1038/s41928-022-00866-z} {\bibfield  {journal} {\bibinfo  {journal} {Nature Electronics}\ }\textbf {\bibinfo {volume} {5}},\ \bibinfo {pages} {735} (\bibinfo {year} {2022})}\BibitemShut {NoStop}%
\bibitem [{\citenamefont {Zhou}\ \emph {et~al.}(2024)\citenamefont {Zhou}, \citenamefont {Feng}, \citenamefont {Zhang}, \citenamefont {\ifmmode~\check{S}\else \v{S}\fi{}mejkal}, \citenamefont {Sinova}, \citenamefont {Mokrousov},\ and\ \citenamefont {Yao}}]{Transport_ruo2}%
  \BibitemOpen
  \bibfield  {author} {\bibinfo {author} {\bibfnamefont {X.}~\bibnamefont {Zhou}}, \bibinfo {author} {\bibfnamefont {W.}~\bibnamefont {Feng}}, \bibinfo {author} {\bibfnamefont {R.-W.}\ \bibnamefont {Zhang}}, \bibinfo {author} {\bibfnamefont {L.}~\bibnamefont {\ifmmode~\check{S}\else \v{S}\fi{}mejkal}}, \bibinfo {author} {\bibfnamefont {J.}~\bibnamefont {Sinova}}, \bibinfo {author} {\bibfnamefont {Y.}~\bibnamefont {Mokrousov}},\ and\ \bibinfo {author} {\bibfnamefont {Y.}~\bibnamefont {Yao}},\ }\bibfield  {title} {\bibinfo {title} {Crystal thermal transport in altermagnetic {${\mathrm{RuO}}_{2}$}},\ }\href {https://doi.org/10.1103/PhysRevLett.132.056701} {\bibfield  {journal} {\bibinfo  {journal} {Phys. Rev. Lett.}\ }\textbf {\bibinfo {volume} {132}},\ \bibinfo {pages} {056701} (\bibinfo {year} {2024})}\BibitemShut {NoStop}%
\bibitem [{\citenamefont {Fedchenko}\ \emph {et~al.}(2024)\citenamefont {Fedchenko}, \citenamefont {Minár}, \citenamefont {Akashdeep}, \citenamefont {D’Souza}, \citenamefont {Vasilyev}, \citenamefont {Tkach}, \citenamefont {Odenbreit}, \citenamefont {Nguyen}, \citenamefont {Kutnyakhov}, \citenamefont {Wind}, \citenamefont {Wenthaus}, \citenamefont {Scholz}, \citenamefont {Rossnagel}, \citenamefont {Hoesch}, \citenamefont {Aeschlimann}, \citenamefont {Stadtmüller}, \citenamefont {Kläui}, \citenamefont {Schönhense}, \citenamefont {Jungwirth}, \citenamefont {Hellenes}, \citenamefont {Jakob}, \citenamefont {Šmejkal}, \citenamefont {Sinova},\ and\ \citenamefont {Elmers}}]{ruo_arpes_2}%
  \BibitemOpen
  \bibfield  {author} {\bibinfo {author} {\bibfnamefont {O.}~\bibnamefont {Fedchenko}}, \bibinfo {author} {\bibfnamefont {J.}~\bibnamefont {Minár}}, \bibinfo {author} {\bibfnamefont {A.}~\bibnamefont {Akashdeep}}, \bibinfo {author} {\bibfnamefont {S.~W.}\ \bibnamefont {D’Souza}}, \bibinfo {author} {\bibfnamefont {D.}~\bibnamefont {Vasilyev}}, \bibinfo {author} {\bibfnamefont {O.}~\bibnamefont {Tkach}}, \bibinfo {author} {\bibfnamefont {L.}~\bibnamefont {Odenbreit}}, \bibinfo {author} {\bibfnamefont {Q.}~\bibnamefont {Nguyen}}, \bibinfo {author} {\bibfnamefont {D.}~\bibnamefont {Kutnyakhov}}, \bibinfo {author} {\bibfnamefont {N.}~\bibnamefont {Wind}}, \bibinfo {author} {\bibfnamefont {L.}~\bibnamefont {Wenthaus}}, \bibinfo {author} {\bibfnamefont {M.}~\bibnamefont {Scholz}}, \bibinfo {author} {\bibfnamefont {K.}~\bibnamefont {Rossnagel}}, \bibinfo {author} {\bibfnamefont {M.}~\bibnamefont {Hoesch}}, \bibinfo {author} {\bibfnamefont {M.}~\bibnamefont {Aeschlimann}}, \bibinfo {author} {\bibfnamefont {B.}~\bibnamefont {Stadtmüller}}, \bibinfo {author} {\bibfnamefont {M.}~\bibnamefont {Kläui}}, \bibinfo {author} {\bibfnamefont {G.}~\bibnamefont {Schönhense}}, \bibinfo {author} {\bibfnamefont {T.}~\bibnamefont {Jungwirth}}, \bibinfo {author} {\bibfnamefont {A.~B.}\ \bibnamefont {Hellenes}}, \bibinfo {author} {\bibfnamefont {G.}~\bibnamefont {Jakob}}, \bibinfo {author} {\bibfnamefont {L.}~\bibnamefont {Šmejkal}}, \bibinfo {author} {\bibfnamefont {J.}~\bibnamefont {Sinova}},\ and\ \bibinfo {author} {\bibfnamefont {H.-J.}\ \bibnamefont {Elmers}},\ }\bibfield  {title} {\bibinfo {title} {Observation of time-reversal symmetry breaking in the band structure of altermagnetic {RuO$_2$}},\ }\href {https://doi.org/10.1126/sciadv.adj4883} {\bibfield  {journal} {\bibinfo  {journal} {Science Advances}\ }\textbf {\bibinfo {volume} {10}},\ \bibinfo {pages} {eadj4883} (\bibinfo {year} {2024})}\BibitemShut {NoStop}%
\bibitem [{\citenamefont {Lin}\ \emph {et~al.}(2024)\citenamefont {Lin}, \citenamefont {Chen}, \citenamefont {Lu}, \citenamefont {Liang}, \citenamefont {Feng}, \citenamefont {Yamagami}, \citenamefont {Osiecki}, \citenamefont {Leandersson}, \citenamefont {Thiagarajan}, \citenamefont {Liu}, \citenamefont {Felser},\ and\ \citenamefont {Ma}}]{ruo_arpes_1}%
  \BibitemOpen
  \bibfield  {author} {\bibinfo {author} {\bibfnamefont {Z.}~\bibnamefont {Lin}}, \bibinfo {author} {\bibfnamefont {D.}~\bibnamefont {Chen}}, \bibinfo {author} {\bibfnamefont {W.}~\bibnamefont {Lu}}, \bibinfo {author} {\bibfnamefont {X.}~\bibnamefont {Liang}}, \bibinfo {author} {\bibfnamefont {S.}~\bibnamefont {Feng}}, \bibinfo {author} {\bibfnamefont {K.}~\bibnamefont {Yamagami}}, \bibinfo {author} {\bibfnamefont {J.}~\bibnamefont {Osiecki}}, \bibinfo {author} {\bibfnamefont {M.}~\bibnamefont {Leandersson}}, \bibinfo {author} {\bibfnamefont {B.}~\bibnamefont {Thiagarajan}}, \bibinfo {author} {\bibfnamefont {J.}~\bibnamefont {Liu}}, \bibinfo {author} {\bibfnamefont {C.}~\bibnamefont {Felser}},\ and\ \bibinfo {author} {\bibfnamefont {J.}~\bibnamefont {Ma}},\ }\href {https://arxiv.org/abs/2402.04995} {\bibinfo {title} {Observation of giant spin splitting and d-wave spin texture in room temperature altermagnet {RuO$_2$}}} (\bibinfo {year} {2024}),\ \Eprint {https://arxiv.org/abs/2402.04995} {arXiv:2402.04995 [cond-mat.mtrl-sci]} \BibitemShut {NoStop}%
\bibitem [{\citenamefont {Chen}\ \emph {et~al.}(2025{\natexlab{b}})\citenamefont {Chen}, \citenamefont {Wang}, \citenamefont {Qin}, \citenamefont {Meng}, \citenamefont {Zhou}, \citenamefont {Wang}, \citenamefont {Liu}, \citenamefont {Zhao}, \citenamefont {Duan}, \citenamefont {Zhang}, \citenamefont {Liu}, \citenamefont {Shao}, \citenamefont {Jiang},\ and\ \citenamefont {Liu}}]{ruo2_Magnetoresistance}%
  \BibitemOpen
  \bibfield  {author} {\bibinfo {author} {\bibfnamefont {H.}~\bibnamefont {Chen}}, \bibinfo {author} {\bibfnamefont {Z.-A.}\ \bibnamefont {Wang}}, \bibinfo {author} {\bibfnamefont {P.}~\bibnamefont {Qin}}, \bibinfo {author} {\bibfnamefont {Z.}~\bibnamefont {Meng}}, \bibinfo {author} {\bibfnamefont {X.}~\bibnamefont {Zhou}}, \bibinfo {author} {\bibfnamefont {X.}~\bibnamefont {Wang}}, \bibinfo {author} {\bibfnamefont {L.}~\bibnamefont {Liu}}, \bibinfo {author} {\bibfnamefont {G.}~\bibnamefont {Zhao}}, \bibinfo {author} {\bibfnamefont {Z.}~\bibnamefont {Duan}}, \bibinfo {author} {\bibfnamefont {T.}~\bibnamefont {Zhang}}, \bibinfo {author} {\bibfnamefont {J.}~\bibnamefont {Liu}}, \bibinfo {author} {\bibfnamefont {D.-F.}\ \bibnamefont {Shao}}, \bibinfo {author} {\bibfnamefont {C.}~\bibnamefont {Jiang}},\ and\ \bibinfo {author} {\bibfnamefont {Z.}~\bibnamefont {Liu}},\ }\bibfield  {title} {\bibinfo {title} {Spin-splitting magnetoresistance in altermagnetic {RuO$_2$} thin films},\ }\href {https://doi.org/https://doi.org/10.1002/adma.202507764} {\bibfield  {journal} {\bibinfo  {journal} {Advanced Materials}\ }\textbf {\bibinfo {volume} {37}},\ \bibinfo {pages} {2507764} (\bibinfo {year} {2025}{\natexlab{b}})}\BibitemShut {NoStop}%
\bibitem [{\citenamefont {Wang}\ \emph {et~al.}(2026)\citenamefont {Wang}, \citenamefont {Shen}, \citenamefont {Lin}, \citenamefont {Hsu}, \citenamefont {Chen}, \citenamefont {Chin}, \citenamefont {Singh}, \citenamefont {Lee}, \citenamefont {Chen}, \citenamefont {Huang},\ and\ \citenamefont {Qu}}]{ruo2_Transport}%
  \BibitemOpen
  \bibfield  {author} {\bibinfo {author} {\bibfnamefont {Y.-C.}\ \bibnamefont {Wang}}, \bibinfo {author} {\bibfnamefont {Z.-Y.}\ \bibnamefont {Shen}}, \bibinfo {author} {\bibfnamefont {C.-H.}\ \bibnamefont {Lin}}, \bibinfo {author} {\bibfnamefont {W.-C.}\ \bibnamefont {Hsu}}, \bibinfo {author} {\bibfnamefont {Y.-S.}\ \bibnamefont {Chen}}, \bibinfo {author} {\bibfnamefont {Y.-Y.}\ \bibnamefont {Chin}}, \bibinfo {author} {\bibfnamefont {A.~K.}\ \bibnamefont {Singh}}, \bibinfo {author} {\bibfnamefont {W.-L.}\ \bibnamefont {Lee}}, \bibinfo {author} {\bibfnamefont {C.-T.}\ \bibnamefont {Chen}}, \bibinfo {author} {\bibfnamefont {S.-Y.}\ \bibnamefont {Huang}},\ and\ \bibinfo {author} {\bibfnamefont {D.}~\bibnamefont {Qu}},\ }\bibfield  {title} {\bibinfo {title} {Absence of transport altermagnetic spin-splitting effect in {RuO$_2$}},\ }\href {https://doi.org/10.1021/acs.nanolett.5c05787} {\bibfield  {journal} {\bibinfo  {journal} {Nano Letters}\ }\textbf {\bibinfo {volume} {26}},\ \bibinfo {pages} {2548} (\bibinfo {year} {2026})}\BibitemShut {NoStop}%
\bibitem [{\citenamefont {Jiang}\ \emph {et~al.}(2025)\citenamefont {Jiang}, \citenamefont {Hu}, \citenamefont {Bai}, \citenamefont {Song}, \citenamefont {Mu}, \citenamefont {Qu}, \citenamefont {Li}, \citenamefont {Zhu}, \citenamefont {Pi}, \citenamefont {Wei}, \citenamefont {Sun}, \citenamefont {Huang}, \citenamefont {Zheng}, \citenamefont {Peng}, \citenamefont {He}, \citenamefont {Li}, \citenamefont {Luo}, \citenamefont {Li}, \citenamefont {Chen}, \citenamefont {Li}, \citenamefont {Weng},\ and\ \citenamefont {Qian}}]{KV2Se2O_ARPES}%
  \BibitemOpen
  \bibfield  {author} {\bibinfo {author} {\bibfnamefont {B.}~\bibnamefont {Jiang}}, \bibinfo {author} {\bibfnamefont {M.}~\bibnamefont {Hu}}, \bibinfo {author} {\bibfnamefont {J.}~\bibnamefont {Bai}}, \bibinfo {author} {\bibfnamefont {Z.}~\bibnamefont {Song}}, \bibinfo {author} {\bibfnamefont {C.}~\bibnamefont {Mu}}, \bibinfo {author} {\bibfnamefont {G.}~\bibnamefont {Qu}}, \bibinfo {author} {\bibfnamefont {W.}~\bibnamefont {Li}}, \bibinfo {author} {\bibfnamefont {W.}~\bibnamefont {Zhu}}, \bibinfo {author} {\bibfnamefont {H.}~\bibnamefont {Pi}}, \bibinfo {author} {\bibfnamefont {Z.}~\bibnamefont {Wei}}, \bibinfo {author} {\bibfnamefont {Y.-J.}\ \bibnamefont {Sun}}, \bibinfo {author} {\bibfnamefont {Y.}~\bibnamefont {Huang}}, \bibinfo {author} {\bibfnamefont {X.}~\bibnamefont {Zheng}}, \bibinfo {author} {\bibfnamefont {Y.}~\bibnamefont {Peng}}, \bibinfo {author} {\bibfnamefont {L.}~\bibnamefont {He}}, \bibinfo {author} {\bibfnamefont {S.}~\bibnamefont {Li}}, \bibinfo {author} {\bibfnamefont {J.}~\bibnamefont {Luo}}, \bibinfo {author} {\bibfnamefont {Z.}~\bibnamefont {Li}}, \bibinfo {author} {\bibfnamefont {G.}~\bibnamefont {Chen}}, \bibinfo {author} {\bibfnamefont {H.}~\bibnamefont {Li}}, \bibinfo {author} {\bibfnamefont {H.}~\bibnamefont {Weng}},\ and\ \bibinfo {author} {\bibfnamefont {T.}~\bibnamefont {Qian}},\ }\bibfield  {title} {\bibinfo {title} {A metallic room-temperature d-wave altermagnet},\ }\href {https://doi.org/10.1038/s41567-025-02822-y} {\bibfield  {journal} {\bibinfo  {journal} {Nature Physics}\ }\textbf {\bibinfo {volume} {21}},\ \bibinfo {pages} {754} (\bibinfo {year} {2025})}\BibitemShut {NoStop}%
\bibitem [{\citenamefont {Zhang}\ \emph {et~al.}(2025{\natexlab{b}})\citenamefont {Zhang}, \citenamefont {Cheng}, \citenamefont {Yin}, \citenamefont {Liu}, \citenamefont {Deng}, \citenamefont {Qiao}, \citenamefont {Shi}, \citenamefont {Zhang}, \citenamefont {Lin}, \citenamefont {Liu}, \citenamefont {Ye}, \citenamefont {Huang}, \citenamefont {Meng}, \citenamefont {Zhang}, \citenamefont {Okuda}, \citenamefont {Shimada}, \citenamefont {Cui}, \citenamefont {Zhao}, \citenamefont {Cao}, \citenamefont {Qiao}, \citenamefont {Liu},\ and\ \citenamefont {Chen}}]{RbVTeO_arpes}%
  \BibitemOpen
  \bibfield  {author} {\bibinfo {author} {\bibfnamefont {F.}~\bibnamefont {Zhang}}, \bibinfo {author} {\bibfnamefont {X.}~\bibnamefont {Cheng}}, \bibinfo {author} {\bibfnamefont {Z.}~\bibnamefont {Yin}}, \bibinfo {author} {\bibfnamefont {C.}~\bibnamefont {Liu}}, \bibinfo {author} {\bibfnamefont {L.}~\bibnamefont {Deng}}, \bibinfo {author} {\bibfnamefont {Y.}~\bibnamefont {Qiao}}, \bibinfo {author} {\bibfnamefont {Z.}~\bibnamefont {Shi}}, \bibinfo {author} {\bibfnamefont {S.}~\bibnamefont {Zhang}}, \bibinfo {author} {\bibfnamefont {J.}~\bibnamefont {Lin}}, \bibinfo {author} {\bibfnamefont {Z.}~\bibnamefont {Liu}}, \bibinfo {author} {\bibfnamefont {M.}~\bibnamefont {Ye}}, \bibinfo {author} {\bibfnamefont {Y.}~\bibnamefont {Huang}}, \bibinfo {author} {\bibfnamefont {X.}~\bibnamefont {Meng}}, \bibinfo {author} {\bibfnamefont {C.}~\bibnamefont {Zhang}}, \bibinfo {author} {\bibfnamefont {T.}~\bibnamefont {Okuda}}, \bibinfo {author} {\bibfnamefont {K.}~\bibnamefont {Shimada}}, \bibinfo {author} {\bibfnamefont {S.}~\bibnamefont {Cui}}, \bibinfo {author} {\bibfnamefont {Y.}~\bibnamefont {Zhao}}, \bibinfo {author} {\bibfnamefont {G.-H.}\ \bibnamefont {Cao}}, \bibinfo {author} {\bibfnamefont {S.}~\bibnamefont {Qiao}}, \bibinfo {author} {\bibfnamefont {J.}~\bibnamefont {Liu}},\ and\ \bibinfo {author} {\bibfnamefont {C.}~\bibnamefont {Chen}},\ }\bibfield  {title} {\bibinfo {title} {Crystal-symmetry-paired spin--valley locking in a layered room-temperature metallic altermagnet candidate},\ }\href {https://doi.org/10.1038/s41567-025-02864-2} {\bibfield  {journal} {\bibinfo  {journal} {Nature Physics}\ }\textbf {\bibinfo {volume} {21}},\ \bibinfo {pages} {760} (\bibinfo {year} {2025}{\natexlab{b}})}\BibitemShut {NoStop}%
\bibitem [{\citenamefont {Liu}\ \emph {et~al.}(2025{\natexlab{b}})\citenamefont {Liu}, \citenamefont {Li}, \citenamefont {Liu}, \citenamefont {Lu}, \citenamefont {Li}, \citenamefont {Liu},\ and\ \citenamefont {Cao}}]{CsVTeO}%
  \BibitemOpen
  \bibfield  {author} {\bibinfo {author} {\bibfnamefont {C.-C.}\ \bibnamefont {Liu}}, \bibinfo {author} {\bibfnamefont {J.}~\bibnamefont {Li}}, \bibinfo {author} {\bibfnamefont {J.-Y.}\ \bibnamefont {Liu}}, \bibinfo {author} {\bibfnamefont {J.-Y.}\ \bibnamefont {Lu}}, \bibinfo {author} {\bibfnamefont {H.-X.}\ \bibnamefont {Li}}, \bibinfo {author} {\bibfnamefont {Y.}~\bibnamefont {Liu}},\ and\ \bibinfo {author} {\bibfnamefont {G.-H.}\ \bibnamefont {Cao}},\ }\bibfield  {title} {\bibinfo {title} {Physical properties and first-principles calculations of an altermagnet candidate {${\mathrm{Cs}}_{1\ensuremath{-}\ensuremath{\delta}}{\mathrm{V}}_{2}{\mathrm{Te}}_{2}\mathrm{O}$}},\ }\href {https://doi.org/10.1103/vch1-4khc} {\bibfield  {journal} {\bibinfo  {journal} {Phys. Rev. B}\ }\textbf {\bibinfo {volume} {112}},\ \bibinfo {pages} {224439} (\bibinfo {year} {2025}{\natexlab{b}})}\BibitemShut {NoStop}%
\bibitem [{\citenamefont {Wang}\ \emph {et~al.}(2025)\citenamefont {Wang}, \citenamefont {Yu}, \citenamefont {Cheng}, \citenamefont {Xiao}, \citenamefont {Ma}, \citenamefont {Quan}, \citenamefont {Song}, \citenamefont {Zhang}, \citenamefont {Zhang}, \citenamefont {Ma}, \citenamefont {Liu}, \citenamefont {Yadav}, \citenamefont {Shi}, \citenamefont {Wang}, \citenamefont {Niu}, \citenamefont {Gao}, \citenamefont {Xiang}, \citenamefont {Liu}, \citenamefont {Wang},\ and\ \citenamefont {Chen}}]{KVSeO_stm}%
  \BibitemOpen
  \bibfield  {author} {\bibinfo {author} {\bibfnamefont {Z.}~\bibnamefont {Wang}}, \bibinfo {author} {\bibfnamefont {S.}~\bibnamefont {Yu}}, \bibinfo {author} {\bibfnamefont {X.}~\bibnamefont {Cheng}}, \bibinfo {author} {\bibfnamefont {X.}~\bibnamefont {Xiao}}, \bibinfo {author} {\bibfnamefont {W.}~\bibnamefont {Ma}}, \bibinfo {author} {\bibfnamefont {F.}~\bibnamefont {Quan}}, \bibinfo {author} {\bibfnamefont {H.}~\bibnamefont {Song}}, \bibinfo {author} {\bibfnamefont {K.}~\bibnamefont {Zhang}}, \bibinfo {author} {\bibfnamefont {Y.}~\bibnamefont {Zhang}}, \bibinfo {author} {\bibfnamefont {Y.}~\bibnamefont {Ma}}, \bibinfo {author} {\bibfnamefont {W.}~\bibnamefont {Liu}}, \bibinfo {author} {\bibfnamefont {P.}~\bibnamefont {Yadav}}, \bibinfo {author} {\bibfnamefont {X.}~\bibnamefont {Shi}}, \bibinfo {author} {\bibfnamefont {Z.}~\bibnamefont {Wang}}, \bibinfo {author} {\bibfnamefont {Q.}~\bibnamefont {Niu}}, \bibinfo {author} {\bibfnamefont {Y.}~\bibnamefont {Gao}}, \bibinfo {author} {\bibfnamefont {B.}~\bibnamefont {Xiang}}, \bibinfo {author} {\bibfnamefont {J.}~\bibnamefont {Liu}}, \bibinfo {author} {\bibfnamefont {Z.}~\bibnamefont {Wang}},\ and\ \bibinfo {author} {\bibfnamefont {X.}~\bibnamefont {Chen}},\ }\href {https://arxiv.org/abs/2512.23290} {\bibinfo {title} {Atomic-scale spin sensing of a 2{D} $d$-wave altermagnet via helical tunneling}} (\bibinfo {year} {2025}),\ \Eprint {https://arxiv.org/abs/2512.23290} {arXiv:2512.23290 [cond-mat.mes-hall]} \BibitemShut {NoStop}%
\bibitem [{\citenamefont {Yang}\ \emph {et~al.}(2026)\citenamefont {Yang}, \citenamefont {Li}, \citenamefont {Wang}, \citenamefont {Zhao}, \citenamefont {Wan}, \citenamefont {Gui}, \citenamefont {Zeng}, \citenamefont {Cao}, \citenamefont {Hu}, \citenamefont {Chen}, \citenamefont {Liu}, \citenamefont {Song}, \citenamefont {Liu}, \citenamefont {Hu}, \citenamefont {Jiao},\ and\ \citenamefont {Yuan}}]{KVSeO_stm2}%
  \BibitemOpen
  \bibfield  {author} {\bibinfo {author} {\bibfnamefont {G.}~\bibnamefont {Yang}}, \bibinfo {author} {\bibfnamefont {C.}~\bibnamefont {Li}}, \bibinfo {author} {\bibfnamefont {C.}~\bibnamefont {Wang}}, \bibinfo {author} {\bibfnamefont {X.}~\bibnamefont {Zhao}}, \bibinfo {author} {\bibfnamefont {Y.}~\bibnamefont {Wan}}, \bibinfo {author} {\bibfnamefont {H.}~\bibnamefont {Gui}}, \bibinfo {author} {\bibfnamefont {G.}~\bibnamefont {Zeng}}, \bibinfo {author} {\bibfnamefont {S.}~\bibnamefont {Cao}}, \bibinfo {author} {\bibfnamefont {C.}~\bibnamefont {Hu}}, \bibinfo {author} {\bibfnamefont {D.}~\bibnamefont {Chen}}, \bibinfo {author} {\bibfnamefont {Y.}~\bibnamefont {Liu}}, \bibinfo {author} {\bibfnamefont {Y.}~\bibnamefont {Song}}, \bibinfo {author} {\bibfnamefont {F.}~\bibnamefont {Liu}}, \bibinfo {author} {\bibfnamefont {L.-H.}\ \bibnamefont {Hu}}, \bibinfo {author} {\bibfnamefont {L.}~\bibnamefont {Jiao}},\ and\ \bibinfo {author} {\bibfnamefont {H.}~\bibnamefont {Yuan}},\ }\href {https://arxiv.org/abs/2603.21969} {\bibinfo {title} {Visualizing spin-polarization of an altermagnet {KV$_2$Se$_2$O} via spin-selective tunneling}} (\bibinfo {year} {2026}),\ \Eprint {https://arxiv.org/abs/2603.21969} {arXiv:2603.21969 [cond-mat.mtrl-sci]} \BibitemShut {NoStop}%
\bibitem [{\citenamefont {Fu}\ \emph {et~al.}(2026)\citenamefont {Fu}, \citenamefont {Yang}, \citenamefont {Shen}, \citenamefont {Xiao}, \citenamefont {Wang}, \citenamefont {Jiang}, \citenamefont {Wang}, \citenamefont {Yao}, \citenamefont {Xue},\ and\ \citenamefont {Li}}]{CsV2Se2O_stm}%
  \BibitemOpen
  \bibfield  {author} {\bibinfo {author} {\bibfnamefont {D.}~\bibnamefont {Fu}}, \bibinfo {author} {\bibfnamefont {L.}~\bibnamefont {Yang}}, \bibinfo {author} {\bibfnamefont {Y.}~\bibnamefont {Shen}}, \bibinfo {author} {\bibfnamefont {K.}~\bibnamefont {Xiao}}, \bibinfo {author} {\bibfnamefont {Y.}~\bibnamefont {Wang}}, \bibinfo {author} {\bibfnamefont {W.}~\bibnamefont {Jiang}}, \bibinfo {author} {\bibfnamefont {Z.}~\bibnamefont {Wang}}, \bibinfo {author} {\bibfnamefont {Y.}~\bibnamefont {Yao}}, \bibinfo {author} {\bibfnamefont {Q.-K.}\ \bibnamefont {Xue}},\ and\ \bibinfo {author} {\bibfnamefont {W.}~\bibnamefont {Li}},\ }\href {https://arxiv.org/abs/2512.24114} {\bibinfo {title} {Atomic-scale visualization of d-wave altermagnetism}} (\bibinfo {year} {2026}),\ \Eprint {https://arxiv.org/abs/2512.24114} {arXiv:2512.24114 [cond-mat.mtrl-sci]} \BibitemShut {NoStop}%
\bibitem [{\citenamefont {Guo}(2026)}]{SanDongGuo:2026}%
  \BibitemOpen
  \bibfield  {author} {\bibinfo {author} {\bibfnamefont {S.-D.}\ \bibnamefont {Guo}},\ }\bibfield  {title} {\bibinfo {title} {Robust $d$-wave altermagnetism in {XC}r$_2${Y}$_2${O} (x=k, rb, cs; y=s, se, t) family},\ }\bibfield  {journal} {\bibinfo  {journal} {Frontiers of Physics}\ }\href {https://doi.org/https://doi.org/10.15302/frontphys.2026.075204} {https://doi.org/10.15302/frontphys.2026.075204} (\bibinfo {year} {2026})\BibitemShut {NoStop}%
\bibitem [{\citenamefont {Ding}\ \emph {et~al.}(2024)\citenamefont {Ding}, \citenamefont {Jiang}, \citenamefont {Chen}, \citenamefont {Tao}, \citenamefont {Liu}, \citenamefont {Li}, \citenamefont {Liu}, \citenamefont {Sun}, \citenamefont {Cheng}, \citenamefont {Liu}, \citenamefont {Yang}, \citenamefont {Zhang}, \citenamefont {Deng}, \citenamefont {Jing}, \citenamefont {Huang}, \citenamefont {Shi}, \citenamefont {Ye}, \citenamefont {Qiao}, \citenamefont {Wang}, \citenamefont {Guo}, \citenamefont {Feng},\ and\ \citenamefont {Shen}}]{CrSb_arpes_3}%
  \BibitemOpen
  \bibfield  {author} {\bibinfo {author} {\bibfnamefont {J.}~\bibnamefont {Ding}}, \bibinfo {author} {\bibfnamefont {Z.}~\bibnamefont {Jiang}}, \bibinfo {author} {\bibfnamefont {X.}~\bibnamefont {Chen}}, \bibinfo {author} {\bibfnamefont {Z.}~\bibnamefont {Tao}}, \bibinfo {author} {\bibfnamefont {Z.}~\bibnamefont {Liu}}, \bibinfo {author} {\bibfnamefont {T.}~\bibnamefont {Li}}, \bibinfo {author} {\bibfnamefont {J.}~\bibnamefont {Liu}}, \bibinfo {author} {\bibfnamefont {J.}~\bibnamefont {Sun}}, \bibinfo {author} {\bibfnamefont {J.}~\bibnamefont {Cheng}}, \bibinfo {author} {\bibfnamefont {J.}~\bibnamefont {Liu}}, \bibinfo {author} {\bibfnamefont {Y.}~\bibnamefont {Yang}}, \bibinfo {author} {\bibfnamefont {R.}~\bibnamefont {Zhang}}, \bibinfo {author} {\bibfnamefont {L.}~\bibnamefont {Deng}}, \bibinfo {author} {\bibfnamefont {W.}~\bibnamefont {Jing}}, \bibinfo {author} {\bibfnamefont {Y.}~\bibnamefont {Huang}}, \bibinfo {author} {\bibfnamefont {Y.}~\bibnamefont {Shi}}, \bibinfo {author} {\bibfnamefont {M.}~\bibnamefont {Ye}}, \bibinfo {author} {\bibfnamefont {S.}~\bibnamefont {Qiao}}, \bibinfo {author} {\bibfnamefont {Y.}~\bibnamefont {Wang}}, \bibinfo {author} {\bibfnamefont {Y.}~\bibnamefont {Guo}}, \bibinfo {author} {\bibfnamefont {D.}~\bibnamefont {Feng}},\ and\ \bibinfo {author} {\bibfnamefont {D.}~\bibnamefont {Shen}},\ }\bibfield  {title} {\bibinfo {title} {Large band splitting in $g$-wave altermagnet {CrSb}},\ }\href {https://doi.org/10.1103/PhysRevLett.133.206401} {\bibfield  {journal} {\bibinfo  {journal} {Phys. Rev. Lett.}\ }\textbf {\bibinfo {volume} {133}},\ \bibinfo {pages} {206401} (\bibinfo {year} {2024})}\BibitemShut {NoStop}%
\bibitem [{\citenamefont {Zeng}\ \emph {et~al.}(2024)\citenamefont {Zeng}, \citenamefont {Zhu}, \citenamefont {Zhu}, \citenamefont {Liu}, \citenamefont {Ma}, \citenamefont {Hao}, \citenamefont {Liu}, \citenamefont {Qu}, \citenamefont {Yang}, \citenamefont {Jiang}, \citenamefont {Yamagami}, \citenamefont {Arita}, \citenamefont {Zhang}, \citenamefont {Shao}, \citenamefont {Dai}, \citenamefont {Shimada}, \citenamefont {Liu}, \citenamefont {Ye}, \citenamefont {Huang}, \citenamefont {Liu},\ and\ \citenamefont {Liu}}]{CrSb_arpes_1}%
  \BibitemOpen
  \bibfield  {author} {\bibinfo {author} {\bibfnamefont {M.}~\bibnamefont {Zeng}}, \bibinfo {author} {\bibfnamefont {M.-Y.}\ \bibnamefont {Zhu}}, \bibinfo {author} {\bibfnamefont {Y.-P.}\ \bibnamefont {Zhu}}, \bibinfo {author} {\bibfnamefont {X.-R.}\ \bibnamefont {Liu}}, \bibinfo {author} {\bibfnamefont {X.-M.}\ \bibnamefont {Ma}}, \bibinfo {author} {\bibfnamefont {Y.-J.}\ \bibnamefont {Hao}}, \bibinfo {author} {\bibfnamefont {P.}~\bibnamefont {Liu}}, \bibinfo {author} {\bibfnamefont {G.}~\bibnamefont {Qu}}, \bibinfo {author} {\bibfnamefont {Y.}~\bibnamefont {Yang}}, \bibinfo {author} {\bibfnamefont {Z.}~\bibnamefont {Jiang}}, \bibinfo {author} {\bibfnamefont {K.}~\bibnamefont {Yamagami}}, \bibinfo {author} {\bibfnamefont {M.}~\bibnamefont {Arita}}, \bibinfo {author} {\bibfnamefont {X.}~\bibnamefont {Zhang}}, \bibinfo {author} {\bibfnamefont {T.-H.}\ \bibnamefont {Shao}}, \bibinfo {author} {\bibfnamefont {Y.}~\bibnamefont {Dai}}, \bibinfo {author} {\bibfnamefont {K.}~\bibnamefont {Shimada}}, \bibinfo {author} {\bibfnamefont {Z.}~\bibnamefont {Liu}}, \bibinfo {author} {\bibfnamefont {M.}~\bibnamefont {Ye}}, \bibinfo {author} {\bibfnamefont {Y.}~\bibnamefont {Huang}}, \bibinfo {author} {\bibfnamefont {Q.}~\bibnamefont {Liu}},\ and\ \bibinfo {author} {\bibfnamefont {C.}~\bibnamefont {Liu}},\ }\bibfield  {title} {\bibinfo {title} {Observation of spin splitting in room-temperature metallic antiferromagnet {CrSb}},\ }\href {https://doi.org/https://doi.org/10.1002/advs.202406529} {\bibfield  {journal} {\bibinfo  {journal} {Advanced Science}\ }\textbf {\bibinfo {volume} {11}},\ \bibinfo {pages} {2406529} (\bibinfo {year} {2024})}\BibitemShut {NoStop}%
\bibitem [{\citenamefont {Reimers}\ \emph {et~al.}(2024)\citenamefont {Reimers}, \citenamefont {Odenbreit}, \citenamefont {{\v{S}}mejkal}, \citenamefont {Strocov}, \citenamefont {Constantinou}, \citenamefont {Hellenes}, \citenamefont {Jaeschke~Ubiergo}, \citenamefont {Campos}, \citenamefont {Bharadwaj}, \citenamefont {Chakraborty}, \citenamefont {Denneulin}, \citenamefont {Shi}, \citenamefont {Dunin-Borkowski}, \citenamefont {Das}, \citenamefont {Kl{\"a}ui}, \citenamefont {Sinova},\ and\ \citenamefont {Jourdan}}]{CrSb_arpes_2}%
  \BibitemOpen
  \bibfield  {author} {\bibinfo {author} {\bibfnamefont {S.}~\bibnamefont {Reimers}}, \bibinfo {author} {\bibfnamefont {L.}~\bibnamefont {Odenbreit}}, \bibinfo {author} {\bibfnamefont {L.}~\bibnamefont {{\v{S}}mejkal}}, \bibinfo {author} {\bibfnamefont {V.~N.}\ \bibnamefont {Strocov}}, \bibinfo {author} {\bibfnamefont {P.}~\bibnamefont {Constantinou}}, \bibinfo {author} {\bibfnamefont {A.~B.}\ \bibnamefont {Hellenes}}, \bibinfo {author} {\bibfnamefont {R.}~\bibnamefont {Jaeschke~Ubiergo}}, \bibinfo {author} {\bibfnamefont {W.~H.}\ \bibnamefont {Campos}}, \bibinfo {author} {\bibfnamefont {V.~K.}\ \bibnamefont {Bharadwaj}}, \bibinfo {author} {\bibfnamefont {A.}~\bibnamefont {Chakraborty}}, \bibinfo {author} {\bibfnamefont {T.}~\bibnamefont {Denneulin}}, \bibinfo {author} {\bibfnamefont {W.}~\bibnamefont {Shi}}, \bibinfo {author} {\bibfnamefont {R.~E.}\ \bibnamefont {Dunin-Borkowski}}, \bibinfo {author} {\bibfnamefont {S.}~\bibnamefont {Das}}, \bibinfo {author} {\bibfnamefont {M.}~\bibnamefont {Kl{\"a}ui}}, \bibinfo {author} {\bibfnamefont {J.}~\bibnamefont {Sinova}},\ and\ \bibinfo {author} {\bibfnamefont {M.}~\bibnamefont {Jourdan}},\ }\bibfield  {title} {\bibinfo {title} {Direct observation of altermagnetic band splitting in {CrSb} thin films},\ }\href {https://doi.org/10.1038/s41467-024-46476-5} {\bibfield  {journal} {\bibinfo  {journal} {Nature Communications}\ }\textbf {\bibinfo {volume} {15}},\ \bibinfo {pages} {2116} (\bibinfo {year} {2024})}\BibitemShut {NoStop}%
\bibitem [{\citenamefont {Yang}\ \emph {et~al.}(2025)\citenamefont {Yang}, \citenamefont {Li}, \citenamefont {Yang}, \citenamefont {Li}, \citenamefont {Zheng}, \citenamefont {Zhu}, \citenamefont {Pan}, \citenamefont {Xu}, \citenamefont {Cao}, \citenamefont {Zhao}, \citenamefont {Jana}, \citenamefont {Zhang}, \citenamefont {Ye}, \citenamefont {Song}, \citenamefont {Hu}, \citenamefont {Yang}, \citenamefont {Fujii}, \citenamefont {Vobornik}, \citenamefont {Shi}, \citenamefont {Yuan}, \citenamefont {Zhang}, \citenamefont {Xu},\ and\ \citenamefont {Liu}}]{CrSb_arpes_4}%
  \BibitemOpen
  \bibfield  {author} {\bibinfo {author} {\bibfnamefont {G.}~\bibnamefont {Yang}}, \bibinfo {author} {\bibfnamefont {Z.}~\bibnamefont {Li}}, \bibinfo {author} {\bibfnamefont {S.}~\bibnamefont {Yang}}, \bibinfo {author} {\bibfnamefont {J.}~\bibnamefont {Li}}, \bibinfo {author} {\bibfnamefont {H.}~\bibnamefont {Zheng}}, \bibinfo {author} {\bibfnamefont {W.}~\bibnamefont {Zhu}}, \bibinfo {author} {\bibfnamefont {Z.}~\bibnamefont {Pan}}, \bibinfo {author} {\bibfnamefont {Y.}~\bibnamefont {Xu}}, \bibinfo {author} {\bibfnamefont {S.}~\bibnamefont {Cao}}, \bibinfo {author} {\bibfnamefont {W.}~\bibnamefont {Zhao}}, \bibinfo {author} {\bibfnamefont {A.}~\bibnamefont {Jana}}, \bibinfo {author} {\bibfnamefont {J.}~\bibnamefont {Zhang}}, \bibinfo {author} {\bibfnamefont {M.}~\bibnamefont {Ye}}, \bibinfo {author} {\bibfnamefont {Y.}~\bibnamefont {Song}}, \bibinfo {author} {\bibfnamefont {L.-H.}\ \bibnamefont {Hu}}, \bibinfo {author} {\bibfnamefont {L.}~\bibnamefont {Yang}}, \bibinfo {author} {\bibfnamefont {J.}~\bibnamefont {Fujii}}, \bibinfo {author} {\bibfnamefont {I.}~\bibnamefont {Vobornik}}, \bibinfo {author} {\bibfnamefont {M.}~\bibnamefont {Shi}}, \bibinfo {author} {\bibfnamefont {H.}~\bibnamefont {Yuan}}, \bibinfo {author} {\bibfnamefont {Y.}~\bibnamefont {Zhang}}, \bibinfo {author} {\bibfnamefont {Y.}~\bibnamefont {Xu}},\ and\ \bibinfo {author} {\bibfnamefont {Y.}~\bibnamefont {Liu}},\ }\bibfield  {title} {\bibinfo {title} {Three-dimensional mapping of the altermagnetic spin splitting in {CrSb}},\ }\href {https://doi.org/10.1038/s41467-025-56647-7} {\bibfield  {journal} {\bibinfo  {journal} {Nature Communications}\ }\textbf {\bibinfo {volume} {16}},\ \bibinfo {pages} {1442} (\bibinfo {year} {2025})}\BibitemShut {NoStop}%
\bibitem [{\citenamefont {Lu}\ \emph {et~al.}(2025)\citenamefont {Lu}, \citenamefont {Feng}, \citenamefont {Wang}, \citenamefont {Chen}, \citenamefont {Lin}, \citenamefont {Liang}, \citenamefont {Liu}, \citenamefont {Feng}, \citenamefont {Yamagami}, \citenamefont {Liu}, \citenamefont {Felser}, \citenamefont {Wu},\ and\ \citenamefont {Ma}}]{CrSb_arpes5}%
  \BibitemOpen
  \bibfield  {author} {\bibinfo {author} {\bibfnamefont {W.}~\bibnamefont {Lu}}, \bibinfo {author} {\bibfnamefont {S.}~\bibnamefont {Feng}}, \bibinfo {author} {\bibfnamefont {Y.}~\bibnamefont {Wang}}, \bibinfo {author} {\bibfnamefont {D.}~\bibnamefont {Chen}}, \bibinfo {author} {\bibfnamefont {Z.}~\bibnamefont {Lin}}, \bibinfo {author} {\bibfnamefont {X.}~\bibnamefont {Liang}}, \bibinfo {author} {\bibfnamefont {S.}~\bibnamefont {Liu}}, \bibinfo {author} {\bibfnamefont {W.}~\bibnamefont {Feng}}, \bibinfo {author} {\bibfnamefont {K.}~\bibnamefont {Yamagami}}, \bibinfo {author} {\bibfnamefont {J.}~\bibnamefont {Liu}}, \bibinfo {author} {\bibfnamefont {C.}~\bibnamefont {Felser}}, \bibinfo {author} {\bibfnamefont {Q.}~\bibnamefont {Wu}},\ and\ \bibinfo {author} {\bibfnamefont {J.}~\bibnamefont {Ma}},\ }\bibfield  {title} {\bibinfo {title} {Signature of topological surface bands in altermagnetic weyl semimetal {CrSb}},\ }\href {https://doi.org/10.1021/acs.nanolett.5c00482} {\bibfield  {journal} {\bibinfo  {journal} {Nano Letters}\ }\textbf {\bibinfo {volume} {25}},\ \bibinfo {pages} {7343} (\bibinfo {year} {2025})}\BibitemShut {NoStop}%
\bibitem [{\citenamefont {Lee}\ \emph {et~al.}(2024)\citenamefont {Lee}, \citenamefont {Lee}, \citenamefont {Jung}, \citenamefont {Jung}, \citenamefont {Kim}, \citenamefont {Lee}, \citenamefont {Seok}, \citenamefont {Kim}, \citenamefont {Park}, \citenamefont {\ifmmode~\check{S}\else \v{S}\fi{}mejkal}, \citenamefont {Kang},\ and\ \citenamefont {Kim}}]{MnTe_arpes_3}%
  \BibitemOpen
  \bibfield  {author} {\bibinfo {author} {\bibfnamefont {S.}~\bibnamefont {Lee}}, \bibinfo {author} {\bibfnamefont {S.}~\bibnamefont {Lee}}, \bibinfo {author} {\bibfnamefont {S.}~\bibnamefont {Jung}}, \bibinfo {author} {\bibfnamefont {J.}~\bibnamefont {Jung}}, \bibinfo {author} {\bibfnamefont {D.}~\bibnamefont {Kim}}, \bibinfo {author} {\bibfnamefont {Y.}~\bibnamefont {Lee}}, \bibinfo {author} {\bibfnamefont {B.}~\bibnamefont {Seok}}, \bibinfo {author} {\bibfnamefont {J.}~\bibnamefont {Kim}}, \bibinfo {author} {\bibfnamefont {B.~G.}\ \bibnamefont {Park}}, \bibinfo {author} {\bibfnamefont {L.}~\bibnamefont {\ifmmode~\check{S}\else \v{S}\fi{}mejkal}}, \bibinfo {author} {\bibfnamefont {C.-J.}\ \bibnamefont {Kang}},\ and\ \bibinfo {author} {\bibfnamefont {C.}~\bibnamefont {Kim}},\ }\bibfield  {title} {\bibinfo {title} {Broken kramers degeneracy in altermagnetic {MnTe}},\ }\href {https://doi.org/10.1103/PhysRevLett.132.036702} {\bibfield  {journal} {\bibinfo  {journal} {Phys. Rev. Lett.}\ }\textbf {\bibinfo {volume} {132}},\ \bibinfo {pages} {036702} (\bibinfo {year} {2024})}\BibitemShut {NoStop}%
\bibitem [{\citenamefont {Krempask{\'y}}\ \emph {et~al.}(2024)\citenamefont {Krempask{\'y}}, \citenamefont {{\v{S}}mejkal}, \citenamefont {D'Souza}, \citenamefont {Hajlaoui}, \citenamefont {Springholz}, \citenamefont {Uhl{\'i}{\v{r}}ov{\'a}}, \citenamefont {Alarab}, \citenamefont {Constantinou}, \citenamefont {Strocov}, \citenamefont {Usanov}, \citenamefont {Pudelko}, \citenamefont {Gonz{\'a}lez-Hern{\'a}ndez}, \citenamefont {Birk~Hellenes}, \citenamefont {Jansa}, \citenamefont {Reichlov{\'a}}, \citenamefont {{\v{S}}ob{\'a}{\v{n}}}, \citenamefont {Gonzalez~Betancourt}, \citenamefont {Wadley}, \citenamefont {Sinova}, \citenamefont {Kriegner}, \citenamefont {Min{\'a}r}, \citenamefont {Dil},\ and\ \citenamefont {Jungwirth}}]{MnTe_arpes_1}%
  \BibitemOpen
  \bibfield  {author} {\bibinfo {author} {\bibfnamefont {J.}~\bibnamefont {Krempask{\'y}}}, \bibinfo {author} {\bibfnamefont {L.}~\bibnamefont {{\v{S}}mejkal}}, \bibinfo {author} {\bibfnamefont {S.~W.}\ \bibnamefont {D'Souza}}, \bibinfo {author} {\bibfnamefont {M.}~\bibnamefont {Hajlaoui}}, \bibinfo {author} {\bibfnamefont {G.}~\bibnamefont {Springholz}}, \bibinfo {author} {\bibfnamefont {K.}~\bibnamefont {Uhl{\'i}{\v{r}}ov{\'a}}}, \bibinfo {author} {\bibfnamefont {F.}~\bibnamefont {Alarab}}, \bibinfo {author} {\bibfnamefont {P.~C.}\ \bibnamefont {Constantinou}}, \bibinfo {author} {\bibfnamefont {V.}~\bibnamefont {Strocov}}, \bibinfo {author} {\bibfnamefont {D.}~\bibnamefont {Usanov}}, \bibinfo {author} {\bibfnamefont {W.~R.}\ \bibnamefont {Pudelko}}, \bibinfo {author} {\bibfnamefont {R.}~\bibnamefont {Gonz{\'a}lez-Hern{\'a}ndez}}, \bibinfo {author} {\bibfnamefont {A.}~\bibnamefont {Birk~Hellenes}}, \bibinfo {author} {\bibfnamefont {Z.}~\bibnamefont {Jansa}}, \bibinfo {author} {\bibfnamefont {H.}~\bibnamefont {Reichlov{\'a}}}, \bibinfo {author} {\bibfnamefont {Z.}~\bibnamefont {{\v{S}}ob{\'a}{\v{n}}}}, \bibinfo {author} {\bibfnamefont {R.~D.}\ \bibnamefont {Gonzalez~Betancourt}}, \bibinfo {author} {\bibfnamefont {P.}~\bibnamefont {Wadley}}, \bibinfo {author} {\bibfnamefont {J.}~\bibnamefont {Sinova}}, \bibinfo {author} {\bibfnamefont {D.}~\bibnamefont {Kriegner}}, \bibinfo {author} {\bibfnamefont {J.}~\bibnamefont {Min{\'a}r}}, \bibinfo {author} {\bibfnamefont {J.~H.}\ \bibnamefont {Dil}},\ and\ \bibinfo {author} {\bibfnamefont {T.}~\bibnamefont {Jungwirth}},\ }\bibfield  {title} {\bibinfo {title} {Altermagnetic lifting of kramers spin degeneracy},\ }\href {https://doi.org/10.1038/s41586-023-06907-7} {\bibfield  {journal} {\bibinfo  {journal} {Nature}\ }\textbf {\bibinfo {volume} {626}},\ \bibinfo {pages} {517} (\bibinfo {year} {2024})}\BibitemShut {NoStop}%
\bibitem [{\citenamefont {Osumi}\ \emph {et~al.}(2024)\citenamefont {Osumi}, \citenamefont {Souma}, \citenamefont {Aoyama}, \citenamefont {Yamauchi}, \citenamefont {Honma}, \citenamefont {Nakayama}, \citenamefont {Takahashi}, \citenamefont {Ohgushi},\ and\ \citenamefont {Sato}}]{MnTe_arpes_2}%
  \BibitemOpen
  \bibfield  {author} {\bibinfo {author} {\bibfnamefont {T.}~\bibnamefont {Osumi}}, \bibinfo {author} {\bibfnamefont {S.}~\bibnamefont {Souma}}, \bibinfo {author} {\bibfnamefont {T.}~\bibnamefont {Aoyama}}, \bibinfo {author} {\bibfnamefont {K.}~\bibnamefont {Yamauchi}}, \bibinfo {author} {\bibfnamefont {A.}~\bibnamefont {Honma}}, \bibinfo {author} {\bibfnamefont {K.}~\bibnamefont {Nakayama}}, \bibinfo {author} {\bibfnamefont {T.}~\bibnamefont {Takahashi}}, \bibinfo {author} {\bibfnamefont {K.}~\bibnamefont {Ohgushi}},\ and\ \bibinfo {author} {\bibfnamefont {T.}~\bibnamefont {Sato}},\ }\bibfield  {title} {\bibinfo {title} {Observation of a giant band splitting in altermagnetic {MnTe}},\ }\href {https://doi.org/10.1103/PhysRevB.109.115102} {\bibfield  {journal} {\bibinfo  {journal} {Phys. Rev. B}\ }\textbf {\bibinfo {volume} {109}},\ \bibinfo {pages} {115102} (\bibinfo {year} {2024})}\BibitemShut {NoStop}%
\bibitem [{\citenamefont {Hajlaoui}\ \emph {et~al.}(2024)\citenamefont {Hajlaoui}, \citenamefont {Wilfred~D'Souza}, \citenamefont {Šmejkal}, \citenamefont {Kriegner}, \citenamefont {Krizman}, \citenamefont {Zakusylo}, \citenamefont {Olszowska}, \citenamefont {Caha}, \citenamefont {Michalička}, \citenamefont {Sánchez-Barriga}, \citenamefont {Marmodoro}, \citenamefont {Výborný}, \citenamefont {Ernst}, \citenamefont {Cinchetti}, \citenamefont {Minar}, \citenamefont {Jungwirth},\ and\ \citenamefont {Springholz}}]{MnTe_arpes_4}%
  \BibitemOpen
  \bibfield  {author} {\bibinfo {author} {\bibfnamefont {M.}~\bibnamefont {Hajlaoui}}, \bibinfo {author} {\bibfnamefont {S.}~\bibnamefont {Wilfred~D'Souza}}, \bibinfo {author} {\bibfnamefont {L.}~\bibnamefont {Šmejkal}}, \bibinfo {author} {\bibfnamefont {D.}~\bibnamefont {Kriegner}}, \bibinfo {author} {\bibfnamefont {G.}~\bibnamefont {Krizman}}, \bibinfo {author} {\bibfnamefont {T.}~\bibnamefont {Zakusylo}}, \bibinfo {author} {\bibfnamefont {N.}~\bibnamefont {Olszowska}}, \bibinfo {author} {\bibfnamefont {O.}~\bibnamefont {Caha}}, \bibinfo {author} {\bibfnamefont {J.}~\bibnamefont {Michalička}}, \bibinfo {author} {\bibfnamefont {J.}~\bibnamefont {Sánchez-Barriga}}, \bibinfo {author} {\bibfnamefont {A.}~\bibnamefont {Marmodoro}}, \bibinfo {author} {\bibfnamefont {K.}~\bibnamefont {Výborný}}, \bibinfo {author} {\bibfnamefont {A.}~\bibnamefont {Ernst}}, \bibinfo {author} {\bibfnamefont {M.}~\bibnamefont {Cinchetti}}, \bibinfo {author} {\bibfnamefont {J.}~\bibnamefont {Minar}}, \bibinfo {author} {\bibfnamefont {T.}~\bibnamefont {Jungwirth}},\ and\ \bibinfo {author} {\bibfnamefont {G.}~\bibnamefont {Springholz}},\ }\bibfield  {title} {\bibinfo {title} {Temperature dependence of relativistic valence band splitting induced by an altermagnetic phase transition},\ }\href {https://doi.org/https://doi.org/10.1002/adma.202314076} {\bibfield  {journal} {\bibinfo  {journal} {Advanced Materials}\ }\textbf {\bibinfo {volume} {36}},\ \bibinfo {pages} {2314076} (\bibinfo {year} {2024})}\BibitemShut {NoStop}%
\bibitem [{\citenamefont {Chilcote}\ \emph {et~al.}(2024)\citenamefont {Chilcote}, \citenamefont {Mazza}, \citenamefont {Lu}, \citenamefont {Gray}, \citenamefont {Tian}, \citenamefont {Deng}, \citenamefont {Moseley}, \citenamefont {Chen}, \citenamefont {Lapano}, \citenamefont {Gardner}, \citenamefont {Eres}, \citenamefont {Ward}, \citenamefont {Feng}, \citenamefont {Cao}, \citenamefont {Lauter}, \citenamefont {McGuire}, \citenamefont {Hermann}, \citenamefont {Parker}, \citenamefont {Han}, \citenamefont {Kayani}, \citenamefont {Rimal}, \citenamefont {Wu}, \citenamefont {Charlton}, \citenamefont {Moore},\ and\ \citenamefont {Brahlek}}]{MnTe_arpes5}%
  \BibitemOpen
  \bibfield  {author} {\bibinfo {author} {\bibfnamefont {M.}~\bibnamefont {Chilcote}}, \bibinfo {author} {\bibfnamefont {A.~R.}\ \bibnamefont {Mazza}}, \bibinfo {author} {\bibfnamefont {Q.}~\bibnamefont {Lu}}, \bibinfo {author} {\bibfnamefont {I.}~\bibnamefont {Gray}}, \bibinfo {author} {\bibfnamefont {Q.}~\bibnamefont {Tian}}, \bibinfo {author} {\bibfnamefont {Q.}~\bibnamefont {Deng}}, \bibinfo {author} {\bibfnamefont {D.}~\bibnamefont {Moseley}}, \bibinfo {author} {\bibfnamefont {A.-H.}\ \bibnamefont {Chen}}, \bibinfo {author} {\bibfnamefont {J.}~\bibnamefont {Lapano}}, \bibinfo {author} {\bibfnamefont {J.~S.}\ \bibnamefont {Gardner}}, \bibinfo {author} {\bibfnamefont {G.}~\bibnamefont {Eres}}, \bibinfo {author} {\bibfnamefont {T.~Z.}\ \bibnamefont {Ward}}, \bibinfo {author} {\bibfnamefont {E.}~\bibnamefont {Feng}}, \bibinfo {author} {\bibfnamefont {H.}~\bibnamefont {Cao}}, \bibinfo {author} {\bibfnamefont {V.}~\bibnamefont {Lauter}}, \bibinfo {author} {\bibfnamefont {M.~A.}\ \bibnamefont {McGuire}}, \bibinfo {author} {\bibfnamefont {R.}~\bibnamefont {Hermann}}, \bibinfo {author} {\bibfnamefont {D.}~\bibnamefont {Parker}}, \bibinfo {author} {\bibfnamefont {M.-G.}\ \bibnamefont {Han}}, \bibinfo {author} {\bibfnamefont {A.}~\bibnamefont {Kayani}}, \bibinfo {author} {\bibfnamefont {G.}~\bibnamefont {Rimal}}, \bibinfo {author} {\bibfnamefont {L.}~\bibnamefont {Wu}}, \bibinfo {author} {\bibfnamefont {T.~R.}\ \bibnamefont {Charlton}}, \bibinfo {author} {\bibfnamefont {R.~G.}\ \bibnamefont {Moore}},\ and\ \bibinfo {author} {\bibfnamefont {M.}~\bibnamefont {Brahlek}},\ }\bibfield  {title} {\bibinfo {title} {Stoichiometry-induced ferromagnetism in altermagnetic candidate {MnTe}},\ }\href {https://doi.org/https://doi.org/10.1002/adfm.202405829} {\bibfield  {journal} {\bibinfo  {journal} {Advanced Functional Materials}\ }\textbf {\bibinfo {volume} {34}},\ \bibinfo {pages} {2405829} (\bibinfo {year} {2024})}\BibitemShut {NoStop}%
\bibitem [{\citenamefont {Amin}\ \emph {et~al.}(2024)\citenamefont {Amin}, \citenamefont {Dal~Din}, \citenamefont {Golias}, \citenamefont {Niu}, \citenamefont {Zakharov}, \citenamefont {Fromage}, \citenamefont {Fields}, \citenamefont {Heywood}, \citenamefont {Cousins}, \citenamefont {Maccherozzi}, \citenamefont {Krempask{\'y}}, \citenamefont {Dil}, \citenamefont {Kriegner}, \citenamefont {Kiraly}, \citenamefont {Campion}, \citenamefont {Rushforth}, \citenamefont {Edmonds}, \citenamefont {Dhesi}, \citenamefont {{\v{S}}mejkal}, \citenamefont {Jungwirth},\ and\ \citenamefont {Wadley}}]{PPEM_MnTe}%
  \BibitemOpen
  \bibfield  {author} {\bibinfo {author} {\bibfnamefont {O.~J.}\ \bibnamefont {Amin}}, \bibinfo {author} {\bibfnamefont {A.}~\bibnamefont {Dal~Din}}, \bibinfo {author} {\bibfnamefont {E.}~\bibnamefont {Golias}}, \bibinfo {author} {\bibfnamefont {Y.}~\bibnamefont {Niu}}, \bibinfo {author} {\bibfnamefont {A.}~\bibnamefont {Zakharov}}, \bibinfo {author} {\bibfnamefont {S.~C.}\ \bibnamefont {Fromage}}, \bibinfo {author} {\bibfnamefont {C.~J.~B.}\ \bibnamefont {Fields}}, \bibinfo {author} {\bibfnamefont {S.~L.}\ \bibnamefont {Heywood}}, \bibinfo {author} {\bibfnamefont {R.~B.}\ \bibnamefont {Cousins}}, \bibinfo {author} {\bibfnamefont {F.}~\bibnamefont {Maccherozzi}}, \bibinfo {author} {\bibfnamefont {J.}~\bibnamefont {Krempask{\'y}}}, \bibinfo {author} {\bibfnamefont {J.~H.}\ \bibnamefont {Dil}}, \bibinfo {author} {\bibfnamefont {D.}~\bibnamefont {Kriegner}}, \bibinfo {author} {\bibfnamefont {B.}~\bibnamefont {Kiraly}}, \bibinfo {author} {\bibfnamefont {R.~P.}\ \bibnamefont {Campion}}, \bibinfo {author} {\bibfnamefont {A.~W.}\ \bibnamefont {Rushforth}}, \bibinfo {author} {\bibfnamefont {K.~W.}\ \bibnamefont {Edmonds}}, \bibinfo {author} {\bibfnamefont {S.~S.}\ \bibnamefont {Dhesi}}, \bibinfo {author} {\bibfnamefont {L.}~\bibnamefont {{\v{S}}mejkal}}, \bibinfo {author} {\bibfnamefont {T.}~\bibnamefont {Jungwirth}},\ and\ \bibinfo {author} {\bibfnamefont {P.}~\bibnamefont {Wadley}},\ }\bibfield  {title} {\bibinfo {title} {Nanoscale imaging and control of altermagnetism in {MnTe}},\ }\href {https://doi.org/10.1038/s41586-024-08234-x} {\bibfield  {journal} {\bibinfo  {journal} {Nature}\ }\textbf {\bibinfo {volume} {636}},\ \bibinfo {pages} {348} (\bibinfo {year} {2024})}\BibitemShut {NoStop}%
\bibitem [{\citenamefont {Sparks}\ and\ \citenamefont {Komoto}(1963)}]{komoto:1963}%
  \BibitemOpen
  \bibfield  {author} {\bibinfo {author} {\bibfnamefont {J.~T.}\ \bibnamefont {Sparks}}\ and\ \bibinfo {author} {\bibfnamefont {T.}~\bibnamefont {Komoto}},\ }\bibfield  {title} {\bibinfo {title} {Neutron diffraction study of nis},\ }\href {https://doi.org/10.1063/1.1729428} {\bibfield  {journal} {\bibinfo  {journal} {Journal of Applied Physics}\ }\textbf {\bibinfo {volume} {34}},\ \bibinfo {pages} {1191} (\bibinfo {year} {1963})}\BibitemShut {NoStop}%
\bibitem [{\citenamefont {Mandal}\ \emph {et~al.}(2025)\citenamefont {Mandal}, \citenamefont {Das},\ and\ \citenamefont {Nanda}}]{Mandal:2025}%
  \BibitemOpen
  \bibfield  {author} {\bibinfo {author} {\bibfnamefont {A.}~\bibnamefont {Mandal}}, \bibinfo {author} {\bibfnamefont {A.}~\bibnamefont {Das}},\ and\ \bibinfo {author} {\bibfnamefont {B.~R.~K.}\ \bibnamefont {Nanda}},\ }\bibfield  {title} {\bibinfo {title} {Deterministic role of chemical bonding in the formation of altermagnetism: Reflection from the correlated electron system nis},\ }\href {https://doi.org/10.1103/sm63-1dcx} {\bibfield  {journal} {\bibinfo  {journal} {Phys. Rev. B}\ }\textbf {\bibinfo {volume} {112}},\ \bibinfo {pages} {014420} (\bibinfo {year} {2025})}\BibitemShut {NoStop}%
\bibitem [{\citenamefont {Guowei}\ \emph {et~al.}(2025)\citenamefont {Guowei}, \citenamefont {Ruihan}, \citenamefont {Changchao}, \citenamefont {Jing}, \citenamefont {Ze}, \citenamefont {Teng}, \citenamefont {Pengyue}, \citenamefont {Liwei}, \citenamefont {Naifu}, \citenamefont {Yu}, \citenamefont {Hao}, \citenamefont {Weifan}, \citenamefont {Yifu}, \citenamefont {Xinying}, \citenamefont {Xin}, \citenamefont {Xiaoping}, \citenamefont {Shengtao}, \citenamefont {Zhe}, \citenamefont {Zhengtai}, \citenamefont {Mao}, \citenamefont {Chao}, \citenamefont {Ming}, \citenamefont {Lunhui}, \citenamefont {Qihang}, \citenamefont {Shan}, \citenamefont {Guanghan}, \citenamefont {Yu},\ and\ \citenamefont {Yang}}]{Liuyang:2025}%
  \BibitemOpen
  \bibfield  {author} {\bibinfo {author} {\bibfnamefont {Y.}~\bibnamefont {Guowei}}, \bibinfo {author} {\bibfnamefont {C.}~\bibnamefont {Ruihan}}, \bibinfo {author} {\bibfnamefont {L.}~\bibnamefont {Changchao}}, \bibinfo {author} {\bibfnamefont {L.}~\bibnamefont {Jing}}, \bibinfo {author} {\bibfnamefont {P.}~\bibnamefont {Ze}}, \bibinfo {author} {\bibfnamefont {H.}~\bibnamefont {Teng}}, \bibinfo {author} {\bibfnamefont {X.}~\bibnamefont {Pengyue}}, \bibinfo {author} {\bibfnamefont {D.}~\bibnamefont {Liwei}}, \bibinfo {author} {\bibfnamefont {Z.}~\bibnamefont {Naifu}}, \bibinfo {author} {\bibfnamefont {T.}~\bibnamefont {Yu}}, \bibinfo {author} {\bibfnamefont {Z.}~\bibnamefont {Hao}}, \bibinfo {author} {\bibfnamefont {Z.}~\bibnamefont {Weifan}}, \bibinfo {author} {\bibfnamefont {X.}~\bibnamefont {Yifu}}, \bibinfo {author} {\bibfnamefont {Z.}~\bibnamefont {Xinying}}, \bibinfo {author} {\bibfnamefont {M.}~\bibnamefont {Xin}}, \bibinfo {author} {\bibfnamefont {W.}~\bibnamefont {Xiaoping}}, \bibinfo {author} {\bibfnamefont {C.}~\bibnamefont {Shengtao}}, \bibinfo {author} {\bibfnamefont {S.}~\bibnamefont {Zhe}}, \bibinfo {author} {\bibfnamefont {L.}~\bibnamefont {Zhengtai}}, \bibinfo {author} {\bibfnamefont {Y.}~\bibnamefont {Mao}}, \bibinfo {author} {\bibfnamefont {C.}~\bibnamefont {Chao}}, \bibinfo {author} {\bibfnamefont {S.}~\bibnamefont {Ming}}, \bibinfo {author} {\bibfnamefont {H.}~\bibnamefont {Lunhui}}, \bibinfo {author} {\bibfnamefont {L.}~\bibnamefont {Qihang}}, \bibinfo {author} {\bibfnamefont {Q.}~\bibnamefont {Shan}}, \bibinfo {author} {\bibfnamefont {C.}~\bibnamefont {Guanghan}}, \bibinfo {author} {\bibfnamefont {S.}~\bibnamefont {Yu}},\ and\ \bibinfo {author} {\bibfnamefont {L.}~\bibnamefont {Yang}},\ }\href@noop {} {\bibinfo {title} {Observation of hidden altermagnetism in cs$_{1-\delta}$v$_2$te$_2$o}} (\bibinfo {year} {2025}),\ \Eprint {https://arxiv.org/abs/2512.00972} {arXiv:2512.00972 [cond-mat.mtrl-sci]} \BibitemShut {NoStop}%
\bibitem [{\citenamefont {Liu}\ \emph {et~al.}(2026)\citenamefont {Liu}, \citenamefont {Xu}, \citenamefont {Bao}, \citenamefont {Lv}, \citenamefont {Li}, \citenamefont {Li}, \citenamefont {Lin}, \citenamefont {Li}, \citenamefont {Lu}, \citenamefont {Zhao}, \citenamefont {Yang}, \citenamefont {Zhang}, \citenamefont {Chen}, \citenamefont {Jiao}, \citenamefont {Liu}, \citenamefont {Zhu},\ and\ \citenamefont {Cao}}]{Caoguanghan:2026}%
  \BibitemOpen
  \bibfield  {author} {\bibinfo {author} {\bibfnamefont {Y.}~\bibnamefont {Liu}}, \bibinfo {author} {\bibfnamefont {C.-C.}\ \bibnamefont {Xu}}, \bibinfo {author} {\bibfnamefont {J.-K.}\ \bibnamefont {Bao}}, \bibinfo {author} {\bibfnamefont {B.-J.}\ \bibnamefont {Lv}}, \bibinfo {author} {\bibfnamefont {H.}~\bibnamefont {Li}}, \bibinfo {author} {\bibfnamefont {J.}~\bibnamefont {Li}}, \bibinfo {author} {\bibfnamefont {Y.-Q.}\ \bibnamefont {Lin}}, \bibinfo {author} {\bibfnamefont {H.-X.}\ \bibnamefont {Li}}, \bibinfo {author} {\bibfnamefont {Y.-M.}\ \bibnamefont {Lu}}, \bibinfo {author} {\bibfnamefont {X.-Y.}\ \bibnamefont {Zhao}}, \bibinfo {author} {\bibfnamefont {W.-Z.}\ \bibnamefont {Yang}}, \bibinfo {author} {\bibfnamefont {Z.-Y.}\ \bibnamefont {Zhang}}, \bibinfo {author} {\bibfnamefont {X.-Y.}\ \bibnamefont {Chen}}, \bibinfo {author} {\bibfnamefont {W.-H.}\ \bibnamefont {Jiao}}, \bibinfo {author} {\bibfnamefont {J.-Y.}\ \bibnamefont {Liu}}, \bibinfo {author} {\bibfnamefont {B.-R.}\ \bibnamefont {Zhu}},\ and\ \bibinfo {author} {\bibfnamefont {G.-H.}\ \bibnamefont {Cao}},\ }\href@noop {} {\bibinfo {title} {Altermagnetism and room-temperature metal-to-insulator transition in {CsCr$_2$S$_2$O}}} (\bibinfo {year} {2026}),\ \Eprint {https://arxiv.org/abs/2604.02114} {arXiv:2604.02114 [cond-mat.mtrl-sci]} \BibitemShut {NoStop}%
\bibitem [{\citenamefont {VERWEY}(1939)}]{VERWEY:1939}%
  \BibitemOpen
  \bibfield  {author} {\bibinfo {author} {\bibfnamefont {E.~J.~W.}\ \bibnamefont {VERWEY}},\ }\bibfield  {title} {\bibinfo {title} {Electronic conduction of magnetite (fe3o4) and its transition point at low temperatures},\ }\href {https://doi.org/10.1038/144327b0} {\bibfield  {journal} {\bibinfo  {journal} {Nature}\ }\textbf {\bibinfo {volume} {144}},\ \bibinfo {pages} {327} (\bibinfo {year} {1939})}\BibitemShut {NoStop}%
\bibitem [{\citenamefont {Senn}\ \emph {et~al.}(2012)\citenamefont {Senn}, \citenamefont {Wright},\ and\ \citenamefont {Attfield}}]{Senn:2012}%
  \BibitemOpen
  \bibfield  {author} {\bibinfo {author} {\bibfnamefont {M.~S.}\ \bibnamefont {Senn}}, \bibinfo {author} {\bibfnamefont {J.~P.}\ \bibnamefont {Wright}},\ and\ \bibinfo {author} {\bibfnamefont {J.~P.}\ \bibnamefont {Attfield}},\ }\bibfield  {title} {\bibinfo {title} {Charge order and three-site distortions in the verwey structure of magnetite},\ }\href {https://doi.org/10.1038/nature10704} {\bibfield  {journal} {\bibinfo  {journal} {Nature}\ }\textbf {\bibinfo {volume} {481}},\ \bibinfo {pages} {173} (\bibinfo {year} {2012})}\BibitemShut {NoStop}%
\bibitem [{\citenamefont {Georges}\ \emph {et~al.}(1996)\citenamefont {Georges}, \citenamefont {Kotliar}, \citenamefont {Krauth},\ and\ \citenamefont {Rozenberg}}]{Georges:1996}%
  \BibitemOpen
  \bibfield  {author} {\bibinfo {author} {\bibfnamefont {A.}~\bibnamefont {Georges}}, \bibinfo {author} {\bibfnamefont {G.}~\bibnamefont {Kotliar}}, \bibinfo {author} {\bibfnamefont {W.}~\bibnamefont {Krauth}},\ and\ \bibinfo {author} {\bibfnamefont {M.~J.}\ \bibnamefont {Rozenberg}},\ }\bibfield  {title} {\bibinfo {title} {Dynamical mean-field theory of strongly correlated fermion systems and the limit of infinite dimensions},\ }\href {https://doi.org/10.1103/RevModPhys.68.13} {\bibfield  {journal} {\bibinfo  {journal} {Rev. Mod. Phys.}\ }\textbf {\bibinfo {volume} {68}},\ \bibinfo {pages} {13} (\bibinfo {year} {1996})}\BibitemShut {NoStop}%
\bibitem [{\citenamefont {Lichtenstein}\ \emph {et~al.}(2001)\citenamefont {Lichtenstein}, \citenamefont {Katsnelson},\ and\ \citenamefont {Kotliar}}]{lichtenstein:2001}%
  \BibitemOpen
  \bibfield  {author} {\bibinfo {author} {\bibfnamefont {A.~I.}\ \bibnamefont {Lichtenstein}}, \bibinfo {author} {\bibfnamefont {M.~I.}\ \bibnamefont {Katsnelson}},\ and\ \bibinfo {author} {\bibfnamefont {G.}~\bibnamefont {Kotliar}},\ }\bibfield  {title} {\bibinfo {title} {Finite-temperature magnetism of transition metals: An ab initio dynamical mean-field theory},\ }\href {https://doi.org/10.1103/PhysRevLett.87.067205} {\bibfield  {journal} {\bibinfo  {journal} {Phys. Rev. Lett.}\ }\textbf {\bibinfo {volume} {87}},\ \bibinfo {pages} {067205} (\bibinfo {year} {2001})}\BibitemShut {NoStop}%
\bibitem [{\citenamefont {Kotliar}\ \emph {et~al.}(2006)\citenamefont {Kotliar}, \citenamefont {Savrasov}, \citenamefont {Haule}, \citenamefont {Oudovenko}, \citenamefont {Parcollet},\ and\ \citenamefont {Marianetti}}]{kotliar:2006}%
  \BibitemOpen
  \bibfield  {author} {\bibinfo {author} {\bibfnamefont {G.}~\bibnamefont {Kotliar}}, \bibinfo {author} {\bibfnamefont {S.~Y.}\ \bibnamefont {Savrasov}}, \bibinfo {author} {\bibfnamefont {K.}~\bibnamefont {Haule}}, \bibinfo {author} {\bibfnamefont {V.~S.}\ \bibnamefont {Oudovenko}}, \bibinfo {author} {\bibfnamefont {O.}~\bibnamefont {Parcollet}},\ and\ \bibinfo {author} {\bibfnamefont {C.~A.}\ \bibnamefont {Marianetti}},\ }\bibfield  {title} {\bibinfo {title} {Electronic structure calculations with dynamical mean-field theory},\ }\href {https://doi.org/10.1103/RevModPhys.78.865} {\bibfield  {journal} {\bibinfo  {journal} {Rev. Mod. Phys.}\ }\textbf {\bibinfo {volume} {78}},\ \bibinfo {pages} {865} (\bibinfo {year} {2006})}\BibitemShut {NoStop}%
\bibitem [{\citenamefont {Haule}\ \emph {et~al.}(2010)\citenamefont {Haule}, \citenamefont {Yee},\ and\ \citenamefont {Kim}}]{Haule:2010}%
  \BibitemOpen
  \bibfield  {author} {\bibinfo {author} {\bibfnamefont {K.}~\bibnamefont {Haule}}, \bibinfo {author} {\bibfnamefont {C.-H.}\ \bibnamefont {Yee}},\ and\ \bibinfo {author} {\bibfnamefont {K.}~\bibnamefont {Kim}},\ }\bibfield  {title} {\bibinfo {title} {Dynamical mean-field theory within the full-potential methods: Electronic structure of {C}e{I}r{I}n$_{5}$, {C}e{C}o{I}n$_{5}$, and {C}e{R}h{I}n$_{5}$},\ }\href {https://doi.org/10.1103/PhysRevB.81.195107} {\bibfield  {journal} {\bibinfo  {journal} {Phys. Rev. B}\ }\textbf {\bibinfo {volume} {81}},\ \bibinfo {pages} {195107} (\bibinfo {year} {2010})}\BibitemShut {NoStop}%
\bibitem [{\citenamefont {Anisimov}\ \emph {et~al.}(2002)\citenamefont {Anisimov}, \citenamefont {Nekrasov}, \citenamefont {Kondakov}, \citenamefont {Rice},\ and\ \citenamefont {Sigrist}}]{Anisimov2002}%
  \BibitemOpen
  \bibfield  {author} {\bibinfo {author} {\bibfnamefont {V.~I.}\ \bibnamefont {Anisimov}}, \bibinfo {author} {\bibfnamefont {I.~A.}\ \bibnamefont {Nekrasov}}, \bibinfo {author} {\bibfnamefont {D.~E.}\ \bibnamefont {Kondakov}}, \bibinfo {author} {\bibfnamefont {T.~M.}\ \bibnamefont {Rice}},\ and\ \bibinfo {author} {\bibfnamefont {M.}~\bibnamefont {Sigrist}},\ }\bibfield  {title} {\bibinfo {title} {Orbital-selective mott-insulator transition in ca2 - xsrxruo4},\ }\href {https://doi.org/10.1140/epjb/e20020021} {\bibfield  {journal} {\bibinfo  {journal} {The European Physical Journal B - Condensed Matter and Complex Systems}\ }\textbf {\bibinfo {volume} {25}},\ \bibinfo {pages} {191} (\bibinfo {year} {2002})}\BibitemShut {NoStop}%
\bibitem [{\citenamefont {Liebsch}(2003)}]{Liebsch:2003}%
  \BibitemOpen
  \bibfield  {author} {\bibinfo {author} {\bibfnamefont {A.}~\bibnamefont {Liebsch}},\ }\bibfield  {title} {\bibinfo {title} {Mott transitions in multiorbital systems},\ }\href {https://doi.org/10.1103/PhysRevLett.91.226401} {\bibfield  {journal} {\bibinfo  {journal} {Phys. Rev. Lett.}\ }\textbf {\bibinfo {volume} {91}},\ \bibinfo {pages} {226401} (\bibinfo {year} {2003})}\BibitemShut {NoStop}%
\bibitem [{\citenamefont {Koga}\ \emph {et~al.}(2004)\citenamefont {Koga}, \citenamefont {Kawakami}, \citenamefont {Rice},\ and\ \citenamefont {Sigrist}}]{Koga:2004}%
  \BibitemOpen
  \bibfield  {author} {\bibinfo {author} {\bibfnamefont {A.}~\bibnamefont {Koga}}, \bibinfo {author} {\bibfnamefont {N.}~\bibnamefont {Kawakami}}, \bibinfo {author} {\bibfnamefont {T.~M.}\ \bibnamefont {Rice}},\ and\ \bibinfo {author} {\bibfnamefont {M.}~\bibnamefont {Sigrist}},\ }\bibfield  {title} {\bibinfo {title} {Orbital-selective mott transitions in the degenerate hubbard model},\ }\href {https://doi.org/10.1103/PhysRevLett.92.216402} {\bibfield  {journal} {\bibinfo  {journal} {Phys. Rev. Lett.}\ }\textbf {\bibinfo {volume} {92}},\ \bibinfo {pages} {216402} (\bibinfo {year} {2004})}\BibitemShut {NoStop}%
\bibitem [{\citenamefont {de'Medici}\ \emph {et~al.}(2005)\citenamefont {de'Medici}, \citenamefont {Georges},\ and\ \citenamefont {Biermann}}]{Medici:2005}%
  \BibitemOpen
  \bibfield  {author} {\bibinfo {author} {\bibfnamefont {L.}~\bibnamefont {de'Medici}}, \bibinfo {author} {\bibfnamefont {A.}~\bibnamefont {Georges}},\ and\ \bibinfo {author} {\bibfnamefont {S.}~\bibnamefont {Biermann}},\ }\bibfield  {title} {\bibinfo {title} {Orbital-selective mott transition in multiband systems: Slave-spin representation and dynamical mean-field theory},\ }\href {https://doi.org/10.1103/PhysRevB.72.205124} {\bibfield  {journal} {\bibinfo  {journal} {Phys. Rev. B}\ }\textbf {\bibinfo {volume} {72}},\ \bibinfo {pages} {205124} (\bibinfo {year} {2005})}\BibitemShut {NoStop}%
\bibitem [{\citenamefont {Ferrero}\ \emph {et~al.}(2005)\citenamefont {Ferrero}, \citenamefont {Becca}, \citenamefont {Fabrizio},\ and\ \citenamefont {Capone}}]{Ferrero:2005}%
  \BibitemOpen
  \bibfield  {author} {\bibinfo {author} {\bibfnamefont {M.}~\bibnamefont {Ferrero}}, \bibinfo {author} {\bibfnamefont {F.}~\bibnamefont {Becca}}, \bibinfo {author} {\bibfnamefont {M.}~\bibnamefont {Fabrizio}},\ and\ \bibinfo {author} {\bibfnamefont {M.}~\bibnamefont {Capone}},\ }\bibfield  {title} {\bibinfo {title} {Dynamical behavior across the mott transition of two bands with different bandwidths},\ }\href {https://doi.org/10.1103/PhysRevB.72.205126} {\bibfield  {journal} {\bibinfo  {journal} {Phys. Rev. B}\ }\textbf {\bibinfo {volume} {72}},\ \bibinfo {pages} {205126} (\bibinfo {year} {2005})}\BibitemShut {NoStop}%
\bibitem [{\citenamefont {Arita}\ and\ \citenamefont {Held}(2005)}]{Arita:2005}%
  \BibitemOpen
  \bibfield  {author} {\bibinfo {author} {\bibfnamefont {R.}~\bibnamefont {Arita}}\ and\ \bibinfo {author} {\bibfnamefont {K.}~\bibnamefont {Held}},\ }\bibfield  {title} {\bibinfo {title} {Orbital-selective mott-hubbard transition in the two-band hubbard model},\ }\href {https://doi.org/10.1103/PhysRevB.72.201102} {\bibfield  {journal} {\bibinfo  {journal} {Phys. Rev. B}\ }\textbf {\bibinfo {volume} {72}},\ \bibinfo {pages} {201102(R)} (\bibinfo {year} {2005})}\BibitemShut {NoStop}%
\bibitem [{\citenamefont {Knecht}\ \emph {et~al.}(2005)\citenamefont {Knecht}, \citenamefont {Bl\"umer},\ and\ \citenamefont {van Dongen}}]{Knecht:2005}%
  \BibitemOpen
  \bibfield  {author} {\bibinfo {author} {\bibfnamefont {C.}~\bibnamefont {Knecht}}, \bibinfo {author} {\bibfnamefont {N.}~\bibnamefont {Bl\"umer}},\ and\ \bibinfo {author} {\bibfnamefont {P.~G.~J.}\ \bibnamefont {van Dongen}},\ }\bibfield  {title} {\bibinfo {title} {Orbital-selective mott transitions in the anisotropic two-band hubbard model at finite temperatures},\ }\href {https://doi.org/10.1103/PhysRevB.72.081103} {\bibfield  {journal} {\bibinfo  {journal} {Phys. Rev. B}\ }\textbf {\bibinfo {volume} {72}},\ \bibinfo {pages} {081103(R)} (\bibinfo {year} {2005})}\BibitemShut {NoStop}%
\bibitem [{\citenamefont {Inaba}\ and\ \citenamefont {Koga}(2007)}]{Inaba:2007}%
  \BibitemOpen
  \bibfield  {author} {\bibinfo {author} {\bibfnamefont {K.}~\bibnamefont {Inaba}}\ and\ \bibinfo {author} {\bibfnamefont {A.}~\bibnamefont {Koga}},\ }\bibfield  {title} {\bibinfo {title} {Metal-insulator transition in the two-orbital hubbard model at fractional band fillings: Self-energy functional approach},\ }\href {https://doi.org/10.1143/JPSJ.76.094712} {\bibfield  {journal} {\bibinfo  {journal} {Journal of the Physical Society of Japan}\ }\textbf {\bibinfo {volume} {76}},\ \bibinfo {pages} {094712} (\bibinfo {year} {2007})}\BibitemShut {NoStop}%
\bibitem [{\citenamefont {Werner}\ and\ \citenamefont {Millis}(2007)}]{Werner:2007}%
  \BibitemOpen
  \bibfield  {author} {\bibinfo {author} {\bibfnamefont {P.}~\bibnamefont {Werner}}\ and\ \bibinfo {author} {\bibfnamefont {A.~J.}\ \bibnamefont {Millis}},\ }\bibfield  {title} {\bibinfo {title} {High-spin to low-spin and orbital polarization transitions in multiorbital mott systems},\ }\href {https://doi.org/10.1103/PhysRevLett.99.126405} {\bibfield  {journal} {\bibinfo  {journal} {Phys. Rev. Lett.}\ }\textbf {\bibinfo {volume} {99}},\ \bibinfo {pages} {126405} (\bibinfo {year} {2007})}\BibitemShut {NoStop}%
\bibitem [{\citenamefont {Jakobi}\ \emph {et~al.}(2009)\citenamefont {Jakobi}, \citenamefont {Bl\"umer},\ and\ \citenamefont {van Dongen}}]{Jakobi:2009}%
  \BibitemOpen
  \bibfield  {author} {\bibinfo {author} {\bibfnamefont {E.}~\bibnamefont {Jakobi}}, \bibinfo {author} {\bibfnamefont {N.}~\bibnamefont {Bl\"umer}},\ and\ \bibinfo {author} {\bibfnamefont {P.}~\bibnamefont {van Dongen}},\ }\bibfield  {title} {\bibinfo {title} {Orbital-selective mott transitions in a doped two-band hubbard model},\ }\href {https://doi.org/10.1103/PhysRevB.80.115109} {\bibfield  {journal} {\bibinfo  {journal} {Phys. Rev. B}\ }\textbf {\bibinfo {volume} {80}},\ \bibinfo {pages} {115109} (\bibinfo {year} {2009})}\BibitemShut {NoStop}%
\bibitem [{\citenamefont {de' Medici}\ \emph {et~al.}(2009)\citenamefont {de' Medici}, \citenamefont {Hassan}, \citenamefont {Capone},\ and\ \citenamefont {Dai}}]{Medici:2009}%
  \BibitemOpen
  \bibfield  {author} {\bibinfo {author} {\bibfnamefont {L.}~\bibnamefont {de' Medici}}, \bibinfo {author} {\bibfnamefont {S.~R.}\ \bibnamefont {Hassan}}, \bibinfo {author} {\bibfnamefont {M.}~\bibnamefont {Capone}},\ and\ \bibinfo {author} {\bibfnamefont {X.}~\bibnamefont {Dai}},\ }\bibfield  {title} {\bibinfo {title} {Orbital-selective mott transition out of band degeneracy lifting},\ }\href {https://doi.org/10.1103/PhysRevLett.102.126401} {\bibfield  {journal} {\bibinfo  {journal} {Phys. Rev. Lett.}\ }\textbf {\bibinfo {volume} {102}},\ \bibinfo {pages} {126401} (\bibinfo {year} {2009})}\BibitemShut {NoStop}%
\bibitem [{\citenamefont {Werner}\ \emph {et~al.}(2009)\citenamefont {Werner}, \citenamefont {Gull},\ and\ \citenamefont {Millis}}]{Werner:2009}%
  \BibitemOpen
  \bibfield  {author} {\bibinfo {author} {\bibfnamefont {P.}~\bibnamefont {Werner}}, \bibinfo {author} {\bibfnamefont {E.}~\bibnamefont {Gull}},\ and\ \bibinfo {author} {\bibfnamefont {A.~J.}\ \bibnamefont {Millis}},\ }\bibfield  {title} {\bibinfo {title} {Metal-insulator phase diagram and orbital selectivity in three-orbital models with rotationally invariant hund coupling},\ }\href {https://doi.org/10.1103/PhysRevB.79.115119} {\bibfield  {journal} {\bibinfo  {journal} {Phys. Rev. B}\ }\textbf {\bibinfo {volume} {79}},\ \bibinfo {pages} {115119} (\bibinfo {year} {2009})}\BibitemShut {NoStop}%
\bibitem [{\citenamefont {Kita}\ \emph {et~al.}(2011)\citenamefont {Kita}, \citenamefont {Ohashi},\ and\ \citenamefont {Kawakami}}]{Kita:2011}%
  \BibitemOpen
  \bibfield  {author} {\bibinfo {author} {\bibfnamefont {T.}~\bibnamefont {Kita}}, \bibinfo {author} {\bibfnamefont {T.}~\bibnamefont {Ohashi}},\ and\ \bibinfo {author} {\bibfnamefont {N.}~\bibnamefont {Kawakami}},\ }\bibfield  {title} {\bibinfo {title} {Mott transition in three-orbital hubbard model with orbital splitting},\ }\href {https://doi.org/10.1103/PhysRevB.84.195130} {\bibfield  {journal} {\bibinfo  {journal} {Phys. Rev. B}\ }\textbf {\bibinfo {volume} {84}},\ \bibinfo {pages} {195130} (\bibinfo {year} {2011})}\BibitemShut {NoStop}%
\bibitem [{\citenamefont {Huang}\ \emph {et~al.}(2012{\natexlab{a}})\citenamefont {Huang}, \citenamefont {Wang},\ and\ \citenamefont {Dai}}]{HuangLi:2012}%
  \BibitemOpen
  \bibfield  {author} {\bibinfo {author} {\bibfnamefont {L.}~\bibnamefont {Huang}}, \bibinfo {author} {\bibfnamefont {Y.}~\bibnamefont {Wang}},\ and\ \bibinfo {author} {\bibfnamefont {X.}~\bibnamefont {Dai}},\ }\bibfield  {title} {\bibinfo {title} {Pressure-driven orbital selective insulator-to-metal transition and spin-state crossover in cubic coo},\ }\href {https://doi.org/10.1103/PhysRevB.85.245110} {\bibfield  {journal} {\bibinfo  {journal} {Phys. Rev. B}\ }\textbf {\bibinfo {volume} {85}},\ \bibinfo {pages} {245110} (\bibinfo {year} {2012}{\natexlab{a}})}\BibitemShut {NoStop}%
\bibitem [{\citenamefont {Huang}\ \emph {et~al.}(2012{\natexlab{b}})\citenamefont {Huang}, \citenamefont {Du},\ and\ \citenamefont {Dai}}]{HuangLi_pd_2012}%
  \BibitemOpen
  \bibfield  {author} {\bibinfo {author} {\bibfnamefont {L.}~\bibnamefont {Huang}}, \bibinfo {author} {\bibfnamefont {L.}~\bibnamefont {Du}},\ and\ \bibinfo {author} {\bibfnamefont {X.}~\bibnamefont {Dai}},\ }\bibfield  {title} {\bibinfo {title} {Complete phase diagram for three-band hubbard model with orbital degeneracy lifted by crystal field splitting},\ }\href {https://doi.org/10.1103/PhysRevB.86.035150} {\bibfield  {journal} {\bibinfo  {journal} {Phys. Rev. B}\ }\textbf {\bibinfo {volume} {86}},\ \bibinfo {pages} {035150} (\bibinfo {year} {2012}{\natexlab{b}})}\BibitemShut {NoStop}%
\bibitem [{\citenamefont {Lanat\`a}\ \emph {et~al.}(2013)\citenamefont {Lanat\`a}, \citenamefont {Strand}, \citenamefont {Giovannetti}, \citenamefont {Hellsing}, \citenamefont {de' Medici},\ and\ \citenamefont {Capone}}]{Capone:2013}%
  \BibitemOpen
  \bibfield  {author} {\bibinfo {author} {\bibfnamefont {N.}~\bibnamefont {Lanat\`a}}, \bibinfo {author} {\bibfnamefont {H.~U.~R.}\ \bibnamefont {Strand}}, \bibinfo {author} {\bibfnamefont {G.}~\bibnamefont {Giovannetti}}, \bibinfo {author} {\bibfnamefont {B.}~\bibnamefont {Hellsing}}, \bibinfo {author} {\bibfnamefont {L.}~\bibnamefont {de' Medici}},\ and\ \bibinfo {author} {\bibfnamefont {M.}~\bibnamefont {Capone}},\ }\bibfield  {title} {\bibinfo {title} {Orbital selectivity in hund's metals: The iron chalcogenides},\ }\href {https://doi.org/10.1103/PhysRevB.87.045122} {\bibfield  {journal} {\bibinfo  {journal} {Phys. Rev. B}\ }\textbf {\bibinfo {volume} {87}},\ \bibinfo {pages} {045122} (\bibinfo {year} {2013})}\BibitemShut {NoStop}%
\bibitem [{\citenamefont {Georges}\ \emph {et~al.}(2013)\citenamefont {Georges}, \citenamefont {Medici},\ and\ \citenamefont {Mravlje}}]{georges:2013}%
  \BibitemOpen
  \bibfield  {author} {\bibinfo {author} {\bibfnamefont {A.}~\bibnamefont {Georges}}, \bibinfo {author} {\bibfnamefont {L.~d.}\ \bibnamefont {Medici}},\ and\ \bibinfo {author} {\bibfnamefont {J.}~\bibnamefont {Mravlje}},\ }\bibfield  {title} {\bibinfo {title} {Strong correlations from {H}und’s coupling},\ }\href {https://doi.org/10.1146/annurev-conmatphys-020911-125045} {\bibfield  {journal} {\bibinfo  {journal} {Annual Review of Condensed Matter Physics}\ }\textbf {\bibinfo {volume} {4}},\ \bibinfo {pages} {137} (\bibinfo {year} {2013})}\BibitemShut {NoStop}%
\bibitem [{\citenamefont {Rinc\'on}\ \emph {et~al.}(2014)\citenamefont {Rinc\'on}, \citenamefont {Moreo}, \citenamefont {Alvarez},\ and\ \citenamefont {Dagotto}}]{Rincon:2014}%
  \BibitemOpen
  \bibfield  {author} {\bibinfo {author} {\bibfnamefont {J.}~\bibnamefont {Rinc\'on}}, \bibinfo {author} {\bibfnamefont {A.}~\bibnamefont {Moreo}}, \bibinfo {author} {\bibfnamefont {G.}~\bibnamefont {Alvarez}},\ and\ \bibinfo {author} {\bibfnamefont {E.}~\bibnamefont {Dagotto}},\ }\bibfield  {title} {\bibinfo {title} {Quantum phase transition between orbital-selective mott states in hund's metals},\ }\href {https://doi.org/10.1103/PhysRevB.90.241105} {\bibfield  {journal} {\bibinfo  {journal} {Phys. Rev. B}\ }\textbf {\bibinfo {volume} {90}},\ \bibinfo {pages} {241105(R)} (\bibinfo {year} {2014})}\BibitemShut {NoStop}%
\bibitem [{\citenamefont {Wang}\ \emph {et~al.}(2016)\citenamefont {Wang}, \citenamefont {Huang}, \citenamefont {Du},\ and\ \citenamefont {Dai}}]{Wang_2016}%
  \BibitemOpen
  \bibfield  {author} {\bibinfo {author} {\bibfnamefont {Y.}~\bibnamefont {Wang}}, \bibinfo {author} {\bibfnamefont {L.}~\bibnamefont {Huang}}, \bibinfo {author} {\bibfnamefont {L.}~\bibnamefont {Du}},\ and\ \bibinfo {author} {\bibfnamefont {X.}~\bibnamefont {Dai}},\ }\bibfield  {title} {\bibinfo {title} {Doping-driven orbital-selective mott transition in multi-band hubbard models with crystal field splitting*},\ }\href {https://doi.org/10.1088/1674-1056/25/3/037103} {\bibfield  {journal} {\bibinfo  {journal} {Chinese Physics B}\ }\textbf {\bibinfo {volume} {25}},\ \bibinfo {pages} {037103} (\bibinfo {year} {2016})}\BibitemShut {NoStop}%
\bibitem [{\citenamefont {Yu}\ and\ \citenamefont {Si}(2017)}]{Yurong:2017}%
  \BibitemOpen
  \bibfield  {author} {\bibinfo {author} {\bibfnamefont {R.}~\bibnamefont {Yu}}\ and\ \bibinfo {author} {\bibfnamefont {Q.}~\bibnamefont {Si}},\ }\bibfield  {title} {\bibinfo {title} {Orbital-selective mott phase in multiorbital models for iron pnictides and chalcogenides},\ }\href {https://doi.org/10.1103/PhysRevB.96.125110} {\bibfield  {journal} {\bibinfo  {journal} {Phys. Rev. B}\ }\textbf {\bibinfo {volume} {96}},\ \bibinfo {pages} {125110} (\bibinfo {year} {2017})}\BibitemShut {NoStop}%
\bibitem [{\citenamefont {Kugler}\ \emph {et~al.}(2019)\citenamefont {Kugler}, \citenamefont {Lee}, \citenamefont {Weichselbaum}, \citenamefont {Kotliar},\ and\ \citenamefont {von Delft}}]{Kugler:2019}%
  \BibitemOpen
  \bibfield  {author} {\bibinfo {author} {\bibfnamefont {F.~B.}\ \bibnamefont {Kugler}}, \bibinfo {author} {\bibfnamefont {S.-S.~B.}\ \bibnamefont {Lee}}, \bibinfo {author} {\bibfnamefont {A.}~\bibnamefont {Weichselbaum}}, \bibinfo {author} {\bibfnamefont {G.}~\bibnamefont {Kotliar}},\ and\ \bibinfo {author} {\bibfnamefont {J.}~\bibnamefont {von Delft}},\ }\bibfield  {title} {\bibinfo {title} {Orbital differentiation in hund metals},\ }\href {https://doi.org/10.1103/PhysRevB.100.115159} {\bibfield  {journal} {\bibinfo  {journal} {Phys. Rev. B}\ }\textbf {\bibinfo {volume} {100}},\ \bibinfo {pages} {115159} (\bibinfo {year} {2019})}\BibitemShut {NoStop}%
\bibitem [{\citenamefont {Kugler}\ and\ \citenamefont {Kotliar}(2022)}]{Kugler:2022}%
  \BibitemOpen
  \bibfield  {author} {\bibinfo {author} {\bibfnamefont {F.~B.}\ \bibnamefont {Kugler}}\ and\ \bibinfo {author} {\bibfnamefont {G.}~\bibnamefont {Kotliar}},\ }\bibfield  {title} {\bibinfo {title} {Is the orbital-selective mott phase stable against interorbital hopping?},\ }\href {https://doi.org/10.1103/PhysRevLett.129.096403} {\bibfield  {journal} {\bibinfo  {journal} {Phys. Rev. Lett.}\ }\textbf {\bibinfo {volume} {129}},\ \bibinfo {pages} {096403} (\bibinfo {year} {2022})}\BibitemShut {NoStop}%
\bibitem [{\citenamefont {Kim}\ \emph {et~al.}(2024)\citenamefont {Kim}, \citenamefont {Choi}, \citenamefont {Brito},\ and\ \citenamefont {Kotliar}}]{Minjae:2024}%
  \BibitemOpen
  \bibfield  {author} {\bibinfo {author} {\bibfnamefont {M.}~\bibnamefont {Kim}}, \bibinfo {author} {\bibfnamefont {S.}~\bibnamefont {Choi}}, \bibinfo {author} {\bibfnamefont {W.~H.}\ \bibnamefont {Brito}},\ and\ \bibinfo {author} {\bibfnamefont {G.}~\bibnamefont {Kotliar}},\ }\bibfield  {title} {\bibinfo {title} {Orbital-selective mott transition effects and nontrivial topology of iron chalcogenide},\ }\href {https://doi.org/10.1103/PhysRevLett.132.136504} {\bibfield  {journal} {\bibinfo  {journal} {Phys. Rev. Lett.}\ }\textbf {\bibinfo {volume} {132}},\ \bibinfo {pages} {136504} (\bibinfo {year} {2024})}\BibitemShut {NoStop}%
\bibitem [{sup()}]{suppl}%
  \BibitemOpen
  \href@noop {} {\bibinfo  {journal} {See Supporting Information at [url] for computational details and more results. The Supplementary Materials include Refs.~\cite{perdew:1996,kresse:1996,kresse:1999,blochl:1994,Anisimov:1997,Blaha:2020,Haule:2015free,Haule:2015,Gull:2011}}\ }\BibitemShut {NoStop}%
\bibitem [{\citenamefont {Brese}\ and\ \citenamefont {O'Keeffe}(1991)}]{Brese:st0462}%
  \BibitemOpen
\bibfield  {journal} {  }\bibfield  {author} {\bibinfo {author} {\bibfnamefont {N.~E.}\ \bibnamefont {Brese}}\ and\ \bibinfo {author} {\bibfnamefont {M.}~\bibnamefont {O'Keeffe}},\ }\bibfield  {title} {\bibinfo {title} {{Bond-valence parameters for solids}},\ }\href {https://doi.org/10.1107/S0108768190011041} {\bibfield  {journal} {\bibinfo  {journal} {Acta Crystallographica Section B}\ }\textbf {\bibinfo {volume} {47}},\ \bibinfo {pages} {192} (\bibinfo {year} {1991})}\BibitemShut {NoStop}%
\bibitem [{\citenamefont {Perdew}\ \emph {et~al.}(1996)\citenamefont {Perdew}, \citenamefont {Burke},\ and\ \citenamefont {Ernzerhof}}]{perdew:1996}%
  \BibitemOpen
  \bibfield  {author} {\bibinfo {author} {\bibfnamefont {J.~P.}\ \bibnamefont {Perdew}}, \bibinfo {author} {\bibfnamefont {K.}~\bibnamefont {Burke}},\ and\ \bibinfo {author} {\bibfnamefont {M.}~\bibnamefont {Ernzerhof}},\ }\bibfield  {title} {\bibinfo {title} {Generalized gradient approximation made simple},\ }\href {https://doi.org/10.1103/PhysRevLett.77.3865} {\bibfield  {journal} {\bibinfo  {journal} {Phys. Rev. Lett.}\ }\textbf {\bibinfo {volume} {77}},\ \bibinfo {pages} {3865} (\bibinfo {year} {1996})}\BibitemShut {NoStop}%
\bibitem [{\citenamefont {Kresse}\ and\ \citenamefont {Furthm\"uller}(1996)}]{kresse:1996}%
  \BibitemOpen
  \bibfield  {author} {\bibinfo {author} {\bibfnamefont {G.}~\bibnamefont {Kresse}}\ and\ \bibinfo {author} {\bibfnamefont {J.}~\bibnamefont {Furthm\"uller}},\ }\bibfield  {title} {\bibinfo {title} {Efficient iterative schemes for \textit{ab initio} total-energy calculations using a plane-wave basis set},\ }\href {https://doi.org/10.1103/PhysRevB.54.11169} {\bibfield  {journal} {\bibinfo  {journal} {Phys. Rev. B}\ }\textbf {\bibinfo {volume} {54}},\ \bibinfo {pages} {11169} (\bibinfo {year} {1996})}\BibitemShut {NoStop}%
\bibitem [{\citenamefont {Kresse}\ and\ \citenamefont {Joubert}(1999)}]{kresse:1999}%
  \BibitemOpen
  \bibfield  {author} {\bibinfo {author} {\bibfnamefont {G.}~\bibnamefont {Kresse}}\ and\ \bibinfo {author} {\bibfnamefont {D.}~\bibnamefont {Joubert}},\ }\bibfield  {title} {\bibinfo {title} {From ultrasoft pseudopotentials to the projector augmented-wave method},\ }\href {https://doi.org/10.1103/PhysRevB.59.1758} {\bibfield  {journal} {\bibinfo  {journal} {Phys. Rev. B}\ }\textbf {\bibinfo {volume} {59}},\ \bibinfo {pages} {1758} (\bibinfo {year} {1999})}\BibitemShut {NoStop}%
\bibitem [{\citenamefont {Bl\"ochl}(1994)}]{blochl:1994}%
  \BibitemOpen
  \bibfield  {author} {\bibinfo {author} {\bibfnamefont {P.~E.}\ \bibnamefont {Bl\"ochl}},\ }\bibfield  {title} {\bibinfo {title} {Projector augmented-wave method},\ }\href {https://doi.org/10.1103/PhysRevB.50.17953} {\bibfield  {journal} {\bibinfo  {journal} {Phys. Rev. B}\ }\textbf {\bibinfo {volume} {50}},\ \bibinfo {pages} {17953} (\bibinfo {year} {1994})}\BibitemShut {NoStop}%
\bibitem [{\citenamefont {Anisimov}\ \emph {et~al.}(1997)\citenamefont {Anisimov}, \citenamefont {Aryasetiawan},\ and\ \citenamefont {Lichtenstein}}]{Anisimov:1997}%
  \BibitemOpen
  \bibfield  {author} {\bibinfo {author} {\bibfnamefont {V.~I.}\ \bibnamefont {Anisimov}}, \bibinfo {author} {\bibfnamefont {F.}~\bibnamefont {Aryasetiawan}},\ and\ \bibinfo {author} {\bibfnamefont {A.~I.}\ \bibnamefont {Lichtenstein}},\ }\bibfield  {title} {\bibinfo {title} {First-principles calculations of the electronic structure and spectra of strongly correlated systems: the {LDA}+u method},\ }\href {https://doi.org/10.1088/0953-8984/9/4/002} {\bibfield  {journal} {\bibinfo  {journal} {Journal of Physics: Condensed Matter}\ }\textbf {\bibinfo {volume} {9}},\ \bibinfo {pages} {767} (\bibinfo {year} {1997})}\BibitemShut {NoStop}%
\bibitem [{\citenamefont {Blaha}\ \emph {et~al.}(2020)\citenamefont {Blaha}, \citenamefont {Schwarz}, \citenamefont {Tran}, \citenamefont {Laskowski}, \citenamefont {Madsen},\ and\ \citenamefont {Marks}}]{Blaha:2020}%
  \BibitemOpen
  \bibfield  {author} {\bibinfo {author} {\bibfnamefont {P.}~\bibnamefont {Blaha}}, \bibinfo {author} {\bibfnamefont {K.}~\bibnamefont {Schwarz}}, \bibinfo {author} {\bibfnamefont {F.}~\bibnamefont {Tran}}, \bibinfo {author} {\bibfnamefont {R.}~\bibnamefont {Laskowski}}, \bibinfo {author} {\bibfnamefont {G.~K.~H.}\ \bibnamefont {Madsen}},\ and\ \bibinfo {author} {\bibfnamefont {L.~D.}\ \bibnamefont {Marks}},\ }\bibfield  {title} {\bibinfo {title} {{WIEN}2k: An {APW}+lo program for calculating the properties of solids},\ }\href {https://doi.org/10.1063/1.5143061} {\bibfield  {journal} {\bibinfo  {journal} {The Journal of Chemical Physics}\ }\textbf {\bibinfo {volume} {152}},\ \bibinfo {pages} {074101} (\bibinfo {year} {2020})}\BibitemShut {NoStop}%
\bibitem [{\citenamefont {Haule}\ and\ \citenamefont {Birol}(2015)}]{Haule:2015free}%
  \BibitemOpen
  \bibfield  {author} {\bibinfo {author} {\bibfnamefont {K.}~\bibnamefont {Haule}}\ and\ \bibinfo {author} {\bibfnamefont {T.}~\bibnamefont {Birol}},\ }\bibfield  {title} {\bibinfo {title} {Free energy from stationary implementation of the $\mathrm{DFT}+\mathrm{DMFT}$ functional},\ }\href {https://doi.org/10.1103/PhysRevLett.115.256402} {\bibfield  {journal} {\bibinfo  {journal} {Phys. Rev. Lett.}\ }\textbf {\bibinfo {volume} {115}},\ \bibinfo {pages} {256402} (\bibinfo {year} {2015})}\BibitemShut {NoStop}%
\bibitem [{\citenamefont {Haule}(2015)}]{Haule:2015}%
  \BibitemOpen
  \bibfield  {author} {\bibinfo {author} {\bibfnamefont {K.}~\bibnamefont {Haule}},\ }\bibfield  {title} {\bibinfo {title} {Exact double counting in combining the dynamical mean field theory and the density functional theory},\ }\href {https://doi.org/10.1103/PhysRevLett.115.196403} {\bibfield  {journal} {\bibinfo  {journal} {Phys. Rev. Lett.}\ }\textbf {\bibinfo {volume} {115}},\ \bibinfo {pages} {196403} (\bibinfo {year} {2015})}\BibitemShut {NoStop}%
\bibitem [{\citenamefont {Gull}\ \emph {et~al.}(2011)\citenamefont {Gull}, \citenamefont {Millis}, \citenamefont {Lichtenstein}, \citenamefont {Rubtsov}, \citenamefont {Troyer},\ and\ \citenamefont {Werner}}]{Gull:2011}%
  \BibitemOpen
  \bibfield  {author} {\bibinfo {author} {\bibfnamefont {E.}~\bibnamefont {Gull}}, \bibinfo {author} {\bibfnamefont {A.~J.}\ \bibnamefont {Millis}}, \bibinfo {author} {\bibfnamefont {A.~I.}\ \bibnamefont {Lichtenstein}}, \bibinfo {author} {\bibfnamefont {A.~N.}\ \bibnamefont {Rubtsov}}, \bibinfo {author} {\bibfnamefont {M.}~\bibnamefont {Troyer}},\ and\ \bibinfo {author} {\bibfnamefont {P.}~\bibnamefont {Werner}},\ }\bibfield  {title} {\bibinfo {title} {Continuous-time monte carlo methods for quantum impurity models},\ }\href {https://doi.org/10.1103/RevModPhys.83.349} {\bibfield  {journal} {\bibinfo  {journal} {Rev. Mod. Phys.}\ }\textbf {\bibinfo {volume} {83}},\ \bibinfo {pages} {349} (\bibinfo {year} {2011})}\BibitemShut {NoStop}%
\end{thebibliography}%

\end{document}